\documentclass[english,nofootinbib, aps, floatfix, superscriptaddress, prd, twocolumn]{revtex4-1}
\usepackage[T1]{fontenc}
\usepackage[latin9]{inputenc}
\usepackage{float}
\usepackage{bm}
\usepackage{amsmath}
\usepackage{amssymb}
\usepackage{graphicx}
\global\long\def\QED{\operatorname{QED_{3}}}

\global\long\def\QEDGN{\operatorname{QED_{3}-GN}}
\global\long\def\QEDZ{\operatorname{QED_{3}-Z_{2}GN}}
\global\long\def\QEDcHGN{\operatorname{QED_{3}-cHGN}}
\global\long\def\GN{\operatorname{GN}}
\global\long\def\ON{\operatorname{O}(N)}
\global\long\def\CP{\operatorname{CP}}

\global\long\def\cHGN{\operatorname{cHGN}}
\global\long\def\CFT{\operatorname{CFT}}
\global\long\def\QFT{\operatorname{QFT}}

\global\long\def\eff{\operatorname{eff}}
\global\long\def\OO{\operatorname{O}}
\global\long\def\SU{\operatorname{SU}}
\global\long\def\U{\operatorname{U}}
\global\long\def\SO{\operatorname{SO}}
\global\long\def\OPE{\operatorname{OPE}}
\global\long\def\RR{\mathcal{\mathbb{R}}}

\makeatletter

\newcommand{\lyxmathsym}[1]{\ifmmode\begingroup\def\b@ld{bold}
  \text{\ifx\math@version\b@ld\bfseries\fi#1}\endgroup\else#1\fi}

\providecommand{\tabularnewline}{\\}

\bibpunct{[}{]}{;}{n}{}{}
\usepackage{amsfonts}
\usepackage{amsmath}
\usepackage{subfigure}

\usepackage{physics}
\usepackage{slashed}
\usepackage{bbm}
\numberwithin{equation}{section}

\usepackage[colorlinks=true,citecolor=blue,
linkcolor=blue,urlcolor=blue]{hyperref}

\usepackage{comment}

\makeatother

\usepackage{babel}
\begin{document}

\title{Anomalous dimensions of monopole operators at the transitions between
Dirac and topological spin liquids}

\author{\'Eric Dupuis}

\affiliation{D\'epartement de physique, Universit\'e de Montr\'eal, Montr\'eal (Qu\'ebec), H3C 3J7, Canada}

\author{Rufus Boyack}

\affiliation{D\'epartement de physique, Universit\'e de Montr\'eal, Montr\'eal (Qu\'ebec), H3C 3J7, Canada}

\author{William Witczak-Krempa}

\affiliation{D\'epartement de physique, Universit\'e de Montr\'eal, Montr\'eal (Qu\'ebec), H3C 3J7, Canada}
\affiliation{Institut Courtois, Universit\'e de Montr\'eal, Montr\'eal (Qu\'ebec), H2V 0B3, Canada}

\affiliation{Centre de Recherches Math\'ematiques, Universit\'e de Montr\'eal; P.O. Box 6128, Centre-ville Station; Montr\'eal (Qu\'ebec), H3C 3J7, Canada}
\begin{abstract}
Monopole operators are studied in a large family of quantum critical points between
 Dirac and topological quantum spin liquids (QSLs): chiral and $Z_2$ QSLs. These quantum phase transitions are described by conformal field theories (CFTs): quantum
electrodynamics in 2+1 dimensions with $2N$ flavors of two-component
massless Dirac fermions and a four-fermion interaction. 
For the transition to a chiral spin liquid, it is the Gross-Neveu interaction
($\QEDGN$), while for the  transitions to $Z_2$ QSLs it is a superconducting pairing term with general spin/valley structure (generalized $\QEDZ$). Using the state-operator correspondence, we obtain monopole
scaling dimensions to sub-leading order in $1/N$. For monopoles with
a minimal topological charge $q=1/2$, the scaling dimension is
$2N\times0.26510$ at leading-order, with the quantum correction being $0.118911(7)$ for the chiral spin liquid, and $0.102846(9)$ for the simplest $Z_2$ case (the expression is also given for a general pairing term). Although these two anomalous dimensions are nearly equal, the underlying quantum fluctuations possess distinct origins.
The analogous result in $\QED$ is also obtained and we find a sub-leading
contribution of $-0.038138(5)$, which differs slightly from the
value first obtained in the literature. 
The scaling dimension
of a $\QEDGN$ monopole with minimal charge is very close to the scaling
dimensions of other operators predicted to be equal by a conjectured
duality between $\QEDGN$ with $2N=2$ flavors and the $\CP^{1}$
model. Additionally, non-minimally charged monopoles 
on both sides of the duality have similar scaling dimensions. By studying
the large-$q$ asymptotics of the scaling dimensions in $\QED$, $\QEDGN$, and $\QEDZ$ 
we verify that the constant $O(q^{0})$ coefficient
precisely matches the universal non-perturbative prediction for CFTs with a global
$\U(1)$ symmetry. 
\end{abstract}
\maketitle
\tableofcontents{}

\section{Introduction}

Gauge theories play an important role in modern condensed matter
physics, in part due to their ability to provide a low-energy description
of many quantum phases of matter. Gauge fields emerge as collective
excitations that capture the highly
entangled nature of certain strongly correlated systems. This is notably apparent
in the case of frustrated two-dimensional magnets hosting quantum
spin liquids and deconfined quantum critical points (dQCPs).

In these lattice systems, the emergent gauge field is compact, and,
as a result, has topological excitations created by topological
disorder operators. For a $\U(1)$ gauge field, these objects
are called monopole or instanton operators and they play an essential role in many
physical systems. Crucially, monopole proliferation
confines the gauge field. This is the case in the pure $\U(1)$
gauge theory~\citep{polyakov_compact_1975,polyakov_quark_1977}.
In the presence of massless matter, however, monopoles are screened
and confinement can be avoided if enough flavors of massless matter
are present.

In particular, we will be first considering a transition from a $\U(1)$
Dirac spin liquid (DSL), which is described by $\QED$ with $2N$
flavors of massless two-component Dirac fermions. Realizations of
the $\U(1)$ DSL were formulated for the Kagome Heisenberg
spin-$1/2$ magnet~\citep{hastings_dirac_2000, ran_projected-wave-function_2007, iqbal_projected_2011,  iqbal_gapless_2013, he_signatures_2017,liao_gapless_2017}
and the $J_{1}-J_{2}$ spin-$1/2$ model on the triangular lattice~\citep{kaneko_gapless_2014,iqbal_triangular_2016, hu_dirac_2019}
with $2N=4$ flavors. Dirac spin-orbital liquid with effective spin
$j=3/2$ and $2N=8$ flavors have also been formulated for quantum
magnets on honeycomb~\citep{corboz_spin_2012,calvera_theory_2021}
and triangular~\citep{calvera_theory_2020} lattices. For a large
number of fermion flavors $2N$, it has been shown
through a $1/N$ expansion that monopole operators are irrelevant~\citep{borokhov_topological_2003},
and thus the $\U(1)$ DSL is stable in this limit. Taking
into account next-to-leading order corrections~\cite{pufu_anomalous_2014},
the critical number of fermion flavors was estimated to be $2N_{c}=12$,
beyond which minimally charged monopoles become irrelevant. This result
was confirmed by Monte Carlo computations~\citep{karthik_numerical_2019}
and is consistent with conformal bootstrap bounds~\citep{chester_anomalous_2016,chester_monopole_2018}.
The addition of disorder renders the model more unstable~\citep{dey_destabilization_2020}.

Monopole operators also serve as order parameters in neighboring phases.
For instance, in the $\CP^{1}$ model, which describes the transition
between an antiferromagnetic (AFM) phase and a valence bond solid
(VBS), there are monopoles with lattice quantum numbers and their
condensation results in VBS order~\citep{read_valence_1989,read_spin_1990,senthil_quantum_2004,senthil_deconfined_2005}. It is in this model where the scaling dimension of monopole operators were first obtained~\citep{murthy_action_1990}. Monopoles are also crucial in the $\U(1)$ DSL, a parent
state for many spin liquids. In this fermionic theory, monopoles can
carry different quantum numbers due to the existence of fermion zero
modes~\citep{atiyah_index_1963} which may dress monopoles in various
ways. Monopoles describe various VBS and AFM orders, depending on
which lattice the $\U(1)$ DSL is formulated~\citep{song_unifying_2019}.
By tuning a flavor-dependent Gross-Neveu ($\GN$) interaction, a fermion
mass is generated and monopoles with specific quantum numbers condense~\citep{ghaemi_neel_2006}.
In particular, the $\U(1)$ DSL on the Kagome lattice orders
to an antiferromagnetic $120^{\circ}$ coplanar order as monopoles
dressed with a magnetic spin polarization condense~\citep{hermele_properties_2008,lu_unification_2017}.
This confinement-deconfinement transition is described by the $\QED\!\operatorname{-}\!$ 
chiral Heisenberg $\GN$ model ($\QEDcHGN$), and the scaling
dimensions of monopole operators at the quantum critical point (QCP)
were obtained in Ref.~\citep{dupuis_transition_2019}. In this case,
the activation of the $\cHGN$ interaction breaks the flavor symmetry,
resulting in a hierarchy among monopoles where the scaling dimension
depends on the total magnetic spin of a monopole~\citep{dupuis_proc_2021,dupuis_monopole_2021}.

The quantum criticality of the dQCP and the $\U(1)$
DSL with $2N=4$ fermions were recently given a precise relation.
It was shown that they can be formulated as so-called Stiefel liquids,
which are related to non-linear sigma models in $2+1$ dimensions
with target manifolds $\SO\left(n\right)/\SO\left(4\right)$, where
$n=5$ and $n=6$ for the dQCP and the $\U(1)$ DSL respectively~\citep{zou_2021_stiefel}.
Higher values are conjectured to realize non-lagrangian critical systems,
for instance realizing a phase between a non-coplanar magnet and a
VBS order when $n=7$.

In this work, we will focus on transitions from the $\U(1)$
DSL to two topological quantum spin liquids (QSL): a chiral spin liquid (CSL) as mentioned above, and a general type of $Z_2$ QSL. The first transition is
described by $\QEDGN$~\citep{bhattacharjee_kagome_2015,janssen_critical_2017},
where the CSL results from the condensation of a symmetric fermion mass induced
by the $\GN$ interaction. This transition can be realized for the
Kagome~\citep{he_chiral_2014,gong_emergent_2014,he_distinct_2015,zhu_chiral_2015}
and triangular~\citep{hu_variational_2016,gong_global_2017,wietek_chiral_2017}
Heisenberg magnets with $2N=4$ Dirac cones. The QCP for the non-compact
$\QEDGN$ has been studied in Refs.~\citep*{janssen_critical_2017,bernhard_deconfined_2018,gracey_fermion_2018,zerf_critical_2018,boyack_deconfined_2019,benvenuti_easy_2019}.
Even though a mass gap is condensed in the CSL, there is no confinement-deconfinement
transition taking place in the compact theory. Despite removing the
screening effect of gapless modes, the symmetric condensed mass induces
a Chern-Simons term in the infrared limit which gaps the monopoles
and prevents their proliferation. The spinons in the gapped CSL thus remain
deconfined. The CSL is a topologically ordered state that breaks time-reversal symmetry, and has robust chiral edge modes. 

In contrast, the non-chiral $Z_2$ QSL is obtained when the fermionic spinons undergo a pairing instability to a gapped $s$-wave superconducting state. The U(1) gauge field is gapped through the Higgs mechanism, and gives place to a discrete $Z_2$ gauge field. In this case, fractionalization remains intact. For the simplest case where the pairing interaction is diagonal in flavor space, the corresponding quantum phase transition was studied in Refs.~\citep*{zerf_superconducting_2016,boyack_transition_2018}. Earlier studies~\cite{lu_unification_2017,lu_Z2SL_2011} qualitatively described how a Z$_{2}$ QSL can be obtained from a Dirac QSL through a superconducting transition for the fermions, albeit without a fluctuating scalar field (Cooper pair field). In addition, Ref.~\cite{Roy_supercond_2013} studied a similar model in the context of superconducting criticality in topological insulators. Interestingly, it turns out that, at leading-order in $1/N$, the monopoles have the same scaling dimension at both QCPs as in the $\U(1)$ DSL~\citep{boyack_transition_2018,dupuis_transition_2019}. In this work, we obtain the next-to-leading order correction to monopole scaling dimensions at those QCPs. Futhermore, we also consider a more general class of $Z_{2}$ QSLs where the pairing interaction is not the same for all spin and valley degrees of freedom. We compute the anomalous dimension for the general $Z_{2}$ QSLs, and we also determine their bandstructure and Chern number inside the spin liquid phase.

This study is also motivated by the duality between $\QEDGN$ with
$2N=2$ fermion flavors and $\text{CP}^{N-1}$ with $N=2$ complex
boson flavors conjectured in Ref.~\citep{wang_deconfined_2018} and
further studied in Ref.~\citep{bernhard_deconfined_2018}. This duality
can be checked by comparing the scaling of monopole operators with
various scaling dimensions that are predicted to be equal according
to this duality. The good agreement obtained in the LO result~\citep{dupuis_transition_2019}
is further improved by the scaling dimension correction we obtain
here for the $\QEDGN$ monopoles.

The paper is organized as follows. In the next section, we present
the $\QEDGN$ model and show how the state-operator correspondence
is used to obtain monopole scaling dimensions. In Sec.~\ref{sec:N_infty},
the leading-order computation presented in Ref.~\citep{dupuis_transition_2019}
is reviewed. In Sec.~\ref{sec:-corrections}, $1/N$ corrections
to monopole scaling dimensions are computed. We also verify that the scaling dimensions satisfy a conjectured convexity property. In Sec.~\ref{sec:large-charge},
we compare our results with the large-charge expansion obtained in Ref.~\citep{hellerman_on_2015} for CFTs. In Sec.~\ref{sec:CFTDuality}, we study the $\QEDGN$ $\Leftrightarrow$ $\CP^{1}$ duality~\citep{wang_deconfined_2018}.
In Sec.~\ref{sec:Z2}, we study monopole scaling dimensions at the transition to a $Z_2$ QSL, and obtain distinct values compared to the CSL.
In Sec.~\ref{sec:phases}, we briefly discuss other phase transitions
that could be studied with this formalism, including the $\QED\!\operatorname{-}\!{\U(N)\times\U(N)}\GN$, {$\QED\!\operatorname{-}\!\text{chiral XY} \GN$}, and {$\QEDcHGN$} QCPs. We conclude with a discussion
of our results and an outlook. In Appendix~\ref{app:non_compact},
we review the phase transition from the $\U(1)$ DSL to
the CSL in the non-compact model. In Apps.~\ref{app:scalar_gauge} and \ref{app:gauge_invariance},
we give more details regarding how the kernels appearing in Sec.~\ref{sec:-corrections}
are obtained and simplified with gauge invariance. The expansion of
these kernels in terms of harmonics is detailed in Apps.~\ref{sec:Green}\ and\ \ref{sec:eigenkernels}.
We give detailed simplifications of the kernels used for the case
of minimally charge monopoles in App.\ \ref{sec:min_charge_results}.
In App.~\ref{sec:remainders}, some remainder coefficients used
to analytically approximate sums over angular momenta are
shown. In App.~\ref{sec:Only-zero-modes}, we show how some contributions
of fermion zero modes neglected in the main text vanish. In App.\ \ref{sec:fitting},
the fitting procedure used to alleviate finite-size effects when computing
monopole anomalous dimensions are described. In App.~\ref{sec:scaling_13},
we list monopole anomalous dimensions in $\QED$, $\QEDGN$, and $\QEDZ$ for
topological charges up to $q=13$. 

\section{Monopoles at transition between Dirac \& chiral spin liquids\label{sec:model}}

The action of the $\QEDGN$ model in euclidean flat spacetime is given
by 
\begin{equation}
S=\!\int\! d^{3}r\left[-\overline{\Psi}\gamma^{\mu}\left(\partial_{\mu}-iA_{\mu}\right)\Psi-\frac{h^{2}}{2}\left(\overline{\Psi}\Psi\right)^{2}\right]+\dotsb,\label{eq:GN_action}
\end{equation}
where $\Psi$ is a $2N$ flavor spinor $\Psi=\left(\psi_{1},\psi_{2},\dots\psi_{2N}\right)^{\intercal}$with
each flavor $\psi_{i}$ being a two-component Dirac fermion. For certain
quantum magnets, where fermions emerge as fractionalized quasiparticles,
the $2N$ flavors are related to two magnetic spin polarizations $s=\uparrow,\downarrow$
and $N$ valley nodes per spin, $v=1,\dots,N$. Typical quantum magnets
have $N=2$ or $4$ nodes, but here we keep $N$ general and use it
as an expansion parameter. The adjoint spinor is given by $\overline{\Psi}=\Psi^{\dagger}\gamma_{0}$,
where the gamma matrices are defined in terms of the Pauli matrices by $\gamma_{x,y} = \sigma_{x,y}$, and $\gamma_{0}=\sigma_{z}$. The fermions are coupled to a compact $\U(1)$
gauge field $A_{\mu}$, and have a $\GN$ self-interaction with coupling
strength $h$. The ellipsis denotes an irrelevant Maxwell term and
the contribution of monopole operators $\mathcal{M}_{q}(x)$
that we discuss further in what follows.

In $2+1$ dimensions, $\U(1)$ gauge theories have an extra
global $\U_{\text{top}}(1)$ symmetry associated with the
following conserved current
\begin{equation}
J_{\text{top}}^{\mu}(x)=\frac{1}{2\pi}\epsilon^{\mu\nu\rho}\partial_{\nu}A_{\rho}(x),
\end{equation}
where ``top'' stands for topological. The operators charged under
$\U_{\text{top}}(1)$ are called topological disorder operators
or instantons. In this $2+1$ dimensional context, we refer to them
as monopole operators. These operators create topological configurations
of the gauge field $\mathcal{A}_{\mu}^{q}$ with a quantized flux
$\int dn_{\mu}\epsilon^{\mu\nu\rho}\partial_{\nu}\mathcal{A}_{\rho}^{q}=4\pi q$, where
the topological charge is a half-integer $q\in\mathbb{Z}/2$ as a
result of the Dirac quantization condition~\citep{Heras_2018}. These
kinds of configurations are allowed in the compact formulation of
the $\U(1)$ gauge group, which gives the correct description
for emergent gauge theories in a condensed matter context. The monopole
operators themselves can be defined by the action of the topological
current on them:
\begin{equation}
J_{\text{top}}^{\mu}(x)\mathcal{M}_{q}^{\dagger}(0)\sim\frac{q}{2\pi}\frac{x^{\mu}}{\left|x\right|^{3}}\mathcal{M}_{q}^{\dagger}(0)+\cdots,
\end{equation}
where the ellipsis denotes less singular terms in the operator-product
expansion ($\OPE$)~\cite{borokhov_topological_2003}. The resulting
factor in front of the monopole operator corresponds to the magnetic
field of a charge-$q$ Dirac magnetic monopole.

The model in Eq.~\eqref{eq:GN_action} describes a transition from a DSL to a CSL. For a sufficiently
strong coupling, a chiral order develops due to the condensation of
a fermion bilinear: $\langle\overline{\Psi}\Psi\rangle\neq0$. This
may be studied by introducing an auxiliary pseudo-scalar boson $\phi$.
The effective action at the quantum critical point (QCP), denoted
by $S_{\eff}^{\text{c}}$, is 
\begin{equation}
S_{\eff}^{\text{c}}=-2 N\ln\det(\slashed{\partial}-i\slashed{A}+\phi),\label{eq:S_eff_c}
\end{equation}
where $\phi$ is an auxiliary boson decoupling the $\GN$ interaction.
More details are shown in App.~\ref{app:non_compact}.

In the compact version of $\QEDGN$, monopole operators are also present
at the QCP. The main goal of this paper is to compute their scaling
dimension $\Delta_{\mathcal{M}_{q}}$ , which controls the scaling
of the monopole two-point correlation function:
\begin{equation}
\langle\mathcal{M}_{q}(x)\mathcal{M}_{q}^{\dagger}(y)\rangle\sim\frac{1}{\left|x-y\right|^{2\Delta_{\mathcal{M}_{q}}}}.
\end{equation}
Since the $\QEDGN$ model at the QCP is a conformal field theory ($\CFT$),
the state-operator correspondence can be used to obtain these scaling
dimensions~\citep{borokhov_topological_2003}. This correspondence
relies on a radial quantization of the $\CFT$ and a conformal transformation
mapping the dilatation operator $\hat{D}$ on $\mathbb{R}^{3}$ to
a Hamiltonian $\hat{H}$ on $S^{2}\times\mathbb{R}$. Denoting the
usual radius on $\mathbb{R}^{3}$ as $r=e^{\tau}$,\footnote{We work in natural units where the two-sphere radius is $R=1$.}
the related Weyl transformation of the spacetime is written as
\begin{align}
\left(ds^{2}\right)_{S^{2}\times\mathbb{R}} & =e^{-2\tau}\left(ds^{2}\right)_{\mathbb{R}^{3}}=d\tau^{2}+d\theta^{2}+\sin^{2}\theta d\phi^{2}.\label{eq:weyl_rescaling}
\end{align}
The scaling dimension of an operator $\mathcal{O}(x)$
then corresponds to the energy of some state $\hat{H}\left|\mathcal{O}\right\rangle =\Delta_{\mathcal{O}}\left|\mathcal{O}\right\rangle $
on this compactified spacetime. Specifically, the charge-$q$ operator
with the smallest scaling dimension corresponds to the ground state
of the $\CFT$ on the compactified spacetime $S^{2}\times\mathbb{R}$,
where the sphere $S^{2}$ is pierced by $4\pi q$ flux. To implement
this flux, an external gauge field is coupled to the fermions 
\begin{equation}
\mathcal{A}^{q}=q\left(1-\cos\theta\right)\mathrm{d}\phi,
\end{equation}
or $\mathcal{A}_{\phi}^{q}=\left(1-\cos\theta\right)/\sin\theta$
in component notation. The smallest scaling dimension of monopole
operators in topological sector $q$ is then given by
\footnote{More formally, we could write $\Delta_{q}=\lim_{\beta\to\infty}\left(F_{q}-F_{0}\right)$
as explained in Ref.~\citep{metlitski_monopoles_2008}. However, it turns out that $\lim_{\beta\to\infty}F_{0}=0$.}
\begin{align}
\Delta_{q} & =\lim_{\beta\to\infty}F_{q}=-\lim_{\beta\to\infty}\frac{1}{\beta}\ln Z\left[\mathcal{A}^{q}\right],\label{eq:delta_q_F_q}
\end{align}
where $F_{q}$ is the free energy and $Z\left[\mathcal{A}^{q}\right]$
is the partition function formulated on $S^{2}\times S_{\beta}^{1}$,
i.e., the previous spacetime but now with the ``time'' direction
compactified to a ``thermal'' circle $S_{\beta}^{1}$ with radius
$\beta$. This formulation allows us to introduce the holonomy of
the gauge field along this circle, written as 
\begin{equation}
\alpha=i\beta^{-1}\int_{S_{\beta}^{1}}d\tau\ A_{\tau}.\label{eq:holonomy}
\end{equation}
The holonomy couples to the fermion number operator $\int d^{2}r\sqrt{g(r)}\Psi^{\dagger}\Psi=\hat{N}_{\text{fermions}}$
and acts as a chemical potential~\citep{chester_monopole_2018,dupuis_transition_2019}.
The saddle-point equation of this holonomy constrains the fermion
number to vanish
\begin{equation}
0=\frac{1}{\beta}\left.\frac{\delta\ln Z\left[\mathcal{A}^{q}\right]}{\delta\alpha}\right|_{\text{s.p.}}=\bigl\langle \hat{N}_{\text{fermions}}\bigr\rangle ,
\end{equation}
where ``s.p.'' stands for saddle-point. The holonomy thus serves
as a Lagrange multiplier that ensures that a state with $\langle \hat{N}_{\text{fermions}}\rangle =0$
is selected to correctly represent a gauge-invariant monopole operator~\citep{dupuis_transition_2019}.

The scaling dimension will be obtained using a large-$N$ expansion.
We first note that the partition function can be written as a path
integral:
\begin{equation}
Z\left[\mathcal{A}^{q}\right]=e^{-\beta F_{q}}=\int\mathcal{D}\phi\mathcal{D}A_{\mu}\exp\left(-S_{\eff}\left[\phi,A_{\mu},\mathcal{A}_{\mu}^{q}\right]\right).
\end{equation}
The effective action is now given by
\begin{align}
S_{\eff}\left[\phi,A_{\mu},\mathcal{A}_{\mu}^{q}\right]= & -2N\ln\det\left(\slashed{D}_{A+\mathcal{A}^{q}}+\phi\right),\label{eq:S_eff}
\end{align}
 where $\slashed{D}_{A+\mathcal{A}^{q}}$ is the gauge-covariant derivative
on a curved spacetime including the external gauge field $\mathcal{A}_{\mu}^{q}$
sourcing the $4\pi q$ flux:
\begin{equation}
\label{eq:SlashDeriv}
\slashed{D}_{A+\mathcal{A}^{q}}=e_{b}^{\mu}\gamma^{b}\left(\nabla_{\mu}-iA_{\mu}-i\mathcal{A}_{\mu}^{q}\right).
\end{equation}
The gamma matrices $\gamma^{b}$ still correspond to the Pauli matrices,
as the spacetime index is normalized with a tetrad $e^{\mu}_{b}$
which encapsulates the information about the metric $g_{\mu\nu}e^{\mu}_{b}e^{\nu}_{c}=\delta_{bc}$.
The path integral defining the partition function can be expanded
around the saddle-point values of the auxiliary and gauge bosons:
\begin{equation}
\phi=\langle\phi\rangle+\sigma,\quad A_{\mu}=\langle A_{\mu}\rangle+a_{\mu},
\end{equation}
which are defined by the following saddle-point conditions 
\begin{equation}
\left.\frac{\delta F_{q}}{\delta\phi}\right|_{\phi=\langle\phi\rangle,A_{\mu}=\langle A_{\mu}\rangle}=\left.\frac{\delta F_{q}}{\delta A_{\mu}}\right|_{\phi=\langle\phi\rangle,A_{\mu}=\langle A_{\mu}\rangle}=0.\label{eq:gap_eqns}
\end{equation}
 Taking the fluctuations to scale as $1/\sqrt{2N}$, the saddle-point
expansion of the partition function is then
\begin{equation}
\int\mathcal{D}\phi\mathcal{D}Ae^{-S_{\eff}}=e^{-\left.S_{\eff}\right|_{\text{s.p.}}}\int\mathcal{D}\sigma\mathcal{D}ae^{-S_{\eff}^{(2)}},
\end{equation}
where $S_{\eff}^{\left(2\right)}$ is the second variation of the
action. Integrating over the quadratic fluctuations, this gives us
the $1/N$ expansion of the free energy: 
\begin{align}
2N F_{q}^{(0)} & =\frac{1}{\beta}\left.S_{\eff}\right|_{\text{s.p.}},\\
F_{q}^{(1)} & =\frac{1}{\beta}\times\frac{1}{2}\left.\ln\det\left[\frac{\delta^{2}S_{\eff}}{\delta\left(\sigma,a\right)\delta\left(\sigma,a\right)}\right]\right|_{\text{s.p.}}.\label{eq:Fq_exp}
\end{align}
Using the relation in Eq.~(\ref{eq:delta_q_F_q}), which follows
from the state-operator correspondence, these first two terms of the
free energy give the scaling dimension at next-to-leading order in
$1/N$.\footnote{The expansion is in terms of the total number of fermion flavors, $2N$, such that $F_q = 2NF_q^{(0)}+F_{q}^{(1)}+O(1/N)$.}

Since the fermionic mass condensed in the ordered phase is flavor-symmetric
$\left\langle \overline{\Psi}\Psi\right\rangle $, the global flavor
symmetry remains unbroken and monopole operators are organized as
representations of $\SU(2N)$. Just as for the various
fermion bilinears and monopole correlation functions in the $\U(1)$
DSL~\citep{rantner_spin_2002,hermele_algebraic_2005,hermele_properties_2008},
monopole correlation functions at the QCP between $\U(1)$
DSL and CSL related by this $\SU(2N )$ symmetry share the
same scaling dimension.\footnote{This degeneracy we described is among a flavor symmetry multiplet composed of monopoles with the smallest scaling dimension, which we focus on. We emphasize that monopoles with larger scaling dimensions can be built  by dressing fermion modes with higher energies.  For instance, a splitting of monopoles  was obtained for $\QED$ $q=1$ monopoles as their scaling dimensions increases with Lorentz spin \citep{chester_monopole_2018}. Notably, for a monopole with Lorentz spin of order $\sqrt N$, there is a $O(N^0)$ additional positive correction.} Depending on the lattice, various magnetic
and VBS correlation functions will be described by minimally charged
monopole operators~\citep{song_unifying_2019}, but they all share
the same scaling dimension $2N\times0.26510+0.118911(7)+O\left(N^{-1}\right)$,
where the leading-order was found in Ref.~\citep{dupuis_transition_2019}
and the next-to-leading order is one of the main results of this work
shown in Eq.~(\ref{eq:QED3GN_q0.5}). For typical quantum magnets,
we have $2N=4$ fermion flavors. The way that monopole scaling dimensions
control observable correlation functions could also be compared at
this QCP and deep in the $\U(1)$ DSL phase. In this latter
case, scaling dimensions are those of monopoles in $\QED$.

\section{Review of $N=\infty$ theory \label{sec:N_infty}}

First, we review the computation of monopole scaling dimensions in
$\QEDGN$ at leading-order in $1/N$~\citep{dupuis_transition_2019}.
At this order, the free energy is given by the effective action in
(\ref{eq:S_eff}) at its saddle-point corresponding to a global minimum
\begin{equation}
F_{q}^{(0)}=-\frac{1}{\beta}\ln\det\left(\slashed{D}_{-i\alpha d\tau+\mathcal{A}^{q}}+\left\langle \phi\right\rangle \right),\label{eq:S_eff_sp}
\end{equation}
where the trace over the $2N$ flavors has been taken and has canceled
a prefactor of $(2N)^{-1}.$ The expectation value of the pseudo-scalar
field is taken to be homogeneous. The gauge field is also constant
at the saddle-point, with a possible non-vanishing holonomy $\alpha$
on the thermal circle described in (\ref{eq:holonomy}). The determinant
operator is diagonalized by introducing monopole harmonics $Y_{q,\ell,m}(\hat{n})$,
which are a generalization of spherical harmonics for a space with
a charge at the center~\citep{wu_dirac_1976,wu_properties_1977}.
For a fixed charge $q$, these functions form a complete basis. One
important difference with these harmonics is that their angular momentum
is now bounded below by this charge $q$. Using these functions to
build appropriate eigenspinors, the eigenvalues of this determinant
operator on $S^{2}\times S_{\beta}^{1}$ in Eq.~(\ref{eq:S_eff_sp})
are shown to be~\citep{borokhov_topological_2003,dupuis_transition_2019}
\begin{align}
-i\times\begin{cases}
\omega_{n}-i\alpha+i\varepsilon_{q} & \ell=q,\\
\pm\sqrt{\left(\omega_{n}-i\alpha\right)^{2}+\varepsilon_{\ell}^{2}} & \ell=q+1,q+2,\dots
\end{cases}
\end{align}
where, for simplicity, we suppose $q>0$ throughout. Here,
$\omega_{n}=2\pi\beta^{-1}\left(n+1/2\right)$, for $n\in\mathbb{Z}$,
are the fermionic Matsubara frequencies, and $\varepsilon_{\ell}$
are the energies of the modes for the quantized theory on $S^{2}\times\mathbb{R}$:
\begin{align}
\varepsilon_{\ell} & =\sqrt{\ell^{2}-q^{2}+\langle\phi\rangle^{2}}.
\end{align}
More details on the diagonalization are presented in App.~(\ref{subsec:GF}).
Note that the energies are dimensionless, as we work in units where
the radius of the sphere is $1$. Each mode has the usual degeneracy
that comes from the azimuthal symmetry, $d_{\ell}=2\ell$. The free
energy at leading-order then becomes
\begin{align}
\begin{split}F_{q}^{(0)} & =-\frac{1}{\beta}\sum_{n=-\infty}^{\infty}\biggl\{ d_{q}\ln\left[\omega_{n}-i\alpha+i\left\langle \phi\right\rangle \right]\\
 & \quad{}+\sum_{\ell=q+1}^{\infty}d_{\ell}\ln\left[\left(\omega_{n}-i\alpha\right)^{2}+\varepsilon_{\ell}^{2}\right]\biggr\} .
\end{split}
\label{eq:S_eff_sp_pr}
\end{align}
The saddle-point equation for the holonomy, given in Eq.~(\ref{eq:gap_eqns}),
yields the condition 
\begin{equation}
-d_{q}\tanh\left(\frac{\beta}{2}(\alpha-\left\langle \phi\right\rangle) \right)-\sum_{\ell=q+1}^{\infty}\frac{2d_{\ell}\sinh\left(\beta\alpha\right)}{\cosh\left(\beta\varepsilon_{\ell}\right)+\cosh\left(\beta\alpha\right)}=0,
\end{equation}
which is solved for $\alpha=\left\langle \phi\right\rangle $ in the
$\beta\to\infty$ limit. With this result, the second gap equation
at leading-order in $\beta$ is given by 
\begin{equation}
2\left\langle \phi\right\rangle \sum_{\ell=q+1}^{\infty}d_{\ell}\varepsilon_{\ell}^{-1}=0,
\end{equation}
whose only solution is $\left\langle \phi\right\rangle =0$. Therefore,
the saddle-point values of both fields vanish.

Inserting this result in Eq.~(\ref{eq:S_eff_sp_pr}), the monopole
scaling dimension at leading-order in $1/N$ is obtained from Eq.~(\ref{eq:delta_q_F_q})\footnote{More explicitly, the leading-order free energy is $\lim_{\beta\to\infty}F_{q}^{(0)}=2\sum_{\ell=q+1}^{\infty}d_{\ell}E_{q,\ell}$.
Using zeta regularization, it follows that the $q=0$ case vanishes:
$\lim_{\beta\to\infty}F_{0}^{(0)}=4\sum_{\ell=1}^{\infty}\ell^{2}=4\zeta\left(-2\right)=0$,
as previously claimed.}
\begin{equation}
\Delta_{q}=2N\sum_{\ell=q+1}^{\infty}d_{\ell}E_{q,\ell}+O\left(N^{0}\right),
\end{equation}
where the energies at the saddle-point are defined as
\begin{equation}
E_{q,\ell}=\sqrt{\ell^{2}-q^{2}}.\label{eq:energy_sp}
\end{equation}
This is simply the leading-order scaling dimension of $\QED$~\citep{borokhov_topological_2003}
(which must still be regularized). For example, the scaling dimension
of the monopole with minimal charge is $\Delta_{q=1/2}=2N\times0.265+O\left(N^{0}\right)$.
Here, a supplementary $\GN$ interaction is considered, but it does
not come into play at this level of the expansion since $\langle\phi\rangle=0$.
Thus, monopoles in $\QED$ and $\QEDGN$ have the same scaling dimensions
at leading-order in $1/N$:
\begin{equation}
\Delta_{q,\QED}^{(0)}=\Delta_{q,\QEDGN}^{(0)}.
\end{equation}

\section{$1/N$ corrections \label{sec:-corrections}}

\subsection{Setup}

\subsubsection{Real-space kernels}

We now turn to the next-to-leading-order term in the free-energy expansion
in Eq.~(\ref{eq:Fq_exp}). The free-energy correction is related
to the second variation of the action by
\begin{equation}
\begin{split}\exp\left(-\beta F_{q}^{(1)}\right)= & \int\mathcal{D}\sigma\mathcal{D}a\exp\left[-\frac{\left(2N\right)}{2}\int_{r,r^{\prime}}\begin{pmatrix}\sigma(r) & a_{\mu}(r)\end{pmatrix}\right.\\
 & \times\left.\begin{pmatrix}D^{q}(r,r^{\prime}) & H_{\mu'}^{q}(r,r^{\prime})\\
H_{\mu}^{q}\left(r^{\prime},r\right) & K_{\mu\mu'}^{q}(r,r^{\prime})
\end{pmatrix}\begin{pmatrix}\sigma\left(r^{\prime}\right)\\
a_{\mu'}\left(r^{\prime}\right)
\end{pmatrix}\right],
\end{split}
\label{eq:S2_PI}
\end{equation}
where $\int_{r}\equiv\int d^{3}r\sqrt{g(r)}$ and we defined
the following kernels 
\begin{align}
D^{q}(r,r^{\prime}) & =\frac{1}{2N}\left.\frac{\delta^{2}S_{\eff}}{\delta\sigma(r)\delta\sigma\left(r^{\prime}\right)}\right|_{\text{s.p.}},\label{eq:kernels_S_eff}\\
K_{\mu\mu^{\prime}}^{q}(r,r^{\prime}) & =\frac{1}{2N}\left.\frac{\delta^{2}S_{\eff}}{\delta a^{\mu}(r)\delta a^{\mu^{\prime}}\left(r^{\prime}\right)}\right|_{\text{s.p.}},\\
H_{\mu^{\prime}}^{q}(r,r^{\prime}) & =\frac{1}{2N}\left.\frac{\delta^{2}S_{\eff}}{\delta\sigma(r)\delta a^{\mu^{\prime}}\left(r^{\prime}\right)}\right|_{\text{s.p.}},
\end{align}
where $S_{\eff}$ is defined in Eq.~\eqref{eq:S_eff}. The remaining
scalar-gauge kernel\footnote{Although $\sigma(r)$ is really a pseudo-scalar, we refer
to it as a ``scalar'' when labelling related kernels for simplicity.} with mixed $a_{\mu}(r)$ and $\sigma\left(r^{\prime}\right)$
partial derivatives is obtained by exchanging coordinates $r,r^{\prime}$
in $H_{\mu^{\prime}}^{q}(r,r^{\prime})$, which has mixed
$\sigma(r)$ and $a_{\mu^{\prime}}\left(r^{\prime}\right)$
partial derivatives; thus we write $H_{\mu}^{q}\left(r^{\prime},r\right)$
in Eq.~(\ref{eq:S2_PI}). In terms of the fermions in the original
system, the kernels are given by 
\begin{align}
D^{q}(r,r^{\prime}) & =\left.\left\langle \overline{\psi}(r)\psi(r)\overline{\psi}\left(r^{\prime}\right)\psi\left(r^{\prime}\right)\right\rangle \right|_{\text{s.p.}},\\
K_{\mu\mu^{\prime}}^{q}(r,r^{\prime}) & =-\left.\bigl\langle J_{\mu}(r)J_{\mu^{\prime}}\left(r^{\prime}\right)\bigr\rangle\right|_{\text{s.p.}},\label{eq:Kmunu}\\
H_{\mu^{\prime}}^{q}(r,r^{\prime}) & =-i\left.\bigl\langle\overline{\psi}(r)\psi(r)J_{\mu^{\prime}}\left(r^{\prime}\right)\bigr\rangle\right|_{\text{s.p.}},\label{eq:Fmu}
\end{align}
where $\psi$ is a single fermion flavor and the current is
\begin{equation}
J_{\mu}(r)=\overline{\psi}(r)\gamma_{\mu}\psi(r).
\end{equation}
This can be re-expressed in terms of the fermionic Green's function
$G_{q}(r,r^{\prime})=\left.\left\langle \psi(r)\overline{\psi}\left(r^{\prime}\right)\right\rangle \right|_{\text{s.p.}}$
and its hermitian conjugate $G_{q}^{\dagger}(r,r^{\prime})=-\left.\left\langle \psi\left(r^{\prime}\right)\overline{\psi}(r)\right\rangle \right|_{\text{s.p.}}$.
The Wick expansion of the kernels yields 
\begin{align}
D^{q}(r,r^{\prime}) & =-\text{tr}\left[G_{q}(r,r^{\prime})G_{q}^{\dagger}(r,r^{\prime})\right],\label{eq:D_rr'}\\
K_{\mu\mu^{\prime}}^{q}(r,r^{\prime}) & =\text{tr}\left[\gamma_{\mu}G_{q}(r,r^{\prime})\gamma_{\mu^{\prime}}G_{q}^{\dagger}(r,r^{\prime})\right],\label{eq:K_rr'}\\
H_{\mu^{\prime}}^{q}(r,r^{\prime}) & =i\text{tr}\left[G_{q}(r,r^{\prime})\gamma_{\mu^{\prime}}G_{q}^{\dagger}(r,r^{\prime})\right],\label{eq:F_rr'}
\end{align}
where the cyclicity of the trace was used. The remaining prefactor $2N$ in Eq.~\eqref{eq:S2_PI} is cancelled as the fluctuation fields are rescaled $\sigma,a_\mu \to \sigma/\sqrt{2N}, a_\mu/\sqrt{2N}$ to control the expansion. The free-energy correction is then obtained by integrating the field fluctuations in Eq.~(\ref{eq:S2_PI}).
It is convenient to subtract the $q=0$ correction $F_{0}^{(1)}=0$,\footnote{Since the scaling dimension of the identity operator vanishes, we
have $\lim_{\beta\to\infty}F_{0}^{(1)}=0$. Computing $F_{q}^{(1)}-F_{0}^{(1)}$
requires less regularization procedures and automatically takes care
of gauge fixing subtleties since the Fadeev-Popov ghost contribution
is independent of the background flux $4\pi q$ in $\QED$~ \citep{dyer_monopole_2013}.} therefore we write the general correction as
\begin{equation}
F_{q}^{(1)}=\frac{1}{2}\ln\left(\frac{\det^{\prime}M^{q}}{\det^{\prime}M^{0}}\right),\label{eq:F_q_1}
\end{equation}
where we define the matrix kernel
\begin{equation}
M^{q}(r,r^{\prime})=\left(\begin{array}{cc}
D^{q}(r,r^{\prime}) & H_{\mu^{\prime}}^{q}(r,r^{\prime})\\
H_{\mu}^{q}\left(r^{\prime},r\right) & K_{\mu\mu^{\prime}}^{q}(r,r^{\prime})
\end{array}\right).
\end{equation}

\subsubsection{Fourier transform}

To compute the determinant operator, the kernels are expanded in terms
of harmonics. For the gauge-gauge kernels, the vector spherical harmonics
are introduced:
\begin{align}
\mathfrak{a}_{\mu,\ell m}^{T}(\hat{n}) & =\delta_{\mu}^{0}Y_{\ell m}(\hat{n}),\label{eq:vector_harm_basis_1}\\
\mathfrak{a}_{\mu,\ell m}^{E}(\hat{n}) & =\frac{1}{\sqrt{\ell(\ell+1)}}\nabla_{\mu}Y_{\ell m}(\hat{n}),\label{eq:vector_harm_basis_2}\\
\mathfrak{a}_{\ell m}^{\mu,B}(\hat{n}) & =\frac{1}{\sqrt{\ell(\ell+1)}}\frac{\epsilon^{0\mu\nu}}{\sqrt{g(r)}}\nabla_{\nu}Y_{\ell m}(\hat{n}).\label{eq:vector_harm_basis_3}
\end{align}
As suggested by the notation, the $B$ mode has zero divergence: $\nabla\cdot\vec{\mathfrak{a}}_{\ell m}^{B}(\hat{n})=0$,
and the $E$ mode has zero curl: $\nabla\boldsymbol{\times}\vec{\mathfrak{a}}_{\ell m}^{E}(\hat{n})=0$.
It is also useful to introduce 4-dimensional eigenfunctions, 
\begin{align}
\mathbb{Y}_{\ell m}^{D}(\hat{n}) & =\begin{pmatrix}Y_{\ell m}(\hat{n})\\
0^{\mu}
\end{pmatrix},\quad\mathbb{Y}_{\ell m}^{X}(\hat{n})=\begin{pmatrix}0\\
\mathfrak{a}_{\ell m}^{\mu,X}(\hat{n})
\end{pmatrix},
\end{align}
where $X\in\{T,E,B\}$. In this basis, the matrix kernel $M^{q}(r,r^{\prime})$
can be expanded as 
\begin{align}
M^{q}(r,r^{\prime}) & =\int_{-\infty}^{\infty}\frac{d\omega}{2\pi}\sum_{\ell=0}^{\infty}\sum_{m=-\ell}^{\ell}e^{-i\omega(\tau-\tau^{\prime})}\label{eq:M_rr'}\\
 & \quad{}\times\begin{pmatrix}\mathbb{Y}_{\ell m}^{D}(\hat{n})\\
\mathbb{Y}_{\ell m}^{T}(\hat{n})\\
\mathbb{Y}_{\ell m}^{E}(\hat{n})\\
\mathbb{Y}_{\ell m}^{B}(\hat{n})
\end{pmatrix}^{\intercal}M_{\ell}^{q}(\omega)\begin{pmatrix}\mathbb{Y}_{\ell m}^{D}(\hat{n}^{\prime})^{\dagger}\\
\mathbb{Y}_{\ell m}^{T}(\hat{n}^{\prime})^{\dagger}\\
\mathbb{Y}_{\ell m}^{E}(\hat{n}^{\prime})^{\dagger}\\
\mathbb{Y}_{\ell m}^{B}(\hat{n}^{\prime})^{\dagger}
\end{pmatrix},\nonumber 
\end{align}
where we directly work on $S^{2}\times\mathbb{R}$ (i.e., taking the
limit $\beta\to\infty$ now) and where 
\begin{equation}
M_{\ell}^{q}(\omega)=\begin{pmatrix}D_{\ell}^{q} & H_{\ell}^{q,T} & H_{\ell}^{q,E} & H_{\ell}^{q,B}\\
-H_{\ell}^{q,T*} & K_{\ell}^{q,TT} & K_{\ell}^{q,TE} & K_{\ell}^{q,TB}\\
-H_{\ell}^{q,E*} & K_{\ell}^{q,TE*} & K_{\ell}^{q,EE} & K_{\ell}^{q,EB}\\
-H_{\ell}^{q,B*} & K_{\ell}^{q,TB*} & K_{\ell}^{q,EB*} & K_{\ell}^{q,BB}
\end{pmatrix}.\label{eq:M_o}
\end{equation}
All of the arguments of the functions appearing in the matrix are
$\omega$. Note that the scalar-gauge kernel is imaginary, hence the
reason for the signs in the first column of the matrix. This last
point is shown explicitly in App.~\ref{app:scalar_gauge}.

This kernel can be simplified by using $\mathcal{CT}$ invariance.
The auxiliary boson $\phi$, a pseudo-scalar, and the $E,T$ modes
of the gauge field are antisymmetric under $\mathcal{CT}$, while
the \textbf{$B$} modes are symmetric under $\mathcal{CT}$. This
implies that the following kernels vanish \footnote{This result was also checked explicitly with the same method giving
the values of non-vanishing kernels.}
\begin{align}
K_{\ell}^{q,TB}(\omega) & =K_{\ell}^{q,EB}(\omega)=H_{\ell}^{q,B}(\omega)=0.
\end{align}
The $\U(1)$ gauge invariance also enables the kernel to
be simplified. Using the conservation of the $\U(1)$ current,
$\nabla_{\mu}J^{\mu}(r)=0$, in Eqs.~(\ref{eq:Kmunu}-\ref{eq:Fmu}),
it follows that 
\begin{align}
K_{\ell}^{q,TE}(\omega) & =\frac{i\omega}{\sqrt{\ell(\ell+1)}}K_{\ell}^{q,TT}(\omega),\label{eq:gauge_c1}\\
K_{\ell}^{q,EE}(\omega) & =\frac{\omega^{2}}{\ell(\ell+1)}K_{\ell}^{q,TT}(\omega),\label{eq:gauge_c2}\\
H_{\ell}^{q,E}(\omega) & =\frac{i\omega}{\sqrt{\ell(\ell+1)}}H_{\ell}^{q,T}(\omega).\label{eq:gauge_c3}
\end{align}
It should also be noted that among vector spherical harmonics, only
$\mathfrak{a}_{\mu,\ell m}^{T}(\hat{n})$ is defined for
$\ell=0.$ In this case, the only remaining gauge-gauge kernel is
$K_{0}^{q,TT}(\omega)$, and it vanishes by gauge invariance.
The computations to obtain these gauge invariance conditions are shown
in App.~\ref{app:gauge_invariance}. Using all these simplifications,
the monopole scaling dimension correction is given by
\begin{widetext}
\begin{align}
\Delta_{q, \QEDGN}^{(1)} & =\frac{1}{2}\int_{\omega}\left\{ \ln\left(\frac{D_{0}^{q}(\omega)}{D_{0}^{0}(\omega)}\right)+\sum_{\ell=1}^{\infty}\left(2\ell+1\right)\ln\left[\dfrac{K_{\ell}^{q,B}(\omega)\left(D_{\ell}^{q}(\omega)K_{\ell}^{q,E}(\omega)+\left(1+\frac{\omega^{2}}{\ell(\ell+1)}\right)\left|H_{\ell}^{q,T}(\omega)\right|^{2}\right)}{K_{\ell}^{0,B}(\omega)D_{\ell}^{0}(\omega)K_{\ell}^{0,E}(\omega)}\right]\right\} ,\label{eq:delta_QED3GN}
\end{align}
\end{widetext} 
where $\int_{\omega}\equiv\int_{-\infty}^{\infty}d\omega/\left(2\pi\right)$
and we defined
\begin{align}
K_{\ell}^{q,E}(\omega) & \equiv K_{\ell}^{q,TT}(\omega)+K_{\ell}^{q,EE}(\omega),\label{eq:KE_def}\\
K_{\ell}^{q,B}(\omega) & \equiv K_{\ell}^{q,BB}(\omega).\label{eq:KB_def}
\end{align}
Note that $H_{\ell}^{0,T}(\omega)=0$ and thus it does
not appear in the denominator.

By turning off the $\GN$ interaction in Eq.~(\ref{eq:delta_QED3GN}),
the scalar-scalar kernel $D_{\ell}^{q}(\omega)$ and the
scalar-gauge kernel $H_{\ell}^{q,T}(\omega)$ do not contribute,
and the monopole scaling dimension correction in $\QED$~\citep{pufu_anomalous_2014}
is recovered:
\begin{equation}
\Delta_{q,\QED}^{(1)}=\frac{1}{2}\int_{\omega}\sum_{\ell=1}^{\infty}\left(2\ell+1\right)\ln\left[\dfrac{K_{\ell}^{q,B}(\omega)K_{\ell}^{q,E}(\omega)}{K_{\ell}^{0,B}(\omega)K_{\ell}^{0,E}(\omega)}\right],\label{eq:delta_QED3}
\end{equation}
One can alternatively deactivate the gauge field, keeping only the
scalar-scalar kernels, and obtain the pure $\GN$ model. Despite the
absence of a gauge field in this model, one can still introduce an
external gauge field with the $4\pi q$ flux, define a correlation
function on this background configuration, and obtain the related
critical exponent. This was notably achieved for the $\ON$
model in Ref.~\citep{pufu_monopoles_2013}. In a forthcoming publication,
we shall also explore this avenue in the pure-$\GN$ model.

The relevant kernel Fourier coefficients to compute the monopole scaling
dimensions in (\ref{eq:delta_QED3GN}) and (\ref{eq:delta_QED3})
are found by inverting Eq.~(\ref{eq:M_rr'}):
\begin{align}
D_{\ell}^{q}(\omega) & =\frac{4\pi}{2\ell+1}\int_{r}e^{i\omega\tau}D^{q}\left(r,0\right)\sum_{m}Y_{\ell m}^{*}(\hat{n})Y_{\ell m}\left(\hat{z}\right),\label{eq:Dlo}\\
H_{\ell}^{q,T}(\omega) & =\frac{4\pi}{2\ell+1}\int_{r}e^{i\omega\tau}H_{0}^{q}\left(r,0\right)\sum_{m}Y_{\ell m}^{*}(\hat{n})Y_{\ell m}\left(\hat{z}\right),\label{eq:FTlo}\\
K_{\ell}^{q,E}(\omega) & =\frac{4\pi}{2\ell+1}\int_{r}e^{i\omega\tau}\bigg[K_{00}^{q}\left(r,0\right)\sum_{m}Y_{\ell m}^{*}(\hat{n})Y_{\ell m}\left(\hat{z}\right)\nonumber \\
 & {}\quad+K_{aa^{\prime}}^{q}\left(r,0\right)\sum_{m}\mathfrak{a}_{\ell m}^{a,E*}(\hat{n})\mathfrak{a}_{\ell m}^{a^{\prime},E}\left(\hat{z}\right)\bigg],\label{eq:KElo}\\
K_{\ell}^{q,B}(\omega) & =\frac{4\pi}{2\ell+1}\int_{r}e^{i\omega\tau}K_{aa^{\prime}}^{q}\left(r,0\right)\sum_{m}\mathcal{\mathfrak{a}}_{\ell m}^{a,B*}(\hat{n})\mathcal{\mathfrak{a}}_{\ell m}^{a^{\prime},B}\left(\hat{z}\right),\label{eq:KBlo}
\end{align}
where the second coordinates are fixed to $\tau^{\prime}=0$ and $\hat{n}^{\prime}=\hat{z}$
without loss of generality, and normalized coordinates $a,a^{\prime}$
are introduced.

\subsection{Anomalous dimensions \label{subsec:Anomalous-dimensions}}

The anomalous dimensions of monopole operators (\ref{eq:delta_QED3GN}, \ref{eq:delta_QED3})
are computed in this section. To do so, the kernel coefficients in
Eqs.~(\ref{eq:Dlo}, \ref{eq:KElo}, \ref{eq:KBlo}) must be obtained.
These coefficients are built with real-space kernels (\ref{eq:D_rr'}-\ref{eq:F_rr'})
that depend on the fermionic Green's function at the saddle-point.
The Green's function is defined by the action of the Dirac operator
on it:
\begin{equation}
i\slashed{D}_{\mathcal{A}^{q}}^{S^{2}\times\mathbb{R}}(r)G_{q}(r,r^{\prime})=-\delta(r-r^{\prime}).\label{eq:GF_EOM}
\end{equation}
The eigenkernels in Eqs.~(\ref{eq:Dlo}, \ref{eq:KElo}, \ref{eq:KBlo}),
involving the sums on spherical harmonics or vector spherical harmonics,
will also be needed.

\subsubsection{$q=0$ kernels\label{subsec:q0_kernels}}

We first compute the expressions in the denominator of the scaling-dimension
corrections (\ref{eq:delta_QED3GN}, \ref{eq:delta_QED3}), that is,
the $q=0$ kernel coefficients. The eigenkernel in the scalar-scalar
kernel (\ref{eq:Dlo}) is just the sum of spherical harmonics, which
is given by the addition theorem
\begin{equation}
\sum_{m}Y_{\ell m}^{*}(\hat{n})Y_{\ell m}(\hat{n}^{\prime})=\frac{2\ell+1}{4\pi}P_{\ell}\left(\cos\gamma\right),\label{eq:addition_theorem}
\end{equation}
where 
\begin{equation}
\cos\gamma\equiv\hat{n}\cdot\hat{n}^{\prime}=\cos\theta\cos\theta^{\prime}+\sin\theta\sin\theta^{\prime}\cos\left(\phi-\phi^{\prime}\right).
\end{equation}
When working with $\hat{n}^{\prime}=\hat{z}$, this is replaced by
\begin{equation}
x\equiv\cos\theta.
\end{equation}
For the sums on vector spherical harmonics appearing in the gauge-gauge
kernels (\ref{eq:KElo}, \ref{eq:KBlo}), a similar result is obtained
in spherical coordinates $a=\hat{\theta},\text{\ensuremath{\hat{\phi},\hat{\tau}}}$
in Eqs.~(\ref{eq:sum_aEaE}, \ref{eq:sum_aBaB}) in App.~\ref{sec:eigenkernels}
and is formulated with the same Legendre polynomial and its first
and second derivatives. This reproduces a result from Ref.~\citep{dyer_scaling_2015}.

The real-space kernel for $q=0$ is also needed. In this case, the
Green's function takes a simple form which is simply the conformally
transformed $3D$ flat space Green's function~\citep{pufu_anomalous_2014}:
\begin{equation}
G_{0}\left(\tau-\tau^{\prime},\hat{n},\hat{n}^{\prime}\right)=\frac{i}{4\pi X^{3}}\vec{\gamma}\cdot\left(e^{\frac{1}{2}(\tau-\tau^{\prime})}\hat{n}-e^{-\frac{1}{2}(\tau-\tau^{\prime})}\hat{n}^{\prime}\right),
\end{equation}
where 
\begin{equation}
X\equiv [ 2\cosh(\tau-\tau')-2\cos\gamma ]^{1/2} .
\end{equation}
The real-space kernels can then be obtained in normalized spherical
coordinates. Inserting this Green's function, along with the eigenkernels,
in Eqs.~(\ref{eq:Dlo}, \ref{eq:KElo}, \ref{eq:KBlo}), the resulting
$q=0$ kernel coefficients are (setting $\tau'=0$)
\begin{align}
D_{\ell}^{0}(\omega) & =-\frac{1}{8\pi^{2}}\int_{r}e^{i\omega\tau}P_{\ell}(x)\frac{1}{X^{4}},\\
K_{\ell}^{0,E}(\omega) & =\frac{1}{32\pi^{2}}\int_{r}e^{i\omega\tau}P_{\ell}(x)\Big(-\nabla_{S^{2}}^{2}\nonumber \\
 & \quad+\frac{1}{\ell(\ell+1)}\nabla_{S^{2}}^{2}\partial_{\tau}^{2}\Big)\frac{1}{X^{2}},\\
K_{\ell}^{0,B}(\omega) & =\frac{1}{16\pi^{2}\ell(\ell+1)}\int_{r}e^{i\omega\tau}P_{\ell}(x)\nabla_{S^{2}}^{2}\frac{1}{X^{4}},
\end{align}
where integration by parts was used to eliminate derivatives of $P_{\ell}(x)$.
The differential operators acting on $e^{i\omega\tau}P_{\ell}(x)$
can be replaced with the corresponding eigenvalues $\nabla_{S^{2}}^{2}\to-\ell(\ell+1)$
and $\partial_{\tau}^{2}\to-\omega^{2}$ with further integration
by parts. The remaining expressions contain Fourier transforms of
the form $\int_{r}e^{i\omega\tau}P_{\ell}(x)X^{p}$ which were obtained
in the appendix of Ref.~\citep{dyer_scaling_2015}. Using these results,
the $q=0$ kernel coefficients are simplified to
\begin{align}
D_{\ell}^{0}(\omega) & =\left(\ell^{2}+\omega^{2}\right)\mathcal{D}_{\ell-1}(\omega),\label{eq:D_0_ana}\\
K_{\ell}^{0,E}(\omega) & =\frac{1}{2}\left(\ell(\ell+1)+\omega^{2}\right)\mathcal{D}_{\ell}(\omega),\label{eq:KE_0_ana}\\
K_{\ell}^{0,B}(\omega) & =\frac{1}{2}\left(\ell^{2}+\omega^{2}\right)\mathcal{D}_{\ell-1}(\omega),\label{eq:KB_0_ana}
\end{align}
where 
\begin{equation}
\mathcal{D}_{\ell}(\omega)=\left|\frac{\Gamma\left(\frac{1+\ell+i\omega}{2}\right)}{4\Gamma\left(\frac{2+\ell+i\omega}{2}\right)}\right|^{2}.\label{eq:D_0_cal}
\end{equation}
Note that we have reproduced the gauge-gauge coefficients $K_{\ell}^{0,E}(\omega)$
and $K_{\ell}^{0,B}(\omega)$ given in Ref.~\citep{pufu_anomalous_2014}
by using the methods of Ref.~\citep{dyer_scaling_2015}.

\subsubsection{Anomalous dimensions for $q=1/2$ \label{subsec:min_charge}}

For the minimal magnetic charge, the eigenkernels in Eqs.~(\ref{eq:Dlo}-\ref{eq:KBlo})
are formulated using the same expression (\ref{eq:addition_theorem}, \ref{eq:sum_aEaE}, \ref{eq:sum_aBaB})
as in the last section. In particular, the gauge-gauge kernels will
be worked out in normalized spherical coordinates. As for the real-space
kernels (\ref{eq:D_rr'}-\ref{eq:F_rr'}), they depend on the $q=1/2$
fermionic Green's function defined through Eq.~(\ref{eq:GF_EOM}).
The spectral decomposition of the Green's function in terms of spinors
with monopole harmonics components is shown in App.~\ref{subsec:GF}.
A generalized addition theorem for monopole harmonics involving the
Jacobi Polynomials $P_{\ell}^{\left(0,2q\right)}(x)$ is
then needed. Specifically, after taking the sum over the azimuthal
quantum number, the Green's function for general $q$ is given by~\citep{pufu_anomalous_2014}\footnote{There is a sign error in the first term of the Green's function in
Ref.~\citep{pufu_anomalous_2014} that we corrected here. This sign
does not affect the conclusions in Ref.~\citep{pufu_anomalous_2014}.}\begin{widetext}
\begin{equation}
\begin{split}G_{q}\left(\tau,\hat{n};\tau^{\prime},\hat{n}^{\prime}\right) & =\frac{i}{2}e^{-i2q\Theta}\sum_{\ell=q}^{\infty}e^{-E_{q,\ell}\left|\tau-\tau^{\prime}\right|}\left\{ -\dfrac{E_{q,\ell}}{1-x}Q_{q,\ell}(x)\left(\hat{n}-\hat{n}^{\prime}\right)\cdot\vec{\gamma}\right.\\
 & \quad{}\left.+\text{sgn}(\tau-\tau^{\prime})\left[qQ_{q,\ell}(x)\mathbb{I}+Q^{\prime}_{q,\ell}(x)\left(\hat{n}+\hat{n}^{\prime}\right)\cdot\vec{\gamma}+i\dfrac{q}{1+x}Q_{q,\ell}(x)\left(\hat{n}\times\hat{n}^{\prime}\right)\cdot\vec{\gamma}\right]\right\} ,
\end{split}
\label{eq:GF}
\end{equation}
\end{widetext}where the energies $E_{q,\ell}$ were defined in Eq.~(\ref{eq:energy_sp})
and where
\begin{equation}
Q_{q,\ell}(x)=\dfrac{\left(1+x\right)^{q}}{\left(4\pi\right)2^{q}}\begin{cases}
P_{\ell-q}^{\left(0,2q\right)}(x)-P_{\ell-1-q}^{\left(0,2q\right)}(x), & \ell>q,\\
1, & \ell=q.
\end{cases}
\end{equation}
The phase $e^{-i2q\Theta}$ comes from the generalized addition theorem
and is defined in Eq.~(\ref{eq:phase}), but it is not involved in
the computation since it is always cancelled by the opposite phase
of the Green's function hermitian conjugate. The Green's function
can be inserted in Eqs.~(\ref{eq:D_rr'}-\ref{eq:F_rr'}) to obtain
the real-space kernels, which, along with the eigenkernels (\ref{eq:addition_theorem}, \ref{eq:sum_aEaE}, \ref{eq:sum_aBaB}),
are inserted in Eqs.~(\ref{eq:Dlo}-\ref{eq:KBlo}) to compute the
four kernel coefficients. Defining $K_{\ell}^{q,D}(\omega)\equiv D_{\ell}^{q}(\omega)$
and $K_{\ell}^{q,T}(\omega)\equiv H_{\ell}^{q,T}(\omega)$,
the kernel coefficients $K_{\ell}^{q,Z}(\omega)$ with
$Z\in\{D,T,E,B\}$ are given by 
\begin{align}
K_{\ell}^{q,Z}(\omega) & =\sum_{\ell^{\prime},\ell^{\prime\prime}}\frac{4\pi A^{Z}\bigl(E_{q,\ell^{\prime}}+E_{q,\ell^{\prime\prime}}\bigr)}{\omega^{2}+\bigl(E_{q,\ell^{\prime}}+E_{q,\ell^{\prime\prime}}\bigr)^{2}}\biggl[\frac{\mathcal{I}_{1}^{Z}}{2}+E_{q,\ell^{\prime}}E_{q,\ell^{\prime\prime}}\mathcal{I}_{2}^{Z}\biggr]\nonumber \\
 & \equiv\sum_{\ell^{\prime},\ell^{\prime\prime}}^{\infty}k_{\ell,\ell^{\prime},\ell^{\prime\prime}}^{q,Z}(\omega),\label{eq:KZ_sum}
\end{align}
where the prefactors are given by
\begin{equation}
A^{Z}=\Bigl\{1,i,\tfrac{1}{\ell(\ell+1)},\tfrac{1}{\ell(\ell+1)}\Bigr\},\ Z\in\{D,T,E,B\},
\end{equation}
 the integrals for scalar-scalar and scalar-gauge kernels are 
\begin{align}
\mathcal{I}_{1}^{D} & =-2\int dxP_{\ell}\Big[q^{2}\frac{1}{1+x}Q_{q,\ell^{\prime}}Q_{q,\ell^{\prime\prime}} \nonumber \\
 & \quad{}+\left(1+x\right)Q^{\prime}{}_{q,\ell^{\prime}}Q^{\prime}{}_{q,\ell^{\prime\prime}}\Big],\\
\mathcal{I}_{2}^{D} & =-\int dx\frac{1}{1-x}P_{\ell}Q_{q,\ell^{\prime}}Q_{q,\ell^{\prime\prime}},\\
\mathcal{I}_{1}^{T} & =-2q\int dxP_{\ell}^{\prime}Q_{q,\ell^{\prime}}Q_{q,\ell^{\prime\prime}},\\
\mathcal{I}_{2}^{T} & =0,
\end{align}
and the integrals for gauge-gauge kernels reproduce the expressions
obtained in Ref.~\citep{pufu_anomalous_2014}\footnote{Here, our definitions for the integrals differ by a factor $(2\ell+1)/(4\pi)$,
so that the Legendre polynomial appears $P_{\ell}(x)$ instead of
$F_{\ell}(x)=[(2\ell+1)/(4\pi)]P_{\ell}(x)$ as in Ref.~ \citep{pufu_anomalous_2014}.}

\begin{align}
\mathcal{I}_{1}^{E} & =2\int dx\left\{ \left[\frac{2\ell(\ell+1)P_{\ell}+\left(1-x\right)P_{\ell}^{\prime}}{1+x}\right]q^{2}Q_{q,\ell^{\prime}}Q_{q,\ell^{\prime\prime}}\right.\nonumber \\
 & \quad{}\left.-\left(1-x^{2}\right)P_{\ell}^{\prime}Q^{\prime}{}_{q,\ell^{\prime}}Q^{\prime}{}_{q,\ell^{\prime\prime}}\right\} ,\\
\mathcal{I}_{2}^{E} & =-\int dx\left(\frac{1+x}{1-x}\right)P_{\ell}^{\prime}Q_{q,\ell^{\prime}}Q_{q,\ell^{\prime\prime}},\\
\mathcal{I}_{1}^{B} & =2\int dx\left\{ \left[P_{\ell}^{\prime}-\left(1-x\right)P_{\ell}^{\prime\prime}\right]\right.\left[q^{2}Q_{q,\ell^{\prime}}Q_{q,\ell^{\prime\prime}}\right.\nonumber \\
 & \quad{}\left.\left.-\left(1+x\right)^{2}Q^{\prime}{}_{q,\ell^{\prime}}Q^{\prime}{}_{q,\ell^{\prime\prime}}\right]\right\} ,\\
\mathcal{I}_{2}^{B} & =\int dx\left[P_{\ell}^{\prime}+\left(1+x\right)P_{\ell}^{\prime\prime}\right]Q_{q,\ell^{\prime}}Q_{q,\ell^{\prime\prime}}.
\end{align}

These integrals can be performed exactly, see App.~\ref{sec:min_charge_results}
for more details. In the end, these quantities depend only on the
angular momenta: $\mathcal{I}_{1}^{Z}\left(\ell,\ell^{\prime},\ell^{\prime\prime}\right)$
and $\mathcal{I}_{2}^{Z}\left(\ell,\ell^{\prime},\ell^{\prime\prime}\right)$.
For $\ell^{\prime}=\ell^{\prime\prime}=q$, this computation requires
more care since both energies vanish, and the integral over time leading
to the prefactor in Eq.~(\ref{eq:KZ_sum}) instead yields a Dirac
delta function $\delta(\omega)$. However, for $\ell^{\prime}=\ell^{\prime\prime}=q=1/2$,
the term in the bracket simply vanishes. When only one of $\ell^{\prime}$
and $\ell^{\prime}$ have their minimal value $q=1/2$, there is a
non-vanishing contribution to the anomalous dimension. In this case,
only $\mathcal{I}_{1}^{Z}$ contributes, since the prefactor in front
of $\mathcal{I}_{2}^{Z}$ in Eq.~(\ref{eq:KZ_sum}) vanishes. For
$q=1/2$, the contribution of zero modes in Eq.~(\ref{eq:KZ_sum}) vanishes with $\ell=\omega=0$, otherwise it is given by
\begin{equation}
2\sum_{\ell'=3/2}^{\infty}k_{\ell,\ell^{\prime},1/2}^{1/2,Z}(\omega)=-\frac{1}{4\pi}\frac{\sqrt{\ell(\ell+1)}}{\omega^{2}+\ell(\ell+1)}\times\{1,-i,0,1\}.\label{eq:half_ZM}
\end{equation}
The remaining
contribution consists in a sum on non-zero modes $\ell^{\prime},\ell^{\prime\prime}\ge3/2$.
The summand depends on $\mathcal{I}_{1}^{Z}\left(\ell,\ell^{\prime},\ell^{\prime\prime}\right)$
and $\mathcal{I}_{2}^{Z}\left(\ell,\ell^{\prime},\ell^{\prime\prime}\right)$
which are formed of three-$J$ symbols in $\ell,\ell^{\prime}$ and
$\ell^{\prime\prime}$ (\ref{eq:J0}-\ref{eq:J2}). Thus, one of the
sums, say on $\ell^{\prime\prime}$, can be viewed as finite. Then,
after taking the sum on $\ell^{\prime\prime}$, the remaining summand
tends to a constant for large $\ell^{\prime}$ 
\begin{equation}
\lim_{\ell^{\prime}\to\infty}\sum_{\ell^{\prime\prime}=3/2}^{\infty}k_{\ell,\ell^{\prime},\ell^{\prime\prime}}^{1/2,Z}(\omega)=\alpha^{Z}=-\frac{1}{4\pi}\times\{2,0,1,1\}.\label{eq:ass_coeff}
\end{equation}
Thus, for kernels with a non-zero asymptotic constant, the sum on
$\ell^{\prime}$ will be divergent. This is regularized with a zeta
function regularization $\sum_{\ell=a}^{\infty}\ell^{-p}=\zeta\left(p,a\right)$,
here specifically $\zeta\left(0,3/2\right)=-1$
\begin{equation}
-\alpha^{Z}+\sum_{\ell^{\prime}=3/2}^{\infty}\biggl[-\alpha^{Z}+\sum_{\ell^{\prime\prime}=3/2}^{\infty}k_{\ell,\ell^{\prime},\ell^{\prime\prime}}^{1/2,Z}(\omega)\biggr].\label{eq:half_NZM}
\end{equation}

The sum above is then finite and is computed numerically up to a cutoff
$\ell_{c}^{\prime}$. The remainder is approximated with a large $\ell^{\prime}$
expansion of the summand $-\alpha^{Z}+\sum_{\ell^{\prime\prime}}k_{\ell,\ell^{\prime},\ell^{\prime\prime}}^{1/2,Z}(\omega)=\sum_{p=2}^{k}c_{\ell,p}^{1/2,Z}(\omega)\left(\ell^{\prime}\right)^{-p}+O\left(1/\ell^{\prime\left(k+1\right)}\right)$.
Each power in the expansion is summed analytically from $\ell^{\prime}=\ell_{c}^{\prime}+1$
to $\ell^{\prime}=\infty$ with a zeta function. The coefficients
$c_{\ell,p}^{1/2,Z}(\omega)$ are found by doing the expansion
for a few fixed values of $\ell$ and deducing the general dependence
on $\ell$. It turns out that only even powers of $1/\ell^{\prime}$
have non-vanishing coefficients $\text{\ensuremath{c_{\ell,p}^{1/2,Z}(\omega)}}.$
We obtained the expansion up to $k=18$. With this remainder, we found
that a cutoff $\ell_{c}^{\prime}=300+1/2$ was sufficiently large
to achieve the desired precision goals.
The first few terms of the remainders for general $q$ are shown in
App.~\ref{sec:remainders}.

\begin{figure*}
\hfill{}\centering\subfigure[]{\label{fig:min_charge_QED3}\includegraphics[height=5.35cm]{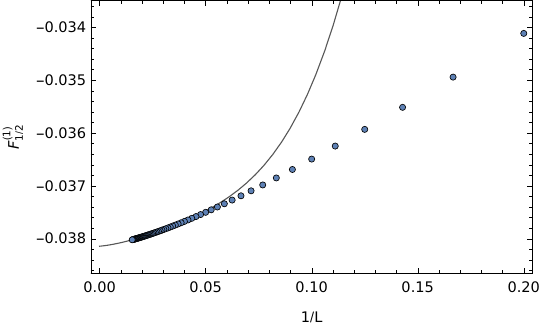}}\hfill{}\subfigure[]{\label{fig:min_charge_QED3GN}\includegraphics[height=5.35cm]{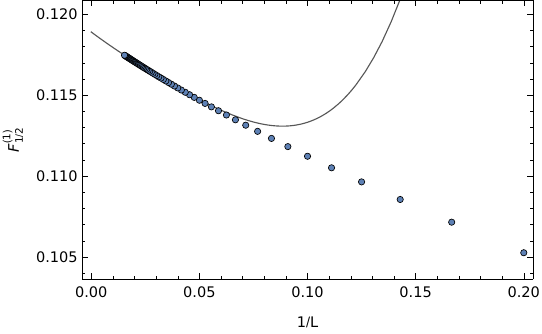}}\hfill{}

\caption{\label{fig:min_charge}Anomalous dimension of the $q\!=\! 1/2$ monopole
$\Delta_{1/2}^{(1)}$ (\ref{eq:delta_QED3GN}, \ref{eq:delta_QED3})
as a function of the relativistic cutoff $L$ (\ref{eq:rel_cutoff})
in (a) $\protect\QED$ ; (b) $\protect\QEDGN$. The points are obtained
by numerically computing Eqs.~(\ref{eq:delta_QED3GN}, \ref{eq:delta_QED3}),
and the solid line is a quartic fit in $1/L$ with the points $L\in[55,  65]$.}
\end{figure*}

After performing the sums in Eqs.~(\ref{eq:half_ZM}, \ref{eq:half_NZM}),
the kernel coefficients in Eq.~(\ref{eq:KZ_sum}) are computed and
inserted in Eq.~(\ref{eq:delta_QED3GN}) (or Eq.~(\ref{eq:delta_QED3})
for the case of $\QED$). The kernel coefficients in the denominator
of the logarithm of the monopole anomalous dimension were obtained
analytically in Eqs.~(\ref{eq:D_0_ana}-\ref{eq:KB_0_ana}). The
remaining sum on $\ell$ and integral on $\omega$ are computed up
to a relativistic cutoff~\citep{pufu_anomalous_2014}
\begin{equation}
\ell(\ell+1)+\omega^{2}\leq L\left(L+1\right).\label{eq:rel_cutoff}
\end{equation}
We obtained the anomalous dimension with a cutoff up to $L_{\max}=65$.
A function of $1/L$ is then fitted to extract the value of the anomalous
dimension as the full sum and integration are taken with $L\to\infty$.
Fig.~\ref{fig:min_charge} shows a quartic function fitted with
the data from $L\in[L_{\max}-10, L_{\max}]$. The anomalous dimension
of a charge $q=1/2$ monopole in $\QEDGN$ extracted from this fit
is $\Delta_{1/2,\QEDGN}^{(1)}=0.11890$, whereas in $\QED$
it is given by $\Delta_{1/2,\QED}^{(1)}=-0.03814$. This
reproduces the result in Ref.~\citep{pufu_anomalous_2014} up to
a difference of order $10^{-4}$.

While we extrapolated the result for $L\to\infty$ with a quartic
fit, based on a cutoff of $L_{\max}=65$, varying the maximal relativistic
cutoff can change the last digit in the result quoted above.  For
instance, $\Delta_{1/2,\QED}^{(1)}|_{L_{\max}=50}=-0.03815$. In App.~\ref{sec:fitting}, we show how we computed anomalous dimensions for  various $L_{\max}$ and used the trend as $L_{\max}\to\infty$ to estimate the anomalous dimensions and their errors. The error we quote in what follows reflects the uncertainties related to the extrapolations
and not the precision of our computation which yields relatively negligible
errors.

Using this method, the anomalous dimension of $q=1/2$ monopoles at
next-to-leading order in the $1/N$ expansion in $\QEDGN$ is given
by
\begin{align}
\Delta_{1/2,\QEDGN}^{(1)} & =0.118911(7).\label{eq:QED3GN_q0.5}
\end{align}
The scaling dimension of $q=1/2$ monopole operators in $\QEDGN$
is then given by $2N\times0.26510+0.118911(7)+O\left(N^{-1}\right)$.
In $\QED$, the correction we found is
\begin{equation}
\Delta_{1/2,\QED}^{(1)}=-0.038138(5).\label{eq:QED3_q0.5}
\end{equation}
With this estimated uncertainty of our result, it is clearer that
there is a small discrepancy when comparing our result with the correction
$-0.0383$ computed in Ref.~\citep{pufu_anomalous_2014}. Trying to
replicate the method in Ref.~\citep{pufu_anomalous_2014}, we used
a cubic fit with data $L\in[5,45]$ and obtained $-0.03823$, which
is closer to $-0.0383$. 

\subsubsection{Anomalous dimensions for general $q$\label{subsec:general_q}}

For larger topological charge $q$, many results used from App.~\ref{sec:min_charge_results}
are not easily generalized. In the previous section, the computations
involved three different Jacobi polynomials that appear after taking
the sum over an appropriate azimuthal quantum number. Here instead,
the real-space kernels and the eigenkernels will be written explicitly
as sums of monopole spherical harmonics, respectively with finite
charge $q$ and vanishing charge. This follows the alternative and
more algorithmic method presented in Ref.~\citep{dyer_monopole_2013}.

For the real-space kernels, the required formulation already appears
as an intermediate step in App.~\ref{subsec:GF} when obtaining the
Green's function in Eq.~(\ref{eq:GF}). The Green's function is a
$2\times2$ matrix acting on particle-hole space with components given
by the product of two monopole harmonics (\ref{eq:GF_monopole_harmonics_1}-\ref{eq:GF_monopole_harmonics_3}).
Consequently, the real-space kernels are formulated as products of
four monopole harmonics. As for the eigenkernels appearing in $D_{\ell}^{q}(\omega)$,
$H_{\ell}^{q,T}(\omega)$, they are already expressed as
the product of two spherical harmonics (\ref{eq:Dlo}, \ref{eq:FTlo}).
Only the gauge-gauge kernels then need a reformulation. In this case,
a different basis $U_{\ell,m}^{\mu}(\hat{n}),V_{\ell,m}^{\mu}(\hat{n}),W_{\ell,m}^{\mu}(\hat{n})$
for the vector spherical harmonics can be introduced. These are eigenfunctions
with respective total spin $j=\ell-1,\ell$ and $\ell+1$~\cite{pufu_anomalous_2014}.
Most importantly, the components of these harmonics are simply given
by spherical harmonics (see App.~\ref{subsec:second_basis}). We
discuss the relation with the previous basis shortly.

Before doing so, we note that, in this new formulation, the kernel
coefficients are expressed as the integral of a product of four monopole
harmonics and two spherical harmonics. To be more precise, half of
these functions are conjugate harmonics, but they can all be expressed
as harmonics with the following relation
\begin{equation}
Y_{q,\ell,m}^{*}(\hat{n})=\left(-1\right)^{q+m}Y_{-q,\ell,-m}(\hat{n}).\label{eq:conjugate}
\end{equation}
Just as we did in Sec.~\ref{subsec:Anomalous-dimensions}, the primed
coordinates can be fixed as $\tau^{\prime}=0$ and $\hat{n}^{\prime}=\hat{z}$
without loss of generality. As a result, half of the six harmonics
are eliminated
\begin{align}
Y_{q,\ell,m}\left(\hat{z}\right) & =\delta_{q,-m}\sqrt{\frac{2\ell+1}{4\pi}}.\label{eq:harmonic_z}
\end{align}
This removes every sum on azimuthal quantum numbers, which greatly
simplifies the computation. There remains an integral over three harmonics
\begin{equation}
\begin{split}\int d\hat{n}Y_{q,\ell,m}(\hat{n})Y_{q^{\prime},\ell^{\prime},m^{\prime}}(\hat{n})Y_{q^{\prime\prime},\ell^{\prime\prime},m^{\prime\prime}}(\hat{n})=\left(-1\right)^{\ell+\ell^{\prime}+\ell^{\prime\prime}}\\
\times\sqrt{\frac{\left(2\ell+1\right)\left(2\ell^{\prime}+1\right)\left(2\ell^{\prime\prime}+1\right)}{4\pi}}\begin{pmatrix}\ell & \ell^{\prime} & \ell^{\prime\prime}\\
q & q^{\prime} & q^{\prime\prime}
\end{pmatrix}\begin{pmatrix}\ell & \ell^{\prime} & \ell^{\prime\prime}\\
m & m^{\prime} & m^{\prime\prime}
\end{pmatrix}.
\end{split}
\end{equation}
The explicit expressions for the kernel coefficients involve the sum
of many such integrals and are not reproduced here.

Returning to the change of basis, the $U,V,W$ vector spherical harmonics
in the $j=\ell$ sector can be related to the harmonics previously
introduced in Eqs.~(\ref{eq:vector_harm_basis_1}-\ref{eq:vector_harm_basis_3})
by
\begin{align}
\begin{pmatrix}U_{\ell+1,m}^{\mu}(\hat{n})\\
W_{\ell-1,m}^{\mu}(\hat{n})\\
V_{\ell,m}^{\mu}(\hat{n})
\end{pmatrix} & =\begin{pmatrix}-\sqrt{\frac{\ell+1}{2\ell+1}} & \sqrt{\frac{\ell}{2\ell+1}} & 0\\
\sqrt{\frac{\ell}{2\ell+1}} & \sqrt{\frac{\ell+1}{2\ell+1}} & 0\\
0 & 0 & i
\end{pmatrix}\begin{pmatrix}\mathfrak{a}_{\ell m}^{T,\mu}(\hat{n})\\
\mathfrak{a}_{\ell m}^{E,\mu}(\hat{n})\\
\mathfrak{a}_{\ell m}^{B,\mu}(\hat{n})
\end{pmatrix}\nonumber \\
 & \equiv\mathcal{R}\begin{pmatrix}\mathfrak{a}_{\ell m}^{T,\mu}(\hat{n})\\
\mathfrak{a}_{\ell m}^{E,\mu}(\hat{n})\\
\mathfrak{a}_{\ell m}^{B,\mu}(\hat{n})
\end{pmatrix}.
\end{align}
The Fourier coefficients can also be transformed in this basis: 
\begin{equation}
\begin{split}\mathcal{R}\begin{pmatrix}K_{\ell}^{q,TT}(\omega) & K_{\ell}^{q,TE}(\omega) & 0\\
K_{\ell}^{q,TE*}(\omega) & K_{\ell}^{q,EE}(\omega) & 0\\
0 & 0 & K_{\ell}^{q,BB}(\omega)
\end{pmatrix}\mathcal{R}^{-1}\\
=\begin{pmatrix}K_{\ell}^{q,UU}(\omega) & K_{\ell}^{q,UW}(\omega) & 0\\
K_{\ell}^{q,UW*}(\omega) & K_{\ell}^{q,WW}(\omega) & 0\\
0 & 0 & K_{\ell}^{q,VV}(\omega)
\end{pmatrix}.
\end{split}
\label{eq:TEB-UWV}
\end{equation}
The matrix of eigenkernels keeps the same structure thanks to the
block-diagonal form of the transformation matrix $\mathcal{R}$. This
is expected, as we could also argue that the kernels $K_{\ell}^{q,UV}(\omega)=K_{\ell}^{q,WV}(\omega)=0$
vanish because of $CT$ invariance, as we did for $K_{\ell}^{q,TB}(\omega)=K_{\ell}^{q,EB}(\omega)=0$.
The relevant relations are then
\begin{align}
K_{\ell}^{q,VV}(\omega) & =K_{\ell}^{q,B}(\omega),\label{eq:KVV}\\
K_{\ell}^{q,UU}(\omega)+K_{\ell}^{q,WW}(\omega) & =K_{\ell}^{q,E}(\omega).
\end{align}
The first relation is found by comparing the bottom-right components
in Eq.~(\ref{eq:TEB-UWV}) and using the definition of $K_{\ell}^{q,B}(\omega)$
(\ref{eq:KB_def}), whereas the second relation is found by taking
the trace of Eq.~(\ref{eq:TEB-UWV}), using the first result in Eq.~(\ref{eq:KVV})
and the definition of $K_{\ell}^{q,E}(\omega)$ (\ref{eq:KE_def}).
The kernels $K_{\ell}^{q,E}(\omega)$ and $K_{\ell}^{q,B}(\omega)$
can then be replaced in the scaling-dimension corrections (\ref{eq:delta_QED3GN}-\ref{eq:delta_QED3})
by their formulation in the new basis.

For general charge, the regularization of the kernels presented in
Eqs.~(\ref{eq:ass_coeff}, \ref{eq:half_NZM}) is still valid : Regulator
terms $-1/\left(2\pi\right)$ and $-1/\left(4\pi\right)$ can be used
respectively for the scalar-scalar and gauge-gauge kernels, while
the scalar-gauge kernel does not require regularization. The contribution
of the zero modes using this method is also very straightforward and
algorithmic. However, there seems to be additional contributions coming
from the combination of zero modes in both Green's function, $\ell^{\prime}=\ell^{\prime\prime}=q$.
As discussed previously, this contribution is proportional to a Dirac
Delta function $\delta(\omega)$ instead of the energy
prefactor in Eq.~(\ref{eq:KZ_sum}). This contribution vanishes once
integrated over $\omega$ (see App.~\ref{sec:Only-zero-modes}).
Numerical sums on $\ell^{\prime}$ are obtained up to $\ell_{c}^{\prime}=200+q$,\footnote{We use a smaller cutoff for the general charge $q$ which is still sufficient for the precision needed and less computationally intensive.}
and the remainder is computed analytically with an expansion up to
$1/\ell^{\prime18}$. In this case, the coefficients of the expansion
also depend on the charge $q$ and are found by fixing $\ell$ and
$q$ for a few values.

\begin{figure*}[t]
\hfill{}\centering\subfigure[]{\label{fig:First-subfigure}\includegraphics[height=5.35cm]{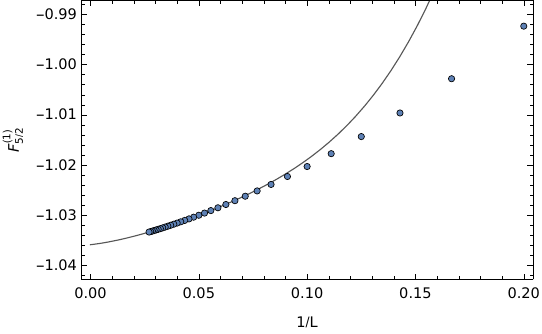}}\hfill{}\subfigure[]{\label{fig:Second-subfigure}\includegraphics[height=5.35cm]{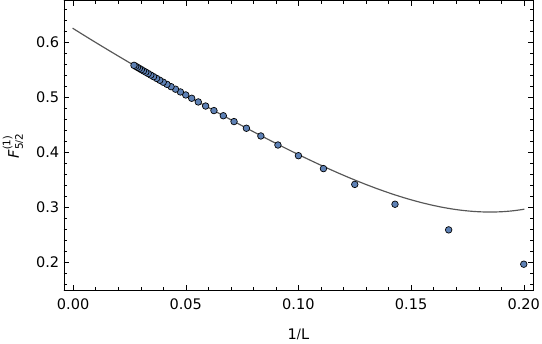}}\hfill{}

\caption{\label{fig:q2.5_dims}Anomalous dimension of the $q=5/2$ monopole
$\Delta_{5/2}^{(1)}$ (\ref{eq:delta_QED3GN}, \ref{eq:delta_QED3})
as a function of the relativistic cutoff $L$ (\ref{eq:rel_cutoff})
in (a) $\protect\QED$ ; (b) $\protect\QEDGN$. The points are obtained
by numerically computing Eqs.~(\ref{eq:delta_QED3GN}, \ref{eq:delta_QED3})
and the solid line is a quartic fit in $1/L$ with the points $L\in[L_{\max}-6, L_{\max}]$,
here $[31, 37]$.}
\end{figure*}

The anomalous dimension $\Delta_{q}^{(1)}$ for each charge
are computed with a relativistic cutoff $L_{\max}=35+\lfloor q\rceil$,
where $\lfloor q\rceil\equiv\text{Round}(q)$. Here, the convention
is that half-integers are rounded to even numbers, e.g. $\lfloor1/2\rceil=0$
and $\lfloor3/2\rceil=2$. The value of $L_{\max}$ is modulated with
the charge $q$ to ensure that a regime with a tail-like behaviour,
as observed in Fig.~\ref{fig:min_charge}, is attained for larger
charges. For the three minimal charges, we used a larger relativistic  cutoff: We use  $L_{\max}=65$ for $q=1/2$ (as in the last section) and $L_{\max}=46,47$ for $q=1, 3/2$.\footnote{In these cases, we use $\ell_{c}^{\prime}=300+q$. For $q=1/2$, we fit with $L\in [L_{\max}-10, L_{\max}]$ (as in the last section) whereas we fit the $q=1, 3/2$ cases with $L \in [L_{\max}-6, L_{\max}]$ (as other charges in this section). The number of points used for the fits is discussed further in App.~\ref{sec:fitting}.} We found that the results are robust as $L_{\max}$ is increased and more precise (see App.~\ref{sec:fitting}). The results for $L\in[L_{\max}-6, L_{\max}]$ are used to fit
a quartic function in $1/L$ to extrapolate the anomalous dimensions
$\Delta_{q}^{(1)}$ as $L\to\infty$. The fits obtained
for $q=5/2$ monopoles in $\QED$ and $\QEDGN$ are shown in Fig.~\ref{fig:q2.5_dims}
and yield scaling-dimension corrections $\Delta_{5/2,\QED}^{(1)}=-1.0359$
and $\Delta_{5/2,\QEDGN}^{(1)}=0.6253$ as $L\to\infty$.

As before, the uncertainty in the scaling dimension is estimated by
varying $L_{\max}$ and estimating the anomalous dimension as $L_{\max}\to\infty$.
More details are shown in App.\ \ref{sec:fitting} . The resulting
scaling dimensions up to $q=7/2$ obtained in this way are shown in
Tab.~\ref{tab:dimension_LO_NLO}. Comparing the $\QED$ monopoles'
anomalous dimensions with the results in Ref.~\citep{dyer_monopole_2013},
discrepancies of order $10^{-4}$ to $10^{-3}$ are again observed
for higher charges.

\begin{table}
\caption{ Leading-order and next-to-leading order in $1/N$ contributions to
monopole scaling dimensions in $\protect\QED$, $\protect\QEDGN$, and $\QEDZ$
models.  The latter model is discussed in Sec.~\ref{sec:Z2}. The leading-order result is the same in all models. The scaling
dimension in a given model is $\Delta_{q}=2N\Delta_{q}^{(0)}+\Delta_{q}^{(1)}+O\left(N^{-1}\right)$.
\label{tab:dimension_LO_NLO}}

\begin{ruledtabular}

\begin{tabular}{ccccc} 
 $q$ & 
$\Delta_{q}^{(0)}$ & 
$\Delta_{q,\QED}^{(1)}$ & 
$\Delta_{q,\QEDGN}^{(1)}$ \vspace{0.25em} &
$\Delta_{q,\QEDZ}^{(1)}$  \\ \hline  
$1/2$ & $0.26510$ & $-0.038138(5)$ & $0.118911(7)$ & $0.102846(9)$ \\ 
$1$ & $0.67315$ & $-0.19340(3)$ & $0.23561(4)$ & $0.18663(4)$ \\ 
$3/2$ & $1.18643$ & $-0.42109(4)$ & $0.35808(6)$ & $0.26528(7)$ \\ 
$2$ & $1.78690$ & $-0.70482(9)$ & $0.4879(2)$ & $0.3426(2)$ \\ 
$5/2$ & $2.46345$ & $-1.0358(2)$ & $0.6254(2)$ & $0.4202(3)$ \\ 
$3$ & $3.20837$ & $-1.4082(2)$ & $0.7705(3)$ & $0.4989(3)$ \\ 
$7/2$ & $4.01591$ & $-1.8181(2)$ & $0.9229(3)$ & $0.5789(4)$ \\ 
\end{tabular}

\end{ruledtabular}
\end{table}

The $q=1/2$ results obtained in Sec.~\ref{subsec:min_charge} are
successfully reproduced with the more general method. Monopole scaling
dimensions up to $q=13$ are shown in App.~\ref{sec:scaling_13}.

The next-to-leading order term in $1/N$ decreases the scaling dimension
of monopoles in $\QED$ whereas it increases for $\QEDGN$. That is,
quantum corrections help to stabilize the $\QEDGN$ model and destabilize
$\QED$. To understand the difference between both cases, it is useful
to write the scaling dimension as
\begin{align}
\Delta_{q,\QEDGN}^{(1)} & =\Delta_{q,\QED}^{(1)}+\Delta_{q,\GN}^{(1)}+\frac{1}{2}\int_{\omega}\sum_{\ell=1}^{\infty}\left(2\ell+1\right)\nonumber \\
 & \quad{}\times\ln\left[1+\dfrac{\left(1+\frac{\omega^{2}}{\ell(\ell+1)}\right)\left|H_{\ell}^{q,T}(\omega)\right|{}^{2}}{D_{\ell}^{q}(\omega)K_{\ell}^{q,E}(\omega)}\right],\label{eq:delta_QED3GN_decompose}
\end{align}
where $\Delta_{q,\GN}^{(1)}=\frac{1}{2}\int_{\omega}\sum_{\ell=0}^{\infty}\left(2\ell+1\right)\log\left[D_{\ell}^{q}(\omega)/D_{\ell}^{0}(\omega)\right]$
is the contribution in the $\QEDGN$ anomalous dimension~(\ref{eq:delta_QED3GN})
coming exclusively from the pseudo-scalar field.\footnote{It is also the expression for the anomalous dimension of monopoles
in a pure $\GN$ model, hence the label.} Computing $\Delta_{q,\GN}^{(1)}$ in the same way as we
did for $\Delta_{q,\text{\ensuremath{\QEDGN}}}^{(1)}$
and $\Delta_{q,\QED}^{(1)}$, we found this contribution
is positive $\Delta_{q,\GN}^{(1)}>0$ and more important
than the contribution coming exclusively from gauge fields $\big|\Delta_{q,\GN}^{(1)}\big|>\big|\Delta_{q,\QED}^{(1)}\big|$.
As for the remaining scalar-gauge contribution in the second line
of Eq.~(\ref{eq:delta_QED3GN_decompose}), it is also positive. To
see this, we must show that the second term in the logarithm is positive.
The numerator is explicitly positive. As for the denominator,  we
note that $\Delta_{q,\GN}^{(1)}$ is real, meaning that
$D_{\ell}^{q}(\omega)$ and the $D_{\ell}^{0}(\omega)$
must have the same sign. The latter $q=0$ kernel is positive, as
seen from Eqs.(\ref{eq:D_0_ana}, \ref{eq:D_0_cal}), meaning that
$D_{\ell}^{q}(\omega)>0$. The same goes for $K_{\ell}^{q,E}(\omega)$,
thus the denominator is positive $D_{\ell}^{q}(\omega)K_{\ell}^{q,E}(\omega)>0$.
This does not come as a surprise as these kernels are diagonal entries
in the Hessian matrix developed around a minimum saddle-point. The
scalar-gauge kernel thus gives a positive contribution to the anomalous
dimension in $\QEDGN$. It then must be that the $\QEDGN$ monopole
anomalous dimension is positive, $\Delta_{q,\QEDGN}^{(1)}>0$,
given what is known about each contribution on the RHS of Eq.~(\ref{eq:delta_QED3GN_decompose}).
It would be desirable to understand heuristically why quantum fluctuations
render monopoles less relevant at the QCP compared to deep in the
Dirac spin liquid.

In the $\CP^{N-1}$ model, similar relations between the different
contributions to the monopole anomalous dimension are found. A positive
contribution coming only from the auxiliary boson was found numerically
in Ref.~\citep{pufu_monopoles_2013}; the correction from the mixed
scalar-gauge kernel can be deduced as positive~\citep{dyer_scaling_2015,dyer_erratum_2016},
and the total anomalous dimension of monopoles in $\CP^{N-1}$ numerically
found in Ref.~\citep{dyer_erratum_2016} is also positive.  

\subsubsection{Convexity conjecture}
It was recently conjectured that CFT operators charged under a global $\U(1)$ symmetry respect the following convexity relation
\begin{equation}
 \Delta((n_1 + n_2)n_0) \geq     \Delta(n_1 n_0) + \Delta(n_2 n_0),
\end{equation}
for some positive integer $n_0$ of order 1~\citep{aharony_convexity_2021}. We test this conjecture using the monopole operators that are charged under  U(1)$_{\rm top}$. 
Here, $n_0,n_1,n_2$ are integers, where in our notation $\Delta(2q) \equiv \Delta_{q}$. Using the scaling dimensions we obtained in Tab.~\ref{tab:scaling_all} and extrapolating to finite $N$, we find this relation is respected for the monopoles under consideration in $\QED$, $\QEDGN$ (and also for the case $\QEDZ$ presented later on) for any $2N \in \mathbb{Z}_+$ starting from $n_0=1$, i.e.\/ the minimal possible value.

\section{Large-charge universality \label{sec:large-charge}}

In CFTs with a global U(1) symmetry, the related charge $q$ can be used as an expansion parameter by using effective-field-theory
methods.
It was shown that the lowest scaling dimension among charge-$q$ operators
has the following expansion at $q \gg 1$~\citep{hellerman_on_2015}:
\begin{equation}
\Delta_{q}=c_{3/2}q^{3/2}+c_{1/2}q^{1/2}+\gamma_{\U(1)}+ \dots ,\label{eq:deltaQ_expansion}
\end{equation}
where the ellipsis denotes negative half-integer and integer powers of $q$~\citep{cuomo_aNote_2021}. 
While $c_{3/2}$ and $c_{1/2}$ depend on the specific $\QFT$ considered,
the $O\left(q^{0}\right)$ coefficient is universal (theory independent)~\citep{hellerman_on_2015,monin_partition_2016}:
\begin{equation}
\gamma_{\U(1)}=-0.0937\dots\label{eq:gamma_expected}
\end{equation}
This coefficient is obtained by computing the Casimir energy of the
$\U(1)$ Goldstone mode. The Goldstone appears in the state-operator
correspondence where the charged operator insertion is mapped to a
state where the saddle-point configuration breaks the $\U(1)$
symmetry. 

This analysis applies to monopole operators in theories with the global
$\U_{\text{top}}(1)$ symmetry group. Given the universality
of the coefficient $\gamma$, no term at $O\left(q^{0}\right)$ should
be present at leading-order in the $1/N$ expansion\footnote{Here, the parameter $N$ is used to designate either $N$ complex
boson flavors or $2N$ fermion flavors.}, since the leading-order term is proportional to $N$ and thus non-universal.
This was indeed observed in $\QED$, and $\OO(2)$- and $\OO(3)$-$\QEDGN$
models\footnote{These models are also known as $\QED-$chiral XY $\GN$ and $\QED-$chiral
Heisenberg $\GN$ models, respectively.}~\citep{dupuis_transition_2019,dupuis_monopole_2021} as well as
in the $\CP^{N-1}$ model~\citep{dyer_scaling_2015,delaFuente_large_2018}.\footnote{While Ref.~\citep{dyer_scaling_2015} discusses only the $O\left(q^{3/2}\right)$
term of the large-$q$ expansion, it is straightforward to use their
analytical results to verify that no $O\left(Nq^{0}\right)$ term
is present.} Since $\QED$ and $\QEDGN$ monopoles have the same leading-order
scaling dimensions, as discussed in Sec.~\ref{sec:N_infty}, this
also applies to $\QEDGN$ monopoles.

\begin{figure}
\begin{centering}
\includegraphics[width=1\linewidth]{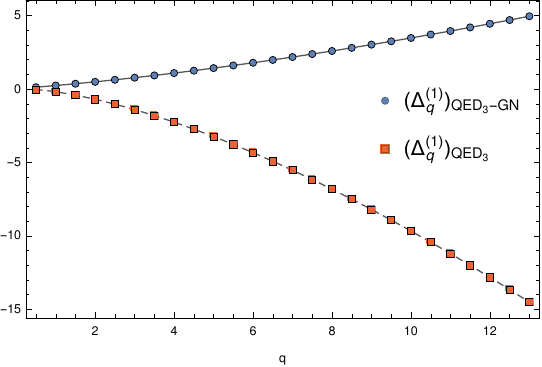}
\par\end{centering}
\caption{Anomalous dimensions of monopoles in $\protect\QEDGN$ and $\protect\QED$
fitted with the large-$q$ expansion~(\ref{eq:deltaQ_expansion}).
The points are the scaling dimension corrections obtained with a quartic
fit in $1/L$. The solid and dashed lines are the fitting functions
for $\protect\QEDGN$ and $\protect\QED$, respectively,  with a minimal power of $q^{-1}$. \label{fig:fitQ}}
\end{figure}

Using the monopole anomalous dimensions $\Delta_{q}^{(1)}$, the 
$O\left(q^{0}\right)$ coefficient $\gamma$ can be computed. This was done for
the $\CP^{N-1}$ model in Ref.~\citep{delaFuente_large_2018}, where
$\Delta_{q}^{(1)}$ was obtained for a hundred charges $q=1/2,1,\dots,50$
and the expected expansion~(\ref{eq:deltaQ_expansion}) is fitted
numerically to extract $\gamma$. A similar computation is performed
here for monopoles in the $\QEDGN$ and $\QED$ models. We fit  all monopole anomalous dimensions
in $\QEDGN$ and $\QED$ shown in Tab.~\ref{tab:scaling_all} by using the fitting function in Eq.~\eqref{eq:deltaQ_expansion} with powers down to $q^{-1}$~\cite{cuomo_aNote_2021}. The
fits and the anomalous dimensions are shown in Fig.~\ref{fig:fitQ};
note that the errors in the values of the anomalous dimension are
smaller than the dots in the figure. Including more powers in the
fitting function would yield significantly larger errors in the estimation
of $\gamma$.

The value of $\gamma$ obtained for each theory is
consistent with the expected universal value (\ref{eq:gamma_expected})
\begin{align}
\gamma_{\QED} & =1.02(4)\times\gamma_{\U(1)},\\
\gamma_{\QEDGN} & =1.01(6)\times\gamma_{\U(1)}.
\end{align}
 This is a nice consistency check of the anomalous scaling dimensions
obtained in the last section.

The universal coefficient of the scaling dimension of $\U(1)$-charged
operators in Eq.~(\ref{eq:deltaQ_expansion}) can also be formulated
with the following sum rule~\citep{hellerman_on_2015}
\begin{align}
q^{2}\Delta_{q}-\left(\frac{q^{2}}{2}+\frac{q}{4}+\frac{3}{16}\right)\Delta_{q-1}-\left(\frac{q^{2}}{2}-\frac{q}{4}+\frac{3}{16}\right)\Delta_{q+1}\nonumber \\
=-\frac{3}{8}\gamma_{\U(1)}+O\bigl(q^{-1/2}\bigr)=0.0351\dots
\end{align}
The RHS results from a cancellation of order $O\left(q^{3/2}\right)$
and $O\left(q^{1/2}\right)$ terms on the LHS. In comparison, the
error on the LHS coming from the triplet $\Delta_{q-1},\Delta_{q},\Delta_{q+1}$
is comparatively large, even more so as the coefficients in front
of the scaling dimensions, which are of order $q^{2}$, become larger
with increasing $q$. The resulting errors are too important to obtain
a reliable fit of the large-$q$ behaviour of this sum rule. Nevertheless,
we did observe a good qualitative agreement for the scaling dimensions
shown in Tab.\ \ref{tab:scaling_all}.

\section{CFT duality: $\protect\QEDGN$ and $\protect\CP^{1}$ models}
\label{sec:CFTDuality}

Another interesting application of our results concerns the duality
between the $\QEDGN$ model with $2N=2$ two-component Dirac fermion
flavors and the $\CP^{N-1}$ model with $N=2$ complex boson flavors~\citep{wang_deconfined_2018}.
Crucially, the duality between these models implies an emergent $\SO(5)$
symmetry. The following $\SO(5)$ multiplet in the $\QEDGN|_{2N=2}$ model
\begin{equation}
\begin{split}\bigg( & \Re(\psi_{1}^{\dagger}\widetilde{\mathcal{M}}_{1/2}^{f}), -\Im\bigl(\psi_{1}^{\dagger}\widetilde{\mathcal{M}}_{1/2}^{f}\bigr),\\
 & \Re(\psi_{2}^{\dagger}\widetilde{\mathcal{M}}_{1/2}^{f}), \Im(\psi_{2}^{\dagger}\widetilde{\mathcal{M}}_{1/2}^{f}), \phi\bigg),
\end{split}
\label{eq:plet_1}
\end{equation}
is dual to the following multiplet in the $\CP^{1}$ model
\begin{align}
 & \left(2\Re(\mathcal{M}_{1/2}^{b}), 2\Im(\mathcal{M}_{1/2}^{b}), z^{\dagger}\sigma_{1}z, z^{\dagger}\sigma_{2}z, z^{\dagger}\sigma_{3}z\right).\label{eq:plet_2}
\end{align}
Here $\widetilde{\mathcal{M}}_{1/2}^{f}$ are the $\QEDGN|_{2N=2}$
minimally charged monopoles which must be dressed with an additional
zero mode $\psi_{1}^{\dagger}$ or $\psi_{2}^{\dagger}$ on top of
a Dirac sea in order to be gauge invariant. These monopoles form an
$\SU(2$) doublet $\mathcal{M}_{1/2}^{f}=\left(\mathcal{M}_{1/2,\uparrow}^{f}, \mathcal{M}_{1/2,\downarrow}^{f}\right)^{\intercal}$.
On the $\CP^{1}$ side, $z$ is an $\SU(2)$ doublet $z=\left(z_{1}, z_{2}\right)^{\intercal}$
where each flavor is a complex boson and $\mathcal{M}_{1/2}^{b}$
is the minimally charged monopole. The $\SO(5)$ symmetry means that
all scaling dimensions within a multiplet should be equal, while operators
identified by the duality should also have the same scaling dimension.
Putting this together, this means that all the operators above should
have the same scaling dimension. A decent agreement was already observed
in Ref.~\citep{dupuis_transition_2019}, but the scaling dimension
of $\QEDGN$ monopoles were obtained only at leading-order in $1/N$,
with $\Delta_{\mathcal{M}_{1/2}^{f}}=0.53$. Updating the comparison
with the next-to-leading order correction, we find $\Delta_{\mathcal{M}_{1/2}^{f}}=0.65,$
which gives an even better agreement. For instance, the scaling dimension
of the $q=1/2$ monopole on the $\CP^{1}$ side obtained at next-to-leading
in $1/N$ is given by $\Delta_{\mathcal{M}_{1/2}^{b}}=0.63$\ \citep{dyer_scaling_2015,dyer_erratum_2016}.
In contrast, if we extrapolate the large-$N$ QED$_3$ result to $2N=2$, we obtain a scaling dimension of 0.49, which is further from the $\CP^1$ result, as expected since the two CFTs are not related by duality.
The scaling dimensions of the other operators in the duality also
show a good agreement coming from both analytical and numerical studies,
as shown in Tab.~\ref{tab:duality}.

\begin{table}
\caption{\label{tab:duality} Operators in the $\protect\SO(5)$ $\boldsymbol{5}$
multiplets (\ref{eq:plet_1}, \ref{eq:plet_2}) and their scaling dimensions.
``VBS'' and ``N\'eel'' make reference to operators whose scaling dimensions
are obtained numerically on lattices. The results for monopole operators
are obtained by using the state-operator correspondence at next-to-leading
order in $1/N$. The scaling dimension of the auxiliary boson $\phi$
in $\protect\QEDGN$ was obtained at order $1/N$ using the mean of
Pad\'e and Pad\'e-Borel $[0/1]$ resummations (non-resummed scaling dimension
are unphysical). The scaling dimension of the fermionic monopole operator
can also be resummed to $(0.59,0.68)$, but not in the bosonic case.
The operator $z^{\dagger}\bm{\sigma}z$ designates any of the boson
bilinears, i.e. flavor spin-1 in the bosonic side. It was obtained
at order $1/N^{2}$ in Ref.~\citep{benvenuti_qeds_2019} and using
functional renormalization group in Ref.~\citep{bartosch_corrections_2013}.}

\centering{}\begin{ruledtabular}%
\begin{tabular}{ccc}
$\mathcal{O}$ & $\Delta_{\mathcal{O}}$ & Ref.\tabularnewline
\hline 
$\mathcal{M}_{1/2}^{f}$ & $0.65$ & This work\tabularnewline
$\mathcal{M}_{1/2}^{b}$ & $0.63$ & \citep{dyer_erratum_2016}\tabularnewline
$\phi$ & $(0.59,0.64)$ & \citep{boyack_deconfined_2019}\tabularnewline
$z^{\dagger}\bm{\sigma}z$ & $0.64$ & \citep{benvenuti_qeds_2019}\tabularnewline
 & $0.61$ & \citep{bartosch_corrections_2013}\tabularnewline
VBS, N\'eel & $[0.60,0.68]$ & \citep{sandvik_evidence_2007,melko_scaling_2008,kaul_lattice_2012,pujari_neel_2013,nahum_deconfined_2015}\tabularnewline
\end{tabular}\end{ruledtabular}
\end{table}

In the same way, monopoles with the second smallest charge $q=1$
were argued to be part of the symmetric traceless $\boldsymbol{14}$
representation of $\SO(5)$~\citep{bernhard_deconfined_2018,benvenuti_qeds_2019}.
The various relevant scaling dimensions obtained with analytical methods
are also compared in Tab.~\ref{tab:duality2}. Again there is a very
good agreement between the scaling dimension of monopole operators,
with $\Delta_{\mathcal{M}_{1}^{f}}=1.58$ and $\Delta_{\mathcal{M}_{1}^{b}}=1.50$.
The agreement is weaker with other operators, but by taking into account
Pad\'e and Pad\'e-Borel resummations the duality prediction seems quite
reasonable. The scaling dimension related to auxiliary bosons $\Delta_{\phi^{2}}$
and $\Delta_{\lambda}$ obtained using the large-$N$ have greater
discrepancy with $\Delta\sim1$, but these expansions are not very
well controlled. However, the same can be said about the monopole
operator on the bosonic side. Overall, the duality for the $\boldsymbol{14}$
representation of $\SO(5)$ is not as convincing as for the $\boldsymbol{5}$,
but still reasonable for perturbative results. The scaling dimension
of the Lagrange field obtained using the functional renormalization
group also agrees reasonably well $\Delta_{\lambda}=1.21$~\citep{bartosch_corrections_2013}.

\begin{table}
\caption{\label{tab:duality2}Operators in the $\protect\SO(5)$ symmetric
traceless $\boldsymbol{14}$ multiplets and their scaling dimensions
predicted to be equal according to the duality between $\protect\QEDGN|_{2N=2}$
and $\protect\CP^{1}$ models. The scaling dimensions presented are
obtained analytically with the large-$N$ expansion. Pad\'e and Pad\'e-Borel
$[0/1]$ resummations are shown in parenthesis (apart from Ref.~\citep{boyack_deconfined_2019}, resummations are not obtained in the references cited). The symbol ``$\times$''
indicates unphysical results, i.e., negative scaling dimensions. The
operator $\lambda$ is the Lagrange multiplier field on the $\protect\CP^{1}$
side. Results for monopole operators are obtained using state-operator
correspondence at order $N^{0},$ while other results were obtained
at order $N^{-1}$. The resummed value for $\Delta_{\bar{\psi}\bm{\sigma}\psi}$
was obtained in Ref.~\citep{boyack_deconfined_2019} and is the same
for $\Delta_{(z^{*}\bm{\sigma}z)(z^{*}\bm{\sigma}z)^{\intercal}}$
at this order. }

\centering{}\begin{ruledtabular}%
\begin{tabular}{cccc}
$\mathcal{O}$ & $\Delta_{\mathcal{O}}$ & $(\Delta_{\mathcal{O}}^{\text{Pad\'e}},\Delta_{\mathcal{O}}^{\text{Pad\'e-Borel}})$ & Ref.\tabularnewline
\hline 
$\mathcal{M}_{1}^{f}$ & $1.58$  & $(1.63,1.75)$ & This work\tabularnewline
$\mathcal{M}_{1}^{b}$ & $1.50$  & $(\times,0.24)$ & \citep{dyer_erratum_2016}\tabularnewline
$\bar{\psi}\bm{\sigma}\psi$ & $1.19$  & $(1.42,1.51)$ & \citep{boyack_deconfined_2019,benvenuti_qeds_2019}\tabularnewline
$\left(z^{\dagger}\bm{\sigma}z\right)\left(z^{\dagger}\bm{\sigma}z\right)^{\intercal}$ & $1.19$ & $(1.42,1.51)$ & \citep{benvenuti_qeds_2019}\tabularnewline
$\phi^{2}$ & $4.43$ & $(\times,1.02)$ & \citep{boyack_deconfined_2019,benvenuti_qeds_2019}\tabularnewline
$\lambda$ & $\times$ & $(0.90,1.11)$ & \citep{halperin_first_1974,kaul_quantum_2008}\tabularnewline
\end{tabular}\end{ruledtabular}
\end{table}

The situation becomes more puzzling when the critical exponents in Tab.~\ref{tab:duality2} are compared to numerical lattice results. The apparent consistency observed in the analytical results (at least for operators that do not need resummation) does not hold for the numerical lattice results. Specifically, we compare the analytical results to the correlation length exponent $\nu$ obtained in many numerical studies of the $\CP^{1}$ model. This exponent is related to the Lagrange field scaling dimension as $\Delta_{\lambda}=3-1/\nu$. Its value has varied greatly among many numerical works. Earlier results indicate that
$\Delta_{\mathcal{\lambda}}\in[1.34,1.67]$~\citep{sandvik_evidence_2007,melko_scaling_2008,jiang_from_2008,lou_antiferromagnetic_2009}
which seems compatible with other scaling dimensions in Tab.~\ref{tab:duality2}. However, unusual scaling behaviour and the ``drifting'' of $\nu$ with increasing
lattice size~\citep{nahum_deconfined_2015} motivated further studies, and lower scaling dimensions have been found. The wide range of values obtained are shown in Tab.~\ref{tab:num}. Notably, a scaling dimension going down to $\Delta_{\lambda}=0.80(1)$ by considering the presence of a second length scale~\citep{sandvik_consistent_2020}.

The varying results among different lattice studies were also interpreted as a hint for a weakly first-order transition. This possibility has been discussed~\citep{benvenuti_qeds_2019,ihrig_abelian_2019} in a field theory context where the dual models $\QEDGN|_{2N=2}$ and $\CP^{N-1}|_{N=2}$
are possibly complex $\CFT$s emerging from the collision of fixed points as the number of matter flavors is lowered below a critical level. On the other hand, our analytical analysis shows there is still consistency among scaling dimensions on both sides of the duality. This may imply that the duality can still give valuable information, even if the CFT is non-unitary.

\begin{table}
\caption{\label{tab:num}Numerical determination of the correlation length
exponent $\nu$ and the related scaling dimension $\Delta_{\lambda}=3-1/\nu$
in lattice studies describing the $\protect\CP^{1}$ side. }

\centering{}\begin{ruledtabular}%
\begin{tabular}{ccc}
$\nu$ & $\Delta_{\mathcal{\lambda}}$ & Ref.\tabularnewline
\hline 
$0.78(3)$ & $1.72(5)$ & \citep{sandvik_evidence_2007}\tabularnewline
$0.68(4)$ & $1.52(9)$ & \citep{melko_scaling_2008}\tabularnewline
$\biggl\{ \!\!$ \begin{tabular}{c} $0.67(1)$ \\ $0.69(2)$ \end{tabular} \phantom{$\biggl\{ \!\!$}& \begin{tabular}{c} $1.51(3)$ \\ $1.55(5)$ \end{tabular} & \citep{lou_antiferromagnetic_2009}\tabularnewline
$0.62(2)$ & $1.39(5)$ & \citep{jiang_from_2008}\tabularnewline
$0.54(5)$ & $1.13(17)$ & \citep{pujari_neel_2013}\tabularnewline
$[0.51,0.69]$ & $[1.04,1.55]$& \citep{pujari_transitions_2015}\tabularnewline
$0.468(6)$ & $0.87(3)$ & \citep{nahum_deconfined_2015}\tabularnewline
$0.455(2)$ & $0.80(1)$ & \citep{sandvik_consistent_2020}\tabularnewline
\end{tabular}\end{ruledtabular}
\end{table}

A similar tension between the results from field theory and lattice
models was observed in Ref.\ \citep{zhijin_on_2021} where the $\QED|_{2N=2}$
model was studied using conformal bootstrap. The duality to
the easy-plane $\CP^{1}$ model  conjectured in Ref.~\citep{wang_deconfined_2018} implies a  self-duality and an emergent $\OO(4)$ symmetry on both sides. While
the conformal bootstrap study of $\QED|_{2N=2}$ is consistent with
the self-duality and the emergent symmetry, it contradicts results
from the lattice study of the easy-plane $\CP^{1}$ model\ \citep{qin_duality_2017}.

An interesting approach to understand these discrepancies could be that of pseudo-criticality, that is a weakly first-order transition with a generically long correlation length. In Ref.~\citep{ma_theory_2020}, a Wess-Zumino-Witten model in $2+\epsilon$ dimensions, with target space $S^{3+\epsilon}$, with global symmetry $\SO(4+\epsilon)$ has been shown to exhibit this behaviour and consistent with numerical results in the literature.  A crucial point was that the physical dimension $d=3$ is close to the critical dimension $d=2.77$ where  fixed points collide. Pseudo-criticality was also found in a  loop model describing the easy-plane N\'eel-VBS  transition~\citep{emergence_serna_2019}. 

\subsection*{Higher charge }

The duality can be tested further by comparing monopoles on both sides
of the duality. First, the relation between minimally charged monopoles
is further discussed. This relation is simpler to see with the appropriate
sub-models. A duality between $\QED|_{2N=2}$ and easy-plane $\CP^{1}$
was formulated by including additional external gauge fields $B_{\mu},B_{\mu}^{\prime}$
and Chern-Simons terms~\citep{karch_particle_2016,wang_deconfined_2018}
\begin{align}
 & \left|D_{b+B}z_{1}\right|^{2}+\left|D_{b+B^{\prime}}z_{2}\right|^{2}-\left|z_{1}\right|^{4}-\left|z_{2}\right|^{4}\nonumber \\
 & \quad{}-\frac{1}{2\pi}bd\left(B+B^{\prime}\right)-\frac{1}{2\pi}BdB^{\prime}-\frac{1}{2\pi}B^{\prime}dB\\
\Leftrightarrow\quad & \overline{\psi}_{1}i\slashed{D}_{a-B}\psi_{1}+\overline{\psi}_{2}i\slashed{D}_{a+B}\psi_{2}+\frac{1}{2\pi}adB^{\prime}\nonumber \\
 & \quad{}+\frac{1}{4\pi}\left(BdB-B^{\prime}dB^{\prime}\right),
\end{align}
where $b_{\mu}$ and $a_{\mu}$ are the dynamical gauge fields in
bosonic and fermionic models, respectively. By inspecting the charges
under the external gauge fields $\left(q_{B},q_{B^{\prime}}\right)$,
we can identify the following bosonic operators 
\begin{equation}
(2\mathcal{M}_{1/2}^{b},2z_{1}^{*}z_{2})\Leftrightarrow\left((\psi_{1}^{\dagger}\widetilde{\mathcal{M}}_{1/2}^{f})^{\dagger},\psi_{2}^{\dagger}\widetilde{\mathcal{M}}_{1/2}^{f}\right).
\end{equation}
Here, the first and second component on both sides have charges $(1,1)$
and $(1,-1)$ under $B_{\mu}$ and $B'_{\mu}$. While an $\SU(2)$
doublet structure is manifest in the RHS with the fermion zero modes,
it is less clear in the LHS and may be seen as a non-trivial corollary
of the duality. This may however be motivated by the self-duality
in the easy-plane $\CP^{1}$ model. The VBS order in the original
model $\mathcal{M}_{1/2}^{b}$ is mapped to the XY order in terms
of the dual bosons $w_{1}^{*}w_{2}$. Conversely, monopoles in the
dual side are mapped to $z_{1}^{*}z_{2}$ in the original model.

\begin{figure}[t]
\begin{centering}
\includegraphics[width=1\linewidth]{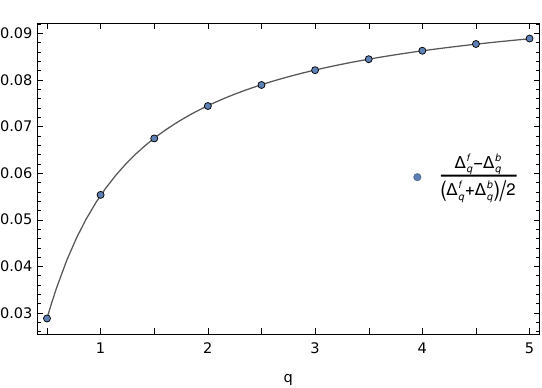}
\par\end{centering}
\caption{Relative difference between the scaling dimensions of monopoles in
$\protect\QEDGN|_{2N=2}$ and $\protect\CP^{1}$ models as a function
of the topological charge. The computation is done with next-to-leading-order
results in both models. The solid line is a fit $f_{0}+f_{-1}q^{-1}+f_{-3/2}q^{-3/2}$,
where the asymptote for large charge is approximately a $10\%$ relative
difference. The powers used in the fitting function are deduced from Eq.~\eqref{eq:deltaQ_expansion}.   \label{fig:duality_check}}
\end{figure}

These relations between operators translate back to the $\QEDGN|_{2N=2}\Leftrightarrow\CP^{1}$
duality. In particular, it is useful to focus on the dual relation
between the $\CP^{1}$ monopole $2\mathcal{M}_{1/2}^{f}$ and the corresponding
dual monopole in $\QEDGN|_{2N=2}$, $\bigl(\psi_{1}^{\dagger}\widetilde{\mathcal{M}}_{1/2}^{f}\bigr)^{\dagger}$.
For convenience, we define the following monopole operator $\mathcal{M}_{1/2}^{f}(x)\equiv2\bigl(\psi_{1}^{\dagger}\widetilde{\mathcal{M}}_{1/2}^{f}\bigr)^{\dagger}$.
Our starting point is then the conjectured dual relation between minimally
charged monopoles in $\QEDGN|_{2N=2}$ $(\mathcal{M}_{1/2}^{f})$
and in $\CP^{1}$ $(\mathcal{M}_{1/2}^{b})$ models 
\begin{equation}
\mathcal{M}_{1/2}^{f}(x)\Leftrightarrow\mathcal{M}_{1/2}^{b}(x).
\end{equation}
Using this relation and the operator product expansion ($\OPE$) 
\begin{equation}
\mathcal{O}_{1}(x)\mathcal{O}_{2} (y)=\sum_{n}c_{n}(x-y)\mathcal{O}_{n}(y),
\end{equation}
the scaling dimensions of higher charge monopoles can also be compared.
The $\OPE$ of two $q=1/2$ monopole operators yields the expansion
over $q=1$ operators
\begin{align}
\lim_{y\to x}\mathcal{M}_{1/2}(x)\mathcal{M}_{1/2}(y) & =\lim_{y\to x}c(x-y)\mathcal{M}_{1}(x)+\cdots
\end{align}
where the ellipsis stands for other primary operators with larger
scaling dimensions. By definition, $\mathcal{M}_{1}(x)$
has the smallest scaling dimension in the $q=1$ topological sector.
We can then identify the scaling dimension of $q=1$ monopole operators
on both sides of the duality $\Delta_{q=1}^{f}=\Delta_{q=1}^{b}.$
This is expected, as these monopoles are conjectured to be components
dual $\SO(5)$ symmetric traceless $\boldsymbol{14}$ multiplet~\citep{bernhard_deconfined_2018,benvenuti_qeds_2019}.
Using the same logic for higher charge monopoles, we find more generally
that 
\begin{equation}
\Delta_{q}^{f}=\Delta_{q}^{b}.
\end{equation}
Comparing our results for $\QEDGN|_{2N=2}$ monopoles in Tab.~\ref{tab:dimension_LO_NLO}
to $\CP^{1}$ monopoles in Ref.~\citep{delaFuente_large_2018}, we
obtain a good agreement for higher charges, as shown in Fig.~\ref{fig:duality_check}.
For larger charges, the relative difference tends to $10\%$. This
is a great improvement compared to the results obtained with only
the leading-order scaling dimensions: the behaviour is similar and
the asymptotic relative difference for large $q$ is $76\%$ instead.

\section{Transition to $Z_{2}$ spin liquids \label{sec:Z2}}

In this section, we consider quantum critical transitions to $Z_2$ spin liquids. The pairing of the spinons gaps out the gauge field through the Higgs mecanism. We begin with the most symmetric pairing interaction, and we then discuss the more general case where the pairing further breaks the flavor symmetry. 

\subsection{Symmetric $Z_{2}$ spin liquid}

The transition out of the U(1) DSL to a $Z_{2}$ spin liquid can also be studied with a gauged Gross-Neveu model~\cite{lu_Z2SL_2011,lu_unification_2017,zerf_superconducting_2016,boyack_transition_2018}. 
The Lagrangian describing this transition is written in euclidean flat spacetime as \begin{align}
\label{eq:LZ2SL}
\mathcal{L}_{\psi}&=\sum_{i=1}^{2N}-\overline{\psi}_{i}(\slashed{\partial}-i\slashed{\mathcal{A}}_{q}-i\slashed{A})\psi_{i}\nonumber\\&\quad+\sum_{i=1}^{2N}(\phi^{*}\psi_{i}^{T}i\gamma_{2}\psi_{i}+\text{h}.\text{c}.),
\end{align}
where $\phi$ is a complex scalar that decouples a quartic superconducting pairing term for the fermions. The interaction term included preserves Lorentz invariance. As $\phi$ describes Cooper pairs, it transforms as $\bar\psi_i i\gamma_2\bar\psi_i^T$ under U(1) gauge transformations. The Yukawa interaction term in the above equation is thus gauge invariant. In the $Z_2$ QSL, $\phi$ acquires an expectation value, which Higgses the gauge field, leading to a gapped $s$-wave superconducting state for the Dirac fermions.

We recall that the Dirac conjugate is defined by $\overline{\psi}_{i}=\psi_{i}^{\dagger}\gamma_{0}$
and $\mathcal{A}^{q}_{\mu}$ is the external gauge field that sources the
flux of $4\pi q$. The gauge-covariant derivative for the external gauge field $\mathcal{A}_{\mu}^{q}$ on a curved spacetime is defined in Eq.~\eqref{eq:SlashDeriv}. We now introduce the Nambu spinor defined as
\begin{equation}
\Upsilon_{i}=\left(\begin{array}{c}
\psi_{i}\\
i\gamma_{2}\overline{\psi}_{i}^{T}
\end{array}\right)=\left(\begin{array}{c}
\psi_{i}\\
C\overline{\psi}_{i}^{T}
\end{array}\right).
\label{eq:NambuSpinor}
\end{equation}
In addition, we define $\mathcal{C}=\text{diag}(C,C)$,
where $C=i\gamma_{2}$. The $C$ operator obeys $C^{2}=-1$, $C^{T}=C^{-1}=-C$,
and $C\gamma_{\mu}C=\gamma_{\mu}^{T}$. The transpose of the Nambu
spinor is given by $\Upsilon_{i}^{T}=(\psi_{i}^{T},\overline{\psi}_{i}C^{T})=(\psi_{i}^{T},-\overline{\psi}_{i}C)$.
Thus, the fermionic action can be expressed as 
\begin{equation}
S_{\Upsilon}=\frac{1}{2}\int_{r,r^{\prime}}\Upsilon_{i}^{T}(r)\mathcal{C}\mathcal{G}^{-1}(r,r^{\prime})\Upsilon_{i}(r^{\prime}),
\end{equation}
where the (inverse) Nambu Green's function is 
\begin{equation}
\label{eq:InvGF}
    \mathcal{G}^{-1}(r,r^{\prime})=
\begin{pmatrix}2\phi^{*}(r) & -\slashed{D}_{-(A+\mathcal{A}_{q})}\\
-\slashed{D}_{A+\mathcal{A}_{q}} & 2 \phi(r)
\end{pmatrix} \delta(r-r^{\prime}).
\end{equation}

As in Sec.~\ref{sec:-corrections}, the fields $\phi$ and $A$ are expanded about their saddle-point
values as $\phi=\left\langle \phi\right\rangle +\sigma/\sqrt{2N}$,
$A=\left\langle A\right\rangle +a/\sqrt{2N}$, where the
fluctuations are suppressed by $1/\sqrt{2N}$. At the
QCP, the saddle-point values are $\left\langle \phi\right\rangle =\left\langle A\right\rangle=0$~\citep{boyack_transition_2018}. 
Thus, in terms of the saddle-point and
fluctuation fields, the inverse Green's function is 
\begin{align}
 \mathcal{G}^{-1}(r,r^{\prime}) & = \mathcal{G}_{0}^{-1}(r,r^{\prime})+\frac{1}{\sqrt{2N}}X_{\sigma}(r)\delta(r-r^{\prime})
 \nonumber\\&\quad+\frac{1}{\sqrt{2N}}X_{a}(r)\delta(r-r^{\prime}).
\end{align}
Here, $\mathcal{G}^{-1}_{0}$ is the bare inverse Green's function, determined from the gauge covariant derivative term involving $\mathcal{A}^{q}_{\mu}$ in Eq.~\eqref{eq:InvGF}, and $X_{\sigma}$ and $X_{a}$ are given by 
\begin{equation}
X_{\sigma}=
2 \begin{pmatrix}\sigma^{*} & 0\\
0 & \sigma
\end{pmatrix},\quad
X_{a}=
\begin{pmatrix}0 & -\slashed{a}\\
\slashed{a} & 0
\end{pmatrix}.
\end{equation}
Integrating out the fermions then gives the effective action as
$\int\mathcal{D}\Upsilon\exp(-S_{\Upsilon})\equiv\exp(-S_{\eff}),$ where $S_{\eff}=-\frac{1}{2}(2N)\text{Tr}\log\mathcal{G}^{-1}$. We let
$\text{Tr}$ denote a ``trace'' over all relevant degrees of freedom, whereas $\text{tr}$ denotes a trace over spinor components.
To compute the effective action, we express the fermionic action as
a quadratic form in the fluctuation fields and perform a Gaussian
functional integral over $a$ and $\sigma$. The linear terms in $a$ and $\sigma$ vanish due to the saddle-point conditions for $A$ and $\phi$. Thus, to quadratic order, the effective action becomes  
\begin{align}
\label{eq:SeffLargeN}
S_{\eff} & = \left.S_{\eff}\right|_{\text{s.p.}}+\frac{1}{4}\text{Tr}\,\mathcal{G}_{0}X_{a}\mathcal{G}_{0}X_{a}+\frac{1}{4}\text{Tr}\,\mathcal{G}_{0}X_{a}\mathcal{G}_{0}X_{\sigma}\nonumber\\ &\quad +\frac{1}{4}\text{Tr}\,\mathcal{G}_{0}X_{\sigma}\mathcal{G}_{0}X_{a}
 +\frac{1}{4}\text{Tr}\,\mathcal{G}_{0}X_{\sigma}\mathcal{G}_{0}X_{\sigma}.
\end{align}

The fluctuations are $O(1/(2N))$, thus they cancel
the prefactor $2N$.  The second and third terms involve both the gauge field and the scalar field. By taking the trace over the Nambu matrix structure, these terms are found to vanish. Indeed, these terms must vanish from gauge invariance. Hence, only the gauge-gauge and scalar-scalar kernels contribute, which is in contrast to the QED$_3$-GN case where mixing between the two sectors exists. 
After performing the trace over the Nambu indices, 
the scalar-scalar kernel is 
\begin{align}
&\frac{1}{4}\text{Tr}\, \mathcal{G}_{0}(r^{\prime},r)X_{\sigma}(r)\mathcal{G}_{0}(r,r^{\prime})X_{\sigma}(r^{\prime}) \nonumber\\
& = \int_{r,r^{\prime}}\sigma^{*}(r) D(r,r^{\prime})\sigma(r^{\prime}),
\end{align}
where the scalar kernel is the same as in Eq.~\eqref{eq:D_rr'}.
Similarly, the gauge-gauge kernel is the same as in QED$_{3}$:
\begin{align}
\label{eq:GaugeGaugeKernel}
&\frac{1}{4}\text{Tr}\, \mathcal{G}_{0}(r^{\prime},r)X_{a}(r)\mathcal{G}_{0}(r,r^{\prime})X_{a}(r^{\prime})\nonumber\\ &= \frac{1}{2}\int_{r,r^{\prime}}a_{\mu}(r)K_{\mu\nu}(r,r^{\prime})a_{\nu}(r^{\prime}),
\end{align}
where the gauge-gauge kernel is the same as in Eq.~\eqref{eq:K_rr'}.
Combining these two results we find that the fluctuation action (obtained after integrating out $\sigma$ and $a$) is just the sum of twice the pure $\GN$ and the $\QED$ results. The anomalous dimension for the minimal charge $q=1/2$ is thus deduced to be 
\begin{equation}
\label{eq:Delta_QED_Z2}
\Delta_{\QEDZ}^{(1)}=\Delta_{\QED}^{(1)}+2\Delta_{\GN}^{(1)}=0.102846(9).
\end{equation}
The value and the error are estimated in the same way as described in Sec.~\ref{sec:-corrections} and App.~\ref{sec:fitting} by finding an expression similar to Eq.~\eqref{eq:delta_QED3GN} for the $\QEDZ$ case. The result is surprisingly close to the $\QEDGN$ case, although the quantum fluctuations possess a different structure at the two transitions. In Table \ref{tab:dimension_LO_NLO}, we give the answer for higher $q$. It can be seen that the values of the anomalous dimensions for $q>1/2$ for the CSL and $Z_2$ QSL are not as close as in the case of the minimal charge. By using anomalous dimensions up to $q=13$ shown in App.~\ref{sec:scaling_13}, one can again confirm the value of the universal coefficient for CFTs with a $\U(1)$ global symmetry as described in Sec.~\ref{sec:large-charge}: In this case, we find $\gamma_{\QEDZ}=0.98(7) \times \gamma_{\U(1)}$.

\subsection{More general $Z_{2}$ spin liquids}

Let us now consider a more general (superconducting) pairing interaction given by 
\begin{equation}
\mathcal{L}_{\mathrm{int}}=\sum_{i=1}^{2N}\phi^{*}_{I}\psi^{T}_{i}CM^{I}_{ij}\psi_{j}+\mathrm{h.c.}.
\label{eq:Z2Gen}
\end{equation}
Here, $C$ is the same as in the previous subsection -- an antisymmetric and unitary matrix with indices denoting the Dirac indices of the spinor $\psi$. The additional term in the interaction $M^{I}$ represents a ``flavor'' matrix, which is symmetric and has indices in the valley and spin spaces. We shall consider some simple concrete examples of $M$ later in this section. The index $I$, which is implicitly summed over, corresponds to the number of charged scalar fields, i.e., the competing pairing channels. In the previous section $I=1$ and $M$ corresponds to the identity operator. 

The analysis for this more general pairing interaction follows the same lines as before. The Nambu spinor is the same as in Eq.~\eqref{eq:NambuSpinor}, however, the inverse Green's function now becomes 
\begin{equation}
\label{eq:InvGF2}
    \mathcal{G}^{-1}(r,r^{\prime})=
\begin{pmatrix}2\phi_I^{*}(r)M^{I} & -\slashed{D}_{-(A+\mathcal{A}_{q})}\\
-\slashed{D}_{A+\mathcal{A}_{q}} & 2 \phi_{I}(r)M^{I}
\end{pmatrix}\delta(r-r^{\prime}).
\end{equation}
Here we suppose that $M$ is hermitian. The inverse Green's function can be expanded again using the large-$N$ formalism, and the effective action can be similarly expressed as in Eq.~\eqref{eq:SeffLargeN}. The terms involving both the gauge field and the scalar field again vanish due to gauge invariance, and the gauge-gauge term is the same as in Eq.~\eqref{eq:GaugeGaugeKernel}. The scalar-scalar kernel is now
\begin{align}
& \frac{1}{4}\text{Tr}\mathcal{G}_{0}(r^{\prime},r)X_{\phi}(r)\mathcal{G}_{0}(r,r^{\prime})X_{\phi}(r^{\prime}) \nonumber\\ & = -\sum_{I,J}\int_{r,r^{\prime}}\phi_I^{*}(r)D(r,r^{\prime})\phi_{I}(r^{\prime})\mathrm{tr}((M^{I})^2)\delta_{IJ}.
\end{align}
where we have assumed that the different channels labelled by $I$ are orthogonal: $\mathrm{tr}(M^I M^J)=0$ if $I\neq J$.
Then, the anomalous dimension is
\begin{equation}
\label{eq:Delta_QED3Z2GN}
\Delta_{\QEDZ'}^{(1)}=\Delta_{\QED}^{(1)}+2\Delta_{\GN}^{(1)} \frac{\sum_I\mathrm{tr}[(M^{I})^2]}{2N}.
\end{equation}
In the previous section, $M^{I}$ was the identity operator and so $\mathrm{tr}[(M^{I})^{2}]=2N$, which leads to the previous result for the anomalous dimension in Eq.~\eqref{eq:Delta_QED_Z2}. As another example, consider the case where the pairing interaction is of the form $\phi_x \psi^T C \sigma_x\psi+ \phi_z \psi^T C \sigma_z\psi$; note that we cannot use $\sigma_y$ since it is not symmetric. In this case $\sum_{I}\mathrm{tr}[(M^{I})^{2}]=4N$, and so the anomalous dimension of the second term in Eq.~\eqref{eq:Delta_QED3Z2GN} is now four times the pure-GN result.

Note that the superconducting pairing generally reduces the flavor (spin/valley) global symmetry. Therefore, monopole operators with different flavor quantum numbers are expected to have different scaling dimensions, resulting in a hierarchy of monopoles~\cite{dupuis_monopole_2021}. As our present formalism selects only the monopole with the smallest scaling dimension, a constraint on the flavor quantum numbers would be needed to describe other monopoles. Moreover, since the pairing field cannot have an expectation value for gauge invariant monopoles, it is expected that next-to-leading corrections are necessary to observe this hierarchy. Generalizing Ref.~\cite{dupuis_monopole_2021} to quantify this effect would be an interesting avenue to explore. 

To understand these more general $Z_{2}$ spin liquids, we analyze the pairing Hamiltonian in further detail. In particular, here we shall focus on the Bogoliubov-de Gennes (BdG) Hamiltonian for the mean-field description of Eq.~\eqref{eq:Z2Gen}. In the preceding section we formulated the theory in terms of a Euclidean Lagrangian description. Since the Hamiltonian $\mathcal{H}$ is the time-component of an energy momentum tensor, it is necessarily a non-Lorentz-invariant entity. Thus, here we shall use Minkowski spacetime to perform the analysis, which enables standard field theory methods to determine $\mathcal{H}$ from $\mathcal{L}$. 

We define $\mathcal{H}=\pi\dot{\psi}-\mathcal{L}$, where $\pi$ is the canonical momentum conjugate to $\psi$. From Eqs.~\eqref{eq:LZ2SL} (without the gauge field) and \eqref{eq:Z2Gen}, we construct the following Hamiltonian 
\begin{equation}
\mathcal{H}=-\left(i\psi^{\dagger}_{i}\gamma^{0}\boldsymbol{\gamma}\cdot\boldsymbol{\nabla}\psi_{i}+\phi^{*}_{I}\psi^{T}_{i}CM^{I}_{ij}\psi_{j}+\mathrm{h.c.}\right).
\end{equation}
Our choice of gamma matrices is given by $\gamma^{\mu}=\left(\tau_{z},i\tau_{x},i\tau_{y}\right)$; this definition is consistent with the Clifford algebra with a mostly minus metric. Here, $C=i\tau_{y}$.  
We define the Nambu spinor $\chi(k)$ by 
\begin{equation}
\chi_{i}(k)=\left(\begin{array}{c}
\psi_{i}(k)\\
C\psi^{*}_{i}(-k)
\end{array}\right).
\label{eq:NambuSpinor}
\end{equation}
In terms of the Nambu spinor $\chi(k)$, the Hamiltonian has the following form in momentum space:
\begin{equation}
\mathcal{H}=\frac{1}{2}\int_{k}\chi_{i}^{\dagger}\left(k\right)\left(\begin{array}{cc}
\delta_{ij}\gamma^{0} \bm{\gamma} \cdot \bm{k} & -2\phi_{ij}\\
-2\phi^{*}_{ij} & -\delta_{ij} \gamma^{0} \bm{\gamma} \cdot \bm{k}
\end{array}\right)\chi_{j}\left(k\right),
\end{equation}
where $\phi_{ij}^* = \phi_I^*M_{ij}^I$. In general, the Hamiltonian can be expressed as $\mathcal{H} = \frac{1}{2} \int_k\chi_{i}^{\dagger}\left(k\right) H(k)\chi_{j}\left(k\right)$, where $H(k)$ is the BdG matrix. In the simplest case we have $2N=2$, that is, there are 2 spin degrees of freedom, and the BdG Hamiltonian is $8\times8$. In the previous section we considered the case where the pairing matrix is the identity, $M^{I}=\Delta\sigma_{0}$, where $\Delta$ is the finite value of the pairing term in the $Z_2$ spin liquid phase. 

Here we contemplate some simple examples of pairing matrices, namely  $M^{I}=\Delta\sigma_{x}$ or $M^{I}=\Delta\sigma_{z}$. 
For these classes of spin liquids, the BdG matrix is of the form
\begin{align}
H(k)&=\sigma_{0}\otimes\left(\begin{array}{cc}
1 & 0\\
0 & -1
\end{array}\right)\otimes\left(-\tau_{y}k_{x}+\tau_{x}k_{y}\right)\nonumber\\
&\quad+\sigma_{l}\otimes\left(\begin{array}{cc}
0 & -2\Delta\\
-2\Delta^{*} & 0
\end{array}\right)\otimes\tau_{0},  
\end{align} 
Here $\sigma_{l}=\sigma_{0},\sigma_{x},$ or $\sigma_{z}$. 
The various matrices appearing above correspond to the spin indices, Nambu indices, and finally the Dirac indices, respectively. For the simple case where $\sigma_{l}=\sigma_{0}$, the eigenvalues are given by $E(k)=\pm\sqrt{{\bf k}^{2}+4\left|\Delta\right|^{2}},$ with a fourfold degeneracy. 
These eigenvalues are exactly the same as in the Fu-Kane model at half filling~\cite{fu_topological_2007}. Indeed, in the Fu-Kane model the Hamiltonian is a $4\times4$ matrix with spin and Nambu indices. Here we have two copies of this model Hamiltonian for each species of spin. In the case where $\sigma_{l}=\sigma_{x}$ or $\sigma_{z}$, we also have the same dispersion. In general, this dispersion describes a gapped $\mathbb{Z}_{2}$ spin liquid. The QCP in the present model is generally thought to be well defined at modest values of $N$; we can incorporate $N$ copies of the valley degrees of freedom and obtain the same (copies) of the eigenvalues. 

For the case where we have two competing channels, $M^{I}=\Delta_{x}\sigma_{x}+\Delta_{z}\sigma_{z}$, the dispersion is given by $E_{\pm,\pm}(k)=\pm\sqrt{{\bf k}^{2}+\left|2\Delta_{x}\right|^{2}+\left|2\Delta_{z}\right|^{2}\pm4\left|\Delta_{x}\Delta_{z}^{*}-\Delta_{x}^{*}\Delta_{z}\right|}$, with twofold degeneracy. In the case where either $\Delta_{x}$ or $\Delta_{z}$ is equal to zero, we recover the previous result for the eigenvalues. Interestingly, for the specific case where $\left|\Delta_{x}\right|=\left|\Delta_{z}\right|$ and  $\mathrm{arg}(\Delta_{x})-\mathrm{arg}(\Delta_{z})=\pm\pi/2$ (mod $\pi$), there are gapless Dirac cones, $\pm k$. For such pairings, we thus have a gapless $Z_2$ spin liquid with massless relativistic fermions, and a gapped $Z_2$ gauge field. 

This last case can be reformulated as a pairing term given by $M^I = |\Delta|\vec{d}\cdot \sigma_y \vec{\sigma}$ with $\vec d = e^{i\varphi}(1, 0, \pm i)$. This expression is similar to the time-reversal-breaking triplet state defined with $\vec d = (1, i, 0)$,  notably used to describe an $\text{La}\text{Ni}\text{C}_2$ compound~\cite{yanagisawa_nonunitary_2012}, as well as a potential order parameter in twisted bilayer  graphene~\cite{christos_superconductivity_2020}. This situation occurs when the general condensates $\Delta_0, \Delta_z$ in $M^I = \Delta_0 \sigma_0 + \Delta_z \sigma_z$ are aligned in the complex plane, $\mathrm{arg}(\Delta_{0})-\mathrm{arg}(\Delta_{z}) \in \{0,\pi\}$.

An important quantity for gapped systems is the Chern number~\cite{bernevig_hughes_topological_2013}, which corresponds to the flux of the Berry curvature in the Brillouin zone (BZ). In the presence of a gap the Chern number is an integer and describes a topological property of the system. Non-zero Chern number indicates broken time reversal symmetry, however, the converse is not always true. For a 2D system, it is defined by 
\begin{equation}
    C=\frac{1}{2\pi}\int d^{2}k\mathcal B(k).
\end{equation}
Here, $\mathcal B$ is the Berry curvature defined by $\mathcal B(k)=i\left(\boldsymbol{\nabla}_{k}\times \boldsymbol{A}(k)\right)_{z}$, where $\boldsymbol{A}=\sum_n\langle u_{n}(k)|\boldsymbol{\nabla}_{k} u_{n}(k)\rangle$ with $u_{n}(k)$ a normalized eigenstate of the Hamiltonian, and the sum running over the occupied bands. Here we are considering a continuum theory where the bandstructure has a power-law dependence on momentum. In a lattice calculation, where the bandstructure is defined in the first BZ and involves trigonometric functions, the Chern number is well defined. 
To incorporate such physics, we extend the continuum model to a simple lattice dispersion where we replace $k_x,k_y$ with $\sin(k_x),\sin(k_y)$. This procedure does necessarily add additional Dirac points in the BZ. For the points in parameter space where there is a gap, we find that the Chern number is well defined and at all such points we obtain $C=0$. This is expected when the system has time-reversal symmetry, which happens when both condensates $\Delta_x$ and $\Delta_z$ are imaginary. Other points that break TRS are connected without gap closing and thus are also expected to have a vanishing Chern number.

\section{Other phase transitions\label{sec:phases}}

The $\QED\!\operatorname{-}\!{\U(N)\times\U(N)}\GN$, {$\QED\!\operatorname{-}\!\text{chiral XY} \GN$}, and {$\QEDcHGN$}
transitions are also described with $\GN$ models, where the fermionic
quartic interaction is respectively decoupled with $N_{b}=1,2,3$
real auxiliary bosons
\begin{equation}
S^{c}=S'=\int d^{3}x\left[-\overline{\Psi}\left(\slashed{D}_{A+\mathcal{A}^{q}}+\phi_{I}\mu_{I}\right)\Psi\right]+\dots,
\end{equation}
where the sum over $1\leq I\leq N_b$ is implicit, and $\mu_{I}$ are Pauli matrices acting on a two-dimensional flavor
subspace where 
\begin{equation}
\phi_{I}\mu_{I}=\begin{cases}
\phi_{z}\mu_{z} &\\
\phi_{x}\mu_{x}+\phi_{y}\mu_{y}\,. &\\
\vec{\phi}\cdot\vec{\mu} & 
\end{cases}
\end{equation}
As mentioned previously, the $\QEDcHGN$ model  describes the transition from a U(1) DSL to an
AFM on the kagome lattice. In this case, the Pauli matrices $\vec{\mu}$
act on magnetic spin subspace. As for the chiral XY interaction, taking
$\mu_{x},\mu_{y}$ to act on a valley subspace, this describes the
transition to a VBS order parameter. These transitions were observed
for Monte-Carlo simulations on a square lattice where by tuning gauge-field
fluctuations, the $\U(1)$ DSL is driven to either an AFM
or VBS order, depending on the number of fermion flavors~\citep{xu_monte_2019, wang_dynamics_2019}.  A theoretical study that elucidated the field theory for the transition to an AFM was performed in Ref.~\citep{zerf_critical_2019} (see also Refs.~\citep{ghaemi_neel_2006,dupuis_transition_2019} for earlier studies of this model), while the field theory for the transition to the VBS was outlined in Refs.~\cite{zerf_xy_2020,janssen_xy_2020,boyack_qed3xy_2021}.
In this work, we used the appellation $\QEDGN$ to designate the model
with $\U(2N)$ symmetry, following the convention of Refs.~\citep{wang_deconfined_2018,boyack_deconfined_2019},
notably. The Pauli matrix in this case acts on valley subspace. However,
the label was also used in the literature to refer to the $\U(N)\times\U(N)$
symmetric model, see Ref.~\citep{gracey_fermion_2018} for instance.
Both variations of $\QED$ were considered in Refs.~\citep{benvenuti_easy_2019,benvenuti_qeds_2019}.

As shown in Ref.~\citep{dupuis_transition_2019}, the auxiliary boson
in these cases has a non-vanishing expectation value: $\left\langle \left|\phi\right|\right\rangle \neq0$
in the monopole background on $S^{2}\times\mathbb{R}$. This is also
true for other choices of Pauli matrices, not only the specific one
prescribed before to describe specific universality classes. For instance,
the $\vec{\mu}$ considered for the $\QEDcHGN$ universality class
could also act on valley subspace, in which case the order parameter
is odd under time reversal. There is still a non-vanishing expectation
value of the auxiliary boson. Consequently, the Green's function in
Eq.~(\ref{eq:GF}) must be modified to include a non-zero mass for
the fermions. Using the addition theorems for spinor monopole harmonics
needed to compute the zero-mass Green's function should be sufficient
for this adaptation. Real-space kernels like in Eqs.~(\ref{eq:D_rr'}-\ref{eq:F_rr'})
would also include Pauli matrices for traces on the magnetic spin
subspace, with the number of kernels to compute increasing accordingly
with the number of auxiliary bosons $N_{b}$.

\section{Conclusion}
We obtained the scaling dimension of monopole operators at the QCP
between a $\U(1)$ DSL and two types of topological spin liquids, namely the CSL and a general class of $Z_2$ QSLs, at next-to-leading order in a $1/N$ expansion.  The most
relevant monopole operator in the CSL case has a minimal charge $q=1/2$ and a scaling
dimension $\Delta_{1/2,\QEDGN}=2N\times0.26510+0.118911(7)$, while the analog scaling dimension in the case of the simplest $Z_2$ QSL is $\Delta_{1/2,\QEDZ}=2N\times0.26510+0.102846(9)$. For the other general $Z_2$ spin liquids, where the spin and valley flavour interaction was included,  we obtained a general expression for the anomalous dimension. Since the spin/valley interaction reduces the size of the flavour group, an interesting question is what type of hierarchy the monopoles will have and how can one observe this in the monopole scaling dimensions.  We also rederived the $\QED$ monopole scaling dimensions and found small
discrepancies, e.g., the $q=1/2$ anomalous dimension is $-0.038138(5)$
instead of $-0.0383$ and so on for other charges up to $q=5/2$~\citep{pufu_anomalous_2014,dyer_monopole_2013}.
This will also lead to corrections to the anomalous dimensions of certain monopole operators in QCD$_3$ with non-abelian gauge groups, such as U$(N_c)$, where the $\QED$ anomalous dimension makes its appearance~\cite{dyer_monopole_2013}.

With these anomalous dimensions, we obtained a fit in the topological charge $q$ and compared the $O(q^{0})$ coefficient with the universal value obtained in a large-charge expansion for operators charged under a global $\U(1)$ symmetry~\citep{hellerman_on_2015}. We obtain the expected value $\gamma=-0.0937$ in $\QED$, $\QEDGN$ and $\QEDZ$. We also revisited the conjectured duality between $\QEDGN|_{2N=2}$ and in $\CP^{1}$ models~\citep{wang_deconfined_2018}. Notably, the $q=1/2$ monopole scaling dimensions in $\QEDGN|_{2N=2}$ agree very well with the scaling dimensions of other operators that are predicted to be equal under the duality. Specifically, the anomalous dimension obtained in this work greatly improves this agreement. We also argued that all monopoles with equal charges should have the same scaling dimensions in the $\QEDGN|_{2N=2}$ and $\CP^{1}$ models. Using next-to-leading order results for both models, we obtain an agreement that is better for a minimally charged monopole, with a relative difference of $3\%$. As the topological charge increases, this difference increases and eventually saturates at $10\%$ for $q\to\infty$.

It would be interesting to study monopole operators in the other gauged $\GN$ models that we briefly discussed, notably the model describing the transition to an AFM~\citep{dupuis_transition_2019,dupuis_proc_2021,dupuis_monopole_2021}.
Another interesting aspect to consider that was not included in this
work is the case of monopole operators in the pure-$\GN$ model. It
is a straightforward adaptation to write out the monopole anomalous
dimensions in this case and use the results of this work to obtain
them. Although there is no $\U(1)_{\text{top}}$ due to
the absence of a gauge field in this model, these objects still have
useful applications. This notably motivated the study of monopoles
in the bosonic $\ON$ model~\citep{sachdev_compressible_2012,pufu_monopoles_2013}.
A study of the $\GN$ global monopoles and some of their applications
will appear in a forthcoming work.

\begin{acknowledgements}
We thank Silviu Pufu for useful discussions as well as for clarifying key points in his QED$_3$ and QCD$_3$ calculations. We also thank Ofer Aharony, Shai Chester,  Joseph Maciejko, and Subir Sachdev for helpful comments. 
{\'E.D.} was funded by an Alexander Graham Bell CGS from NSERC. 
{W.W.-K.} and {R.B.} were funded by a Discovery Grant from NSERC, a Canada Research Chair, a grant from the Fondation Courtois, and a ``\'Etablissement de nouveaux chercheurs et de nouvelles chercheuses universitaires'' grant from FRQNT. 
\end{acknowledgements}

\onecolumngrid
\appendix
\numberwithin{equation}{section}

\section{Large $N$ non-compact quantum phase transition\label{app:non_compact}}

An auxiliary boson $\phi$ can be introduced to decouple the $\GN$
term in the action in Eq.~(\ref{eq:GN_action}) through a Hubbard-Stratonovich
transformation
\begin{equation}
S=\int d^{3}x\left[-\overline{\Psi}\left(\slashed{\partial}-i\slashed{A}+\phi\right)\Psi+\frac{N}{h^{2}}\phi^{2}\right],\label{eq:action_hubbard}
\end{equation}
where the coupling constant $h^{2}$ was rescaled with $N$, the number
of valley nodes. The fermion part of the action is now quadratic and
can be integrated 
\begin{equation}
S_{{\rm \text{eff}}}=N\left[-\ln\det\left(\slashed{\partial}-i\slashed{A}+\phi\right)+\int d^{3}x\frac{1}{h^{2}}\phi^{2}\right],\label{eq:S_eff_pre}
\end{equation}
where the valley subspace has been traced out. The saddle-point equation
for the gauge field is
\begin{align}
0=\left.\frac{\delta S{}_{{\rm \text{eff}}}}{\delta A_{\mu}}\right|_{\left\langle \phi\right\rangle ,\left\langle A_{\mu}\right\rangle } & =iN\int\frac{d^{3}p}{\left(2\pi\right)^{3}}\text{tr}\left[\frac{\gamma^{\mu}}{-i\slashed{p}-i\left\langle \slashed{A}\right\rangle +\phi}\right]\nonumber \\
 & =iN\int\frac{d^{3}p}{\left(2\pi\right)^{3}}\text{tr}\left(\frac{\gamma^{\mu}\gamma^{\nu}}{\left(p+\left\langle A\right\rangle \right)^{2}+\left\langle \phi\right\rangle ^{2}}\right)\left\langle A_{\nu}\right\rangle \nonumber \\
 & =2iN\int\frac{d^{3}p}{\left(2\pi\right)^{3}}\frac{\left\langle A^{\mu}\right\rangle }{\left(p+\left\langle A\right\rangle \right)^{2}+\left\langle \phi\right\rangle ^{2}},
\end{align}
which is solved for a vanishing gauge field $\left\langle a_{\mu}\right\rangle =0$,
as required by gauge invariance. Taking a homogeneous ansatz for the
pseudo-scalar field, the remaining gap equation is given by
\begin{equation}
0=\left.\frac{\delta S{}_{{\rm \text{eff}}}}{\delta\phi}\right|_{\left\langle \phi\right\rangle ,\left\langle A_{\mu}\right\rangle =0}=2N\left\langle \phi\right\rangle \left[\frac{1}{h^{2}}-\int\frac{d^{3}p}{\left(2\pi\right)^{3}}\frac{1}{p^{2}+\left\langle \phi\right\rangle ^{2}}\right],
\end{equation}
At the QCP, where $\left\langle \phi\right\rangle =0$, the critical
coupling is defined through
\begin{equation}
\frac{1}{h_{c}^{2}}=\int\frac{d^{3}p}{\left(2\pi\right)^{3}}\frac{1}{p^{2}}=0,
\end{equation}
where this result is obtained through zeta-regularization of the integral.
In this scheme, only the determinant operator remains in the effective
action (\ref{eq:S_eff_pre}), i.e., we obtain Eq.~(\ref{eq:S_eff_c}).

\section{Scalar-gauge kernel\label{app:scalar_gauge} }

We noted earlier that $M_{\ell}^{q}(\omega)$ is non-hermitian.
Here we elaborate on this point in more detail. First, we note that
$H_{\mu^{\prime}}^{q}(r,r^{\prime})$ is imaginary. Conjugating
the expression in Eq.~(\ref{eq:F_rr'}), we obtain

\begin{align}
H_{\mu'}^{q}\left(r,r^{\prime}\right) & =i\text{tr}\left[G_{q}(r,r^{\prime})\gamma_{\mu^{\prime}}G_{q}^{\dagger}(r,r^{\prime})\right]\\
\left[H_{\mu^{\prime}}^{q}(r,r^{\prime})\right]^{*} & =-i\text{tr}\left[G_{q}^{*}(r,r^{\prime})\gamma_{\mu^{\prime}}^{*}G_{q}^{\intercal}(r,r^{\prime})\right]=-i\text{tr}\left[G_{q}(r,r^{\prime})\gamma_{\mu^{\prime}}^{\dagger}G_{q}^{\dagger}(r,r^{\prime})\right]
\end{align}
where we used that the trace of a matrix is equal to the trace of
the transposed matrix. Here, the gamma matrices are simply the Pauli
matrices, thus $\gamma_{\mu^{\prime}}^{\dagger}=\gamma_{\mu^{\prime}}$.
As a result, there is an extra sign in the conjugation of $H_{\mu^{\prime}}^{q}(r,r^{\prime})$:
\begin{equation}
\left[H_{\mu^{\prime}}^{q}(r,r^{\prime})\right]^{*}=-H_{\mu^{\prime}}^{q}(r,r^{\prime}).
\end{equation}
Hence, the kernel is imaginary.

Next, we make relevant observation for the kernel Fourier coefficient.
The decomposition of $H_{\tau'}^{q}(r,r^{\prime})$, by definition
(\ref{eq:M_rr'}, \ref{eq:M_o}), is 
\begin{equation}
H_{\tau^{\prime}}^{q}(r,r^{\prime})=\int_{\omega}\sum_{\ell}H_{\ell}^{q,T}(\omega)e^{-i\omega(\tau-\tau^{\prime})}P_{\ell}\left(\hat{n}\cdot\hat{n}^{\prime}\right),
\end{equation}
where we used the addition theorem in Eq.~(\ref{eq:addition_theorem}).
As for the other scalar-gauge kernel, $H_{\tau}^{q}\left(r^{\prime},r\right)$,
it can be defined in the same way, but with a different coefficient,
say $\widetilde{H}_{\ell}^{q,T}(\omega)$. This is then
related to $H_{\ell}^{q,T}(\omega)$ by exchanging coordinates
in the expression above 
\begin{equation}
H_{\tau^{\prime}}^{q}(r,r^{\prime})=\int_{\omega}\sum_{\ell}H_{\ell}^{q,T}(\omega)e^{-i\omega(\tau-\tau^{\prime})}P_{\ell}\left(\hat{n}\cdot\hat{n}^{\prime}\right)=\int_{\omega}\sum_{\ell}H_{\ell}^{q,T}\left(-\omega\right)e^{-i\omega(\tau-\tau^{\prime})}P_{\ell}\left(\hat{n}\cdot\hat{n}^{\prime}\right).
\end{equation}
Thus, $\widetilde{H}_{\ell}^{q,T}(\omega)=H_{\ell}^{q,T}\left(-\omega\right)$.
Since $H_{\tau^{\prime}}^{q}(r,r^{\prime})$ is imaginary,
we have that $H_{\ell}^{q,T}\left(-\omega\right)=-\left[H_{\ell}^{q,T}(\omega)\right]{}^{*}$,
meaning that 
\begin{equation}
\widetilde{H}_{\ell}^{q,T}(\omega)=-\left[H_{\ell}^{q,T}(\omega)\right]^{*}.
\end{equation}
This explains the signs in the first column of Eq.~(\ref{eq:M_o})

\section{Gauge invariance \label{app:gauge_invariance}}

Using conservation of the $\U(1)$ current $\nabla_{\mu}J^{\mu}(r)=0$
in Eqs.~(\ref{eq:Kmunu}, \ref{eq:Fmu}), one can show the gauge invariance
of the kernels
\begin{equation}
\nabla^{\mu}K_{\mu\mu^{\prime}}(r,r^{\prime})=0,\quad\nabla^{\mu^{\prime}}K_{\mu\mu^{\prime}}(r,r^{\prime})=0,\quad\nabla^{\mu^{\prime}}H_{\mu^{\prime}}(r,r^{\prime})=0.
\end{equation}
We re-express these conditions in the Fourier transformed space. To
do so, we take the divergence of the various eigenvectors of the gauge
field 
\begin{align}
\nabla^{\mu}e^{-i\omega\tau}\mathfrak{a}_{\mu,\ell m}^{T}(\hat{n}) & =\nabla^{\mu}\left(\frac{1}{-i\omega}Y_{\ell m}(\hat{n})\nabla_{\mu}e^{-i\omega\tau}\right)=\frac{1}{-i\omega}Y_{\ell m}(\hat{n})\nabla^{\mu}\nabla_{\mu}e^{-i\omega t}=-i\omega e^{-i\omega\tau}Y_{\ell m}(\hat{n}),\\
\nabla^{\mu}e^{-i\omega\tau}\mathfrak{a}_{\mu,\ell m}^{E}(\hat{n}) & =\nabla^{\mu}\left(\frac{e^{-i\omega\tau}}{\sqrt{\ell(\ell+1)}}\nabla_{\mu}Y_{\ell m}(\hat{n})\right)=\frac{e^{-i\omega\tau}}{\sqrt{\ell(\ell+1)}}\nabla^{\mu}\nabla_{\mu}Y_{\ell m}(\hat{n})=-\sqrt{\ell(\ell+1)}e^{-i\omega\tau}Y_{\ell m}(\hat{n}),\\
\nabla^{\mu}e^{-i\omega\tau}\mathfrak{a}_{\mu,\ell m}^{B}(\hat{n}) & =\nabla_{\mu}\left(\frac{e^{-i\omega\tau}}{\sqrt{\ell(\ell+1)}}\frac{\epsilon^{0\mu\nu}}{\sqrt{g(r)}}\nabla_{\nu}Y_{\ell m}(\hat{n})\right)=0e^{-i\omega\tau}Y_{\ell m}(\hat{n}),
\end{align}
which implies the following relation
\begin{equation}
\nabla^{\mu}\begin{pmatrix}\mathfrak{a}_{\mu,\ell m}^{T}(\hat{n}) & \mathfrak{a}_{\mu,\ell m}^{E}(\hat{n}) & \mathfrak{a}_{\mu,\ell m}^{B}(\hat{n})\end{pmatrix}=\begin{pmatrix}-i\omega & -\sqrt{\ell(\ell+1)} & 0\end{pmatrix}e^{-i\omega\tau}Y_{\ell m}(\hat{n}).
\end{equation}
Taking the divergence of the kernels, we obtain 
\begin{align}
\nabla^{\mu}K_{\mu\mu^{\prime}}(r,r^{\prime}) & =\int_{\omega}\sum_{\ell=0}^{\infty}\sum_{m=-\ell}^{\ell}e^{-i\omega(\tau-\tau^{\prime})}Y_{\ell m}(\hat{n})\nonumber \\
 & \quad\times\begin{pmatrix}-i\omega & -\sqrt{\ell(\ell+1)} & 0\end{pmatrix}\begin{pmatrix}K_{\ell}^{q,TT}(\omega) & K_{\ell}^{q,TE}(\omega) & K_{\ell}^{q,TB}(\omega)\\
K_{\ell}^{q,TE*}(\omega) & K_{\ell}^{q,EE}(\omega) & K_{\ell}^{q,EB}(\omega)\\
K_{\ell}^{q,TB*}(\omega) & K_{\ell}^{q,EB*}(\omega) & K_{\ell}^{q,BB}(\omega)
\end{pmatrix}\begin{pmatrix}\mathfrak{a}_{\mu,\ell m}^{T\,\dagger}(\hat{n}^{\prime})\\
\mathfrak{a}_{\mu,\ell m}^{E\,\dagger}(\hat{n}^{\prime})\\
\mathfrak{a}_{\mu,\ell m}^{B\,\dagger}(\hat{n}^{\prime})
\end{pmatrix},\\
\nabla^{\mu}H_{\mu}(r,r^{\prime}) & =\int_{\omega}\sum_{\ell=0}^{\infty}\sum_{m=-\ell}^{\ell}e^{-i\omega(\tau-\tau^{\prime})}Y_{\ell m}(\hat{n})Y_{\ell m}^{*}(\hat{n}^{\prime})\begin{pmatrix}-i\omega & -\sqrt{\ell(\ell+1)} & 0\end{pmatrix}\begin{pmatrix}-H_{\ell}^{q,T*}(\omega)\\
-H_{\ell}^{q,E*}(\omega)\\
-H_{\ell}^{q,T*}(\omega)
\end{pmatrix},
\end{align}
where $\int_{\omega}\equiv\int d\omega/(2\pi)$. Requiring gauge invariance
and setting these divergences to $0$, we obtain the following relations
\begin{align}
-i\omega K_{\ell}^{q,TT}(\omega)-\sqrt{\ell(\ell+1)}K_{\ell}^{q,TE*}(\omega) & =0,\\
-i\omega K_{\ell}^{q,TE}(\omega)-\sqrt{\ell(\ell+1)}K_{\ell}^{q,EE}(\omega) & =0,\\
-i\omega K_{\ell}^{q,TB}(\omega)-\sqrt{\ell(\ell+1)}K_{\ell}^{q,EB}(\omega) & =0,\\
i\omega H_{\ell}^{q,T*}(\omega)+\sqrt{\ell(\ell+1)}H_{\ell}^{q,E*}(\omega) & =0.
\end{align}

\subsection*{Verifications}

Let us check one important relation following from gauge invariance:
\begin{equation}
K_{\ell}^{q,EE}(\omega)=\frac{\omega^{2}}{\ell(\ell+1)}K_{\ell}^{q,TT}(\omega).\label{eq:gauge_inv_cond}
\end{equation}
This is easily verified for $q=0$ where closed forms of the kernels
are easily obtained as discussed in Sec.~\ref{subsec:q0_kernels}.
This is a bit more involved when $q\neq0$. We have expressions where
the dependence on $\omega$ is easily isolated, taking the following
form
\begin{equation}
K_{\ell}^{q,ZZ}(\omega)=\int dx\sum_{\ell^{\prime},\ell^{\prime\prime}}\frac{E_{q,\ell^{\prime}}+E_{q,\ell^{\prime\prime}}}{\omega^{2}+\left(E_{q,\ell^{\prime}}+E_{q,\ell^{\prime\prime}}\right)^{2}}\tilde{k}_{\ell,\ell^{\prime},\ell^{\prime\prime}}^{q,ZZ}(x),\quad Z\in\{T,E\}.
\end{equation}
In the RHS of the gauge invariance condition in Eq.~(\ref{eq:gauge_inv_cond}),
we may reexpress the $\omega$-dependent function as
\begin{equation}
\frac{\omega^{2}}{\omega^{2}+\left(E_{q,\ell^{\prime}}+E_{q,\ell^{\prime\prime}}\right)^{2}}=1-\frac{\left(E_{q,\ell^{\prime}}+E_{q,\ell^{\prime\prime}}\right)^{2}}{\omega^{2}+\left(E_{q,\ell^{\prime}}+E_{q,\ell^{\prime\prime}}\right)^{2}}.
\end{equation}
Upon integration over $\omega$, the first term is a simple divergence
that can be regularized away. In the presence of test function $f(\omega)$,
this contribution is $\int d\omega f(\omega)\times1$ and
it again vanishes provided the test function is convergent with no
poles. The gauge-invariance condition in Eq.~(\ref{eq:gauge_inv_cond})
can then be written by simply comparing the finite parts 
\begin{equation}
\int dx\sum_{\ell^{\prime},\ell^{\prime\prime}}\frac{E_{q,\ell^{\prime}}+E_{q,\ell^{\prime\prime}}}{\omega^{2}+\left(E_{q,\ell^{\prime}}+E_{q,\ell^{\prime\prime}}\right)^{2}}\tilde{k}_{\ell,\ell^{\prime},\ell^{\prime\prime}}^{q,EE}(x)=\int dx\sum_{\ell^{\prime},\ell^{\prime\prime}}\left[-\frac{\left(E_{q,\ell^{\prime}}+E_{q,\ell^{\prime\prime}}\right)^{2}}{\ell(\ell+1)}\right]\times\frac{E_{q,\ell^{\prime}}+E_{q,\ell^{\prime\prime}}}{\omega^{2}+\left(E_{q,\ell^{\prime}}+E_{q,\ell^{\prime\prime}}\right)^{2}}\tilde{k}_{\ell,\ell^{\prime},\ell^{\prime\prime}}^{q,TT}(x).\label{eq:new_condition}
\end{equation}
This last relation following from gauge invariance was verified for
$q=1/2$. Note that gauge invariance also implies that 
\begin{equation}
k_{0,\ell^{\prime},\ell^{\prime\prime}}^{1/2,TT}(x)=0,
\end{equation}
which is also verified by direct computation.

\section{Green's function\label{sec:Green}}

\subsection{Eigenvalues of determinant operator }

In a general basis, the gauge covariant derivative acting on a spin-1/2
spinor on spacetime $\mathfrak{M}$ will take the form 
\begin{equation}
\slashed{D}_{\mathcal{A}^{q}}=e_{b}^{\mu}\gamma^{b}\left[\partial_{\mu}-\Omega_{\mu}-i\mathcal{A}_{\mu}^{q}\right],
\end{equation}
where $\Omega_{\mu}$ is the spin connection transporting the fermion
fields on spacetime $\mathfrak{M}$. On a flat spacetime $\mathfrak{M}=\mathbb{R}^{3}$,
there are also spin connections in spherical coordinates that can
be eliminated with a unitary transformation~\citep{briggs_equivalence_2013}.
In this case, the covariant derivate $\nabla_{\mu=r,\theta,\phi}$
can be traded for a normal derivative $\partial_{\mu=r,\theta,\phi}$
\begin{equation}
\slashed{D}_{\mathcal{A}^{q}}^{\mathbb{R}^{3}}=(e^{\RR^{3}})_{b}^{\mu}\gamma^{b}\left[\partial_{\mu}-i\mathcal{A}_{\mu}^{q}\right].
\end{equation}
Proceeding with the Weyl transformation $\psi\to e^{-\tau}\psi$,
$g_{\mu\nu}\to e^{-2\tau}$ discussed in Eq.~(\ref{eq:weyl_rescaling}),
the Dirac operator on $S^{2}\times\RR$ is given by~\citep{borokhov_topological_2003}
\begin{equation}
\slashed{D}_{\mathcal{A}^{q}}^{S^{2}\times\mathbb{R}}=\left(e^{S^{2}\times\RR}\right)_{b}^{\mu}\gamma^{b}\left[\partial_{\mu}-\frac{1}{R}\delta_{\mu}^{\tau}-i\mathcal{A}_{\mu}^{q}\right].
\end{equation}

To diagonalize this operator, we introduce spinor monopole harmonics 

\begin{equation}
S_{q,\ell^{\prime},m^{\prime}}^{\pm}=\begin{pmatrix}\pm\alpha_{\pm}Y_{q,\ell^{\prime},m^{\prime}}\\
\alpha_{\mp}Y_{q,\ell^{\prime},m^{\prime}+1}
\end{pmatrix},\quad\alpha_{\pm}=\sqrt{\frac{\ell^{\prime}+1/2\pm\left(m^{\prime}+1/2\right)}{2\ell^{\prime}+1}}.\label{eq:spinor_monopole}
\end{equation}
These spinors diagonalize the following generalized total spin and
angular momentum operators $J_{q}^{2}, J_{q}^{z}, L_{q}^{2}$. In particular,
the spinor monopole harmonics $S_{q,\ell,m}^{\pm}$ have a total spin
$j=\ell\pm1/2$. In the $j=\ell-1/2$ basis, the Dirac operator mixes
the two types of spinors and simply becomes a matrix with c-number
entries~\citep{borokhov_topological_2003}:
\begin{equation}
\left[\begin{array}{l}
i\slashed{D}_{\mathcal{A}^{q}}e^{-i\omega\tau}S_{q,\ell-1,m}^{+}\\
i\slashed{D}_{\mathcal{A}^{q}}e^{-i\omega\tau}S_{q,\ell,m}^{-}
\end{array}\right]=\boldsymbol{N}_{q,\ell}\left(\omega+i\boldsymbol{M}_{q,\ell}\right)\left[\begin{array}{l}
e^{-i\omega\tau}S_{q,\ell-1,m}^{+}\\
e^{-i\omega\tau}S_{q,\ell,m}^{-}
\end{array}\right],
\end{equation}
where 
\begin{equation}
\boldsymbol{N}_{q,\ell}=-\frac{1}{\ell}\left(q\tau_{z}+E_{q;\ell}\tau_{x}\right),\quad\boldsymbol{M}_{q,\ell}=\frac{E_{q;\ell}}{\ell}\left(E_{q;\ell}\tau_{z}-q\tau_{x}\right),\quad E_{q;\ell}=\sqrt{\ell^{2}-q^{2}}.
\end{equation}
Here, the $\tau_{i}$ are the Pauli matrices acting in the $j=\ell-1/2$
basis, i.e., they mix the components $S_{q,\ell-1,m}^{+}$ and $S_{q,\ell,m}^{-}$.
For $\ell=q$ (we suppose a positive magnetic charge $q>0$), only
$S_{q,q,m}^{-}$ exists and corresponds to a zero mode of the Dirac
operator. By diagonalizing this matrix, we retrieve the eigenvalues
used in Sec.~\ref{sec:N_infty}. 

\subsection{Green's function\label{subsec:GF}}

The Green's function can be obtained with the spectral decomposition
\begin{equation}
G_{q}(r,r^{\prime})=-\sum_{\lambda}\frac{\psi_{\lambda}(r)\psi_{\lambda}^{\dagger}\left(r^{\prime}\right)}{E_{\lambda}},
\end{equation}
where $\psi_{\lambda}(r)$ are eigenspinors of the Dirac
operator $i\slashed{D}_{\mathcal{A}^{q}}\psi_{\lambda}=E_{\lambda}\psi_{\lambda}$
forming a complete basis $\sum_{\lambda}\psi_{\lambda}(r)\psi_{\lambda}^{\dagger}(r^{\prime})=\delta(r-r^{\prime}).$
With this formulation, the Green's function respects its defining
equation of motion (\ref{eq:GF_EOM}). We can simply keep working
in the spinor monopole harmonics basis instead of further diagonalizing.
The spectral decomposition of the Green's function in this basis is
then
\begin{equation}
G_{q}(r,r^{\prime})=-\widetilde{\psi}_{\lambda}(r)\left(\widetilde{E}^{-1}\right)_{\lambda\lambda^{\prime}}\widetilde{\psi}_{\lambda^{\prime}}^{\dagger}\left(r^{\prime}\right).
\end{equation}
The eigenvalue matrix is block diagonal, separating each $j=\ell-1/2$
sectors. To obtain a Green's function which has two particle-hole
indices but which is a scalar with respect to the $+/-$ structure
described above, we take the left eigenspinor as a row vector in the
$+/-$ space, $\left[e^{-i\omega\tau}S_{q,\ell-1,m}^{+}(\hat{n}),e^{-i\omega\tau}S_{q,\ell,m}^{-}(\hat{n})\right]$.
The action of the Dirac operator on the left eigenspinor is then given
by 
\begin{align}
\left[\begin{array}{l}
i\slashed{D}_{\mathcal{A}^{q}}\bigl(e^{-i\omega\tau}S_{q,\ell-1,m}^{+}\bigr)\\
i\slashed{D}_{\mathcal{A}^{q}}\bigl(e^{-i\omega\tau}S_{q,\ell,m}^{-}\bigr)
\end{array}\right]^{T} & =\left(\boldsymbol{N}_{q,\ell}\left(\omega+i\boldsymbol{M}_{q,\ell}\right)\left[\begin{array}{l}
e^{-i\omega\tau}S_{q,\ell-1,m}^{+}\\
e^{-i\omega\tau}S_{q,\ell,m}^{-}
\end{array}\right]\right)^{T}\\
 & =\left[e^{-i\omega\tau}S_{q,\ell-1,m}^{+}\quad e^{-i\omega\tau}S_{q,\ell,m}^{-}\right]\left[\boldsymbol{N}_{q,\ell}\left(\omega-i\boldsymbol{M}_{q,\ell}\right)\right],
\end{align}
where we used that $\boldsymbol{N}_{q,\ell}^{T}=\boldsymbol{N}_{q,\ell},\,\boldsymbol{M}_{q,\ell}^{T}=\boldsymbol{M}_{q,\ell}$
and $\boldsymbol{M}_{q,\ell}\boldsymbol{N}_{q,\ell}=-\boldsymbol{N}_{q,\ell}\boldsymbol{M}_{q,\ell}$.
We can read the eigenvalue matrix from this relation and write Green's
function\footnote{The inverse eigenvalue matrix here is $\bigl[\boldsymbol{N}_{q,\ell}\bigl(\omega-i\boldsymbol{M}_{q,\ell}\bigr)\bigr]^{-1}$instead
of $\bigl[\boldsymbol{N}_{q,\ell}\bigl(\omega+i\boldsymbol{M}_{q,\ell}\bigr)\bigr]^{-1}$as
described in Ref.~\citep{pufu_anomalous_2014}. This explains the
benign different sign in our Green's function in what follows.}
\begin{equation}
\begin{aligned}G_{q}(r,r^{\prime})= & -\int\frac{d\omega}{2\pi}\sum_{\ell=q}^{\infty}\sum_{m=-\ell}^{\ell-1}\left[S_{q,\ell-1,m}^{+}(\hat{n})\quad S_{q,\ell,m}^{-}(\hat{n})\right]\times\frac{e^{-i\omega(\tau-\tau^{\prime})}}{\boldsymbol{N}_{q,\ell}\bigl(\omega-i\boldsymbol{M}_{q,\ell}\bigr)}\left[\begin{array}{c}
\left(S_{q,\ell-1,m}^{+}(\hat{n}^{\prime})\right)^{\dagger}\\
\left(S_{q,\ell,m}^{-}(\hat{n}^{\prime})\right)^{\dagger}
\end{array}\right].\end{aligned}
\label{eq:spectral}
\end{equation}
We can note that $\boldsymbol{N}_{q,\ell}^{-1}=\boldsymbol{N}_{q,\ell}$
as this matrix squares to identity $\boldsymbol{N}_{q,\ell}^{2}=\ell^{-2}\left(q^{2}+E_{q;\ell}^{2}\right)\tau_{0}=\tau_{0}$.
Also, by noting that $\left|\omega+i\mathbf{M}_{q,\ell}\right|^{2}=\left(\omega^{2}+E_{q;\ell}^{2}\right)\tau_{0}$,
it follows that $\left(\omega-i\mathbf{M}_{q,\ell}\right)^{-1}=\left(\omega^{2}+E_{q;\ell}^{2}\right)^{-1}\left(\omega+i\mathbf{M}_{q,\ell}\right)$.
Then, the inverse matrix in the spectral decomposition becomes
\begin{eqnarray}
\left(\omega-i\mathbf{M}_{q,\ell}\right)^{-1}\boldsymbol{N}_{q,\ell}^{-1} & = & \frac{1}{\omega^{2}+E_{q;\ell}^{2}}\left(\omega+i\mathbf{M}_{q,\ell}\right)\mathbf{N}_{q,\ell}\\
 & = & \frac{1}{\omega^{2}+E_{q;\ell}^{2}}\left(\omega\mathbf{N}_{q,\ell}+E_{q;\ell}\tau_{y}\right),
\end{eqnarray}
where we used that $\mathbf{M}_{q,\ell}\mathbf{N}_{q,\ell}=-iE_{q;\ell}\tau_{y}$.
The spectral decomposition of the Green's function then becomes 
\begin{equation}
G_{q}(r,r^{\prime})=-\int_{-\infty}^{\infty}\frac{d\omega}{2\pi}\sum_{\ell=q}^{\infty}\sum_{m=-\ell}^{\ell-1}\left[S_{q,\ell-1,m}^{+}\quad S_{q,\ell,m}^{-}\right]\times\frac{e^{-i\omega(\tau-\tau^{\prime})}}{\omega^{2}+E_{q;\ell}^{2}}\left(\omega\mathbf{N}_{q,\ell}+E_{q;\ell}\tau_{y}\right)\left[\begin{array}{c}
\left(S_{q,\ell-1,m}^{+}\right)^{\dagger}\\
\left(S_{q,\ell,m}^{-}\right)^{\dagger}
\end{array}\right].
\end{equation}
The contour integral on $\omega$ is obtained with the residue theorem
\begin{eqnarray}
\int_{-\infty}^{\infty}\frac{d\omega}{2\pi}\frac{e^{-i\omega\left(\tau-\tau'\right)}}{\omega^{2}+E_{q;\ell}^{2}}{\omega \brace 1} & = & -i\text{sgn}(\tau-\tau^{\prime})\frac{e^{-E_{q;\ell}\left|\tau-\tau^{\prime}\right|}}{-2iE_{q;\ell}\text{sgn}(\tau-\tau^{\prime})}{-iE_{q;\ell}\text{sgn}(\tau-\tau^{\prime}) \brace 1}\nonumber \\
 & = & \frac{1}{2}e^{-E_{q;\ell}\left|\tau-\tau^{\prime}\right|}{-i\text{sgn}(\tau-\tau^{\prime}) \brace E_{q;\ell}^{-1}}.
\end{eqnarray}
The spectral decomposition after the $\omega$ integration becomes
\begin{equation}
G_{q}(r,r^{\prime})=\frac{i}{2}\sum_{\ell=q}^{\infty}e^{-E_{q;\ell}\left|\tau-\tau^{\prime}\right|}\sum_{m=-\ell}^{\ell-1}\left[S_{q,\ell-1,m}^{+}\quad S_{q,\ell,m}^{-}\right]\left(\text{sgn}(\tau-\tau^{\prime})\mathbf{N}_{q,\ell}+\left(\begin{array}{cc}
0 & 1\\
-1 & 0
\end{array}\right)\right)\left[\begin{array}{c}
\left(S_{q,\ell-1,m}^{+}\right)^{\dagger}\\
\left(S_{q,\ell,m}^{-}\right)^{\dagger}
\end{array}\right].\label{eq:GF_monopole_harmonics_1}
\end{equation}
By inserting Eq.~(\ref{eq:spinor_monopole}), we obtain a $2\times2$
matrix whose components are pairs of monopole harmonics
\begin{equation}
G_{q}(r,r^{\prime})=\left(2\times2\text{ matrix}\right)_{\tau\tau^{\prime}}\propto\sum_{\ell^{\prime}=q}^{\infty}\sum_{m^{\prime}=-\ell^{\prime}+1}^{\ell^{\prime}}Y_{q,\ell^{\prime}+\delta\ell^{\prime},m^{\prime}+\delta m^{\prime}}(\hat{n})Y_{q,\ell^{\prime}+\widetilde{\delta}\ell^{\prime},m^{\prime}+\widetilde{\delta}m^{\prime}}^{*}(\hat{n}^{\prime}),\label{eq:GF_monopole_harmonics_2}
\end{equation}
where 
\begin{equation}
\begin{cases}
\mathfrak{\delta\ell}^{\prime},\widetilde{\delta}\ell^{\prime} & \in\{-1,0\}\\
\delta m^{\prime},\widetilde{\delta}m^{\prime} & \in\{0,1\}
\end{cases}.\label{eq:GF_monopole_harmonics_3}
\end{equation}
This formulation is used in Sec.~\ref{subsec:general_q}. 

For the minimal charge case, the Green's function can be further simplified
by taking the sum on the azimutal quantum number~\citep{pufu_anomalous_2014},
which yields Eq.~(\ref{eq:GF}) in the main text \footnote{The difference that we noted concerning what inverse matrix is used
in Eq.~(\ref{eq:spectral}) implies an extra sign in the first line
of the Green's function.}. The phase appearing in Eq.~(\ref{eq:GF}) is given by~\citep{pufu_anomalous_2014}

\begin{align}
e^{-i\Theta} & \cos\frac{\gamma}{2}=\cos\frac{\theta}{2}\cos\frac{\theta^{\prime}}{2}+\sin\frac{\theta}{2}\sin\frac{\theta^{\prime}}{2}e^{-i\left(\phi-\phi^{\prime}\right)}.\label{eq:phase}
\end{align}

\section{Eigenkernels \label{sec:eigenkernels}}

\subsection{First basis}

We work in spherical normalized coordinates

\begin{equation}
[e_{\mu=x,y,z}^{a=\hat{\theta},\hat{\phi},\hat{\tau}}]=\begin{pmatrix}\cos\theta\cos\phi & -\sin\theta\sin\phi & \sin\theta\cos\phi\\
\cos\theta\sin\phi & \sin\theta\cos\phi & \sin\theta\sin\phi\\
-\sin\theta & 0 & \cos\theta
\end{pmatrix}.
\end{equation}
Using the definitions of the vector spherical harmonics (\ref{eq:vector_harm_basis_1}),
the eigenkernels in Eqs.~(\ref{eq:KElo}, \ref{eq:KBlo}) can be written
as 
\begin{align}
\sum_{m}\mathfrak{a}_{\ell m}^{a,E*}(\hat{n})\mathfrak{a}_{\ell m}^{a^{\prime},E}(\hat{n}^{\prime}) & =\frac{1}{\ell(\ell+1)}\nabla_{a}\nabla_{a^{\prime}}\left(\sum_{m}Y_{\ell m}^{*}(\hat{n})Y_{\ell m}(\hat{n}^{\prime})\right),\\
\sum_{m}\mathfrak{a}_{\ell m}^{a,B*}(\hat{n})\mathfrak{a}_{\ell m}^{a^{\prime},B}(\hat{n}^{\prime}) & =\frac{1}{\ell(\ell+1)}\frac{\epsilon^{ab}\epsilon^{a^{\prime}b^{\prime}}}{\sqrt{g(r)g\left(r^{\prime}\right)}}\nabla_{b}\nabla_{b^{\prime}}\left(\sum_{m}Y_{\ell m}^{*}(\hat{n})Y_{\ell m}(\hat{n}^{\prime})\right).
\end{align}
Using the addition formula in Eq.~(\ref{eq:addition_theorem}), we
obtain the results in Ref.~\citep{dyer_scaling_2015}.

\begin{align}
\sum_{m}\mathfrak{a}_{\ell m}^{a,E*}(\hat{n})\mathfrak{a}_{\ell m}^{a^{\prime},E}\left(\hat{z}\right)= & \frac{2\ell+1}{4\pi}\frac{1}{\ell(\ell+1)}\begin{pmatrix}-\left(1-x^{2}\right)P_{\ell}^{\prime\prime}(x)+xP_{\ell}^{\prime}(x) & 0\\
0 & P_{\ell}^{\prime}(x)
\end{pmatrix}\begin{pmatrix}\cos\phi & \sin\phi\\
-\sin\phi & \cos\phi
\end{pmatrix},\label{eq:sum_aEaE}\\
\sum_{m}\mathfrak{a}_{\ell m}^{a,B*}(\hat{n})\mathfrak{a}_{\ell m}^{a^{\prime},B}\left(\hat{z}\right)= & \frac{2\ell+1}{4\pi}\frac{1}{\ell(\ell+1)}\begin{pmatrix}P_{\ell}^{\prime}(x) & 0\\
0 & -\left(1-x^{2}\right)P_{\ell}^{\prime\prime}(x)+xP_{\ell}^{\prime}(x)
\end{pmatrix}\begin{pmatrix}\cos\phi & \sin\phi\\
-\sin\phi & \cos\phi
\end{pmatrix}.\label{eq:sum_aBaB}
\end{align}
This may be further simplified by using the differential equation
for Legendre polynomials 
\begin{equation}
P_{\ell}^{\prime\prime}(x)=\frac{1}{1-x^{2}}\left[2xP_{\ell}^{\prime}(x)-\ell(\ell+1)P_{\ell}(x)\right].
\end{equation}

\subsection{Second basis \label{subsec:second_basis}}

By working in helical coordinates,

\begin{equation}
r^{a=+,z,-}=\frac{1}{\sqrt{2}}\left(-x+iy,\sqrt{2}z,x+iy\right),
\end{equation}
the vector spherical harmonics $U_{\ell m}^{a}(\hat{n}),V_{\ell m}^{a}(\hat{n}),W_{\ell m}^{a}(\hat{n})$
introduced in Sec.~\ref{subsec:general_q} take the following form
\begin{align}
U_{\ell m}^{a}(\hat{n}) & =\begin{pmatrix}\sqrt{\frac{\left(\ell-m+1\right)\left(\ell-m+2\right)}{\left(2\ell+2\right)\left(2\ell+3\right)}}Y_{\ell+1,m-1}(\hat{n})\\
-\sqrt{\frac{\left(\ell-m+1\right)\left(\ell+m+1\right)}{(\ell+1)\left(2\ell+3\right)}}Y_{\ell+1,m}(\hat{n})\\
\sqrt{\frac{(\ell+m+1)(\ell+m+2)}{(2\ell+2)(2\ell+3)}}Y_{\ell+1,m+1}(\hat{n})
\end{pmatrix},\\
V_{\ell m}^{a}(\hat{n}) & =\begin{pmatrix}-\sqrt{\frac{\left(\ell-m+1\right)\left(\ell+m\right)}{2\ell(\ell+1)}}Y_{\ell,m-1}(\hat{n})\\
\frac{m}{\sqrt{\ell(\ell+1)}}Y_{\ell.m}(\hat{n})\\
\sqrt{\frac{\left(\ell-m\right)\left(\ell+m+1\right)}{2\ell(\ell+1)}}Y_{\ell,m+1}(\hat{n})
\end{pmatrix},\\
W_{\ell m}^{a}(\hat{n}) & =\begin{pmatrix}\sqrt{\frac{\left(\ell+m-1\right)\left(\ell+m\right)}{2\ell\left(2\ell-1\right)}}Y_{\ell-1,m-1}(\hat{n})\\
\sqrt{\frac{\left(\ell-m\right)\left(\ell+m\right)}{\ell\left(2\ell-1\right)}}Y_{\ell-1,m}(\hat{n})\\
\sqrt{\frac{\left(\ell-m-1\right)\left(\ell-m\right)}{2\ell\left(2\ell-1\right)}}Y_{\ell-1,m+1}(\hat{n})
\end{pmatrix},
\end{align}

Using a transformation matrix 
\begin{equation}
[e_{a=+,z,-}^{\mu=x,y,z}]=\begin{pmatrix}-1/\sqrt{2} & 0 & 1/\sqrt{2}\\
-i/\sqrt{2} & 0 & -i/\sqrt{2}\\
0 & 1 & 0
\end{pmatrix},
\end{equation}
these harmonics can be rotated to cartesian coordinates 
\begin{equation}
e_{a}^{\mu}Z_{\ell,m}^{a}(\hat{n}),\quad Z=U,V,W,
\end{equation}
which corresponds to the harmonics used in the main text.

\subsection{Kernel coefficients for general $q$}

As we turn to compute kernel coefficients, 
\begin{equation}
\left[K_{\ell}^{q}(\omega)\right]_{XZ}=\frac{1}{2\ell+1}\sum_{m}\int d^{3}rd^{3}r^{\prime}\sqrt{g(r)}\sqrt{g\left(r^{\prime}\right)}X_{\ell,m}^{\mu*}(\hat{n})\mathcal{K}_{\mu\mu^{\prime}}^{q}(r,r^{\prime})Z_{\ell,m}^{\mu^{\prime}}(\hat{n}^{\prime})e^{i\omega\bigl(\tau-\tau^{\prime}\bigr)},
\end{equation}
with $X,Z\in\{U,V,W\}$, the real-space kernels can also be worked
out in cartesian coordinates 

\[
K_{\mu\mu^{\prime}}^{q}(r,r^{\prime})=\text{tr}\left[\gamma_{\mu}G_{q}(r,r^{\prime})\gamma_{\mu^{\prime}}G_{q}^{\dagger}(r,r^{\prime})\right],\quad\gamma_{\mu},\gamma_{\mu^{\prime}}=\left(\sigma_{x},\sigma_{y},\sigma_{z}\right).
\]

In the limit $r^{\prime}\to0$, where half of the harmonics can be
eliminated (\ref{eq:harmonic_z}), the various functions at play in
our computation can be rewritten as (\ref{eq:conjugate})

\begin{align}
G_{q}\left(r,0\right) & =\left(2\times2\text{ matrix}\right)_{\tau\tau^{\prime}}\propto\sum_{\ell^{\prime}}Y_{q,\ell^{\prime}+\delta\ell^{\prime},-q+\delta\mathfrak{m}^{\prime}}(\hat{n}),\\
K_{\mu\mu'}^{q}\left(r,0\right) & =\left(3\times3\text{ matrix}\right)_{\mu\mu^{\prime}}\propto\sum_{\ell^{\prime},\ell^{\prime\prime}}Y_{q,\ell^{\prime}+\delta\ell^{\prime},-q+\delta\mathfrak{m}^{\prime}}(\hat{n})Y_{-q,\ell^{\prime\prime}+\delta\ell^{\prime\prime},q+\delta\mathfrak{m}^{\prime\prime}}(\hat{n}),\\
Z_{\ell,m}^{\mu*}(\hat{n})Z_{\ell,m}^{\mu^{\prime}}\left(\hat{z}\right) & =\left(3\times3\text{ matrix}\right)^{\mu\mu^{\prime}}\propto Y_{0,\ell+\delta\ell_{Z},\delta\mathfrak{m}}(\hat{n}),
\end{align}
where
\begin{align}
\delta\ell^{\prime},\ell^{\prime\prime} & \in\{-1,0\},\\
\delta\mathfrak{m}^{\prime},\delta\mathfrak{m}^{\prime\prime} & \in\{-1,0,1\},\\
\delta\ell_{Z} & =\{-1,0,1\},\;Z\in\{W,V,U\},\\
\delta\mathfrak{m} & \in\{-2,-1,0,1,-2\},
\end{align}
As claimed in the main text, the kernel coefficients take the form
\begin{equation}
\int d^{3}r\sqrt{g(r)}K_{\mu\mu^{\prime}}^{q}\left(r,0\right)Z_{\ell,m}^{\mu*}(\hat{n})Z_{\ell,m}^{\mu^{\prime}}\left(\hat{z}\right)\sim\sum_{\ell^{\prime},\ell^{\prime\prime}}\int d\hat{n}Y_{q,\ell^{\prime}+\delta\ell^{\prime},-q+\delta\mathfrak{m}^{\prime}}(\hat{n})Y_{-q,\ell^{\prime\prime}+\delta\ell^{\prime\prime},q+\delta\mathfrak{m}^{\prime\prime}}(\hat{n})Y_{0,\ell+\delta\ell_{Z},\delta\mathfrak{m}}(\hat{n}).
\end{equation}

\section{Results for the $q=1/2$ computations\label{sec:min_charge_results}}

We review specific results that concern the $q=1/2$ computation in
Sec.~\ref{subsec:min_charge}. Using the differential equation defining
a Legendre Polynomial
\begin{equation}
P_{\ell}^{\prime\prime}(x)-\frac{1}{1-x^{2}}\left[2xP_{\ell}^{\prime}(x)-\ell(\ell+1)P_{\ell}(x)\right]=0,
\end{equation}
and the equivalent relation for $Q_{q,\ell}^{\prime\prime}(x)$~\citep{pufu_anomalous_2014} 

\begin{equation}
Q_{q,\ell}^{\prime\prime}(x)+\frac{1}{1+x}Q_{q,\ell}^{\prime}(x)+\frac{1}{1-x^{2}}\left[\ell^{2}-\frac{2q^{2}}{1+x}\right]Q_{q,\ell}(x)=0,
\end{equation}
the integrals appearing in the $q=1/2$ computation in Sec.~\ref{subsec:min_charge}
can be reformulated in the form
\begin{equation}
\int dx\left[a_{\ell,\ell^{\prime},\ell^{\prime\prime}}(x)P_{\ell}(x)+b_{\ell,\ell^{\prime},\ell^{\prime\prime}}(x)P_{\ell}^{\prime}(x)\right]Q_{q,\ell^{\prime}}(x)Q_{q,\ell^{\prime\prime}}(x).
\end{equation}
Specifically, for $q=1/2$, we obtain 
\begin{equation}
\begin{array}{l}
\mathcal{I}_{1}^{D}=J_{0}\left[\ell(\ell+1)-\ell^{\prime2}-\ell^{\prime\prime2}+\frac{1}{2}\right]-J_{1}-J_{2},\\
\mathcal{I}_{2}^{D}=-J_{0},\\
\mathcal{I}_{1}^{T}=-\left(J_{1}-J_{2}\right),\\
\mathcal{I}_{1}^{E}=\left(J_{1}-J_{2}\right)\left[\ell(\ell+1)-\ell^{\prime2}-\ell^{\prime\prime2}+\frac{1}{2}\right],\\
\mathcal{I}_{2}^{E}=-J_{1}-J_{2},\\
\mathcal{I}_{1}^{B}=\ell(\ell+1)\left[\ell(\ell+1)J_{0}-2J_{2}\right]-\left[J_{1}-J_{2}+\ell(\ell+1)J_{0}\right]\left[\ell^{\prime2}+\ell^{\prime\prime2}-\frac{1}{2}\right],\\
\mathcal{I}_{2}^{B}=J_{1}+J_{2}-\ell(\ell+1)J_{0},
\end{array}
\end{equation}
where 
\begin{equation}
\begin{aligned}J_{0}\left(\ell,\ell^{\prime},\ell^{\prime\prime}\right) & =\int_{-1}^{1}dx\frac{1}{1-x}P_{\ell}(x)Q_{1/2,\ell^{\prime}}(x)Q_{1/2,\ell^{\prime\prime}}(x),\\
J_{1}\left(\ell,\ell^{\prime},\ell^{\prime\prime}\right) & =\int_{-1}^{1}dx\frac{1}{1-x}P{}_{\ell}^{\prime}(x)Q_{1/2,\ell^{\prime}}(x)Q_{1/2,\ell^{\prime\prime}}(x),\\
J_{2}\left(\ell,\ell^{\prime},\ell^{\prime\prime}\right) & =\int_{-1}^{1}dx\frac{x}{1-x}P{}_{\ell}^{\prime}(x)Q_{1/2,\ell^{\prime}}(x)Q_{1/2,\ell^{\prime\prime}}(x).
\end{aligned}
\end{equation}
The result for these integrals was obtained in~\citep{pufu_anomalous_2014}
\footnote{Our definitions have for the $J_{i}$ have an extra factor $4\pi/(2\ell+1)$
since we defined them with $P_{\ell}(x)$ and not $F_{\ell}(x)=(2\ell+1)P_{\ell}(x)/(4\pi)$
as in Ref.~\citep{pufu_anomalous_2014}.}
\begin{align}
J_{0}(\ell,\ell_{1}+1/2,\ell_{2}+1/2) & =-\frac{\left(\ell_{1}+1/2\right)\left(\ell_{2}+1/2\right)\begin{pmatrix}\ell & \ell_{1} & \ell_{2}\\
0 & 0 & 0
\end{pmatrix}\begin{pmatrix}\ell+1 & \ell_{1} & \ell_{2}\\
0 & 1 & -1
\end{pmatrix}}{4\pi^{2}\sqrt{\ell_{1}(\ell_{1}+1)\ell_{2}(\ell_{2}+1)}},\label{eq:J0}\\
J_{1}(\ell,\ell_{1}+1/2,\ell_{2}+1/2) & =\frac{\sqrt{\ell(\ell+1)}\left(\ell_{1}+1/2\right)\left(\ell_{2}+1/2\right)\begin{pmatrix}\ell & \ell_{1} & \ell_{2}\\
0 & 0 & 0
\end{pmatrix}}{8\pi^{2}\sqrt{\ell_{1}(\ell_{1}+1)\ell_{2}(\ell_{2}+1)}}\nonumber \\
 & \times\left[\sqrt{(\ell+2)(\ell+3)}\begin{pmatrix}\ell+1 & \ell_{1} & \ell_{2}\\
-2 & 1 & 1
\end{pmatrix}-\sqrt{\ell(\ell+1)}\begin{pmatrix}\ell+1 & \ell_{1} & \ell_{2}\\
0 & 1 & -1
\end{pmatrix}\right]\label{eq:J1}\\
J_{2}(\ell,\ell_{1}+1/2,\ell_{2}+1/2) & =\frac{\sqrt{\ell(\ell+1)}\left(\ell_{1}+1/2\right)\left(\ell_{2}+1/2\right)\begin{pmatrix}\ell & \ell_{1} & \ell_{2}\\
0 & 0 & 0
\end{pmatrix}}{8\pi^{2}\sqrt{\ell_{1}(\ell_{1}+1)\ell_{2}(\ell_{2}+1)}}\nonumber \\
 & \times\left[\sqrt{(\ell-1)(\ell+2)}\begin{pmatrix}\ell+1 & \ell_{1} & \ell_{2}\\
-2 & 1 & 1
\end{pmatrix}-\sqrt{\ell(\ell+1)}\begin{pmatrix}\ell+1 & \ell_{1} & \ell_{2}\\
0 & 1 & -1
\end{pmatrix}\right]\label{eq:J2}
\end{align}
where $\ell'=\ell_{1}+1/2$ and $\ell''=\ell_{2}+1/2$.

\section{Remainder coefficients\label{sec:remainders}}

When computing kernel coefficients in Secs.~\ref{subsec:min_charge}
and~\ref{subsec:general_q}, we deal with regularized sums as 
\begin{equation}
\sum_{\ell^{\prime}=q+1}^{\infty}\biggl[-\alpha^{Z}+\sum_{\ell^{\prime\prime}=q+1}^{\infty}k_{\ell,\ell^{\prime},\ell^{\prime\prime}}^{q,Z}(\omega)\biggr],
\end{equation}
which is the general-$q$ version of the sum in Eq.~(\ref{eq:half_NZM}).
As in the main text, $Z\in\{D,T,E,B\}$. Setting a numerical cutoff
$\ell'_{c}=200+q$, the remainder of the sum is obtained analytically,
as discussed in Sec.\ref{subsec:min_charge}
\begin{align}
\sum_{\ell'=\ell_{c}^{\prime}+1}^{\infty}\biggl[-\alpha^{Z}+\sum_{\ell^{\prime\prime}=q+1}^{\infty}k_{\ell,\ell^{\prime},\ell^{\prime\prime}}^{q,Z}(\omega)\biggr] & =\sum_{\ell^{\prime}=\ell_{c}^{\prime}+1}^{\infty}\biggl[\sum_{p=2}^{k}c_{\ell,p}^{q,Z}(\omega)\left(\ell^{\prime}\right)^{-p}\biggr]=\sum_{p=2}^{k}c_{\ell,p}^{q,Z}(\omega)\zeta\left(p,\ell_{c}^{\prime}+1\right).
\end{align}
In our computations, we obtained the remainders down to order $\left(\ell^{\prime}\right)^{-18}$.
To obtain the coefficients, the expansion in $1/\ell^{\prime}$ must
be carried out, which in turn requires fixing $\ell$ and $q$. The
resulting expansion then yields the analytic dependence on $\omega$,
while the dependence on $\ell$ and $q$ is found by fitting many
coefficients with specific values of $\ell$ and $q$. The coefficients
we find $c_{\ell,p}^{q,Z}(\omega)$ are polynomials of
$\omega^{2}$, $\ell(\ell+1)\equiv\ell_{2}$, $q^{2}$
(the scalar-gauge kernel $Z=T$ has an extra factor of $q$)

\begin{align}
D:\quad & \frac{\zeta\left(2,\ell'_{c}+1\right)}{16\pi}\left(\ell_{2}-4q^{2}+2\omega^{2}\right)+\frac{\zeta\left(4,\ell'_{c}+1\right)}{256\pi}\left(7\ell_{2}^{2}+\ell_{2}\left(24q^{2}+8\omega^{2}-2\right)-8\left(-6q^{2}\omega^{2}+6q^{4}+\omega^{4}\right)\right)\nonumber \\
 & \quad{}+\frac{\zeta\left(6,\ell'_{c}+1\right)}{1024\pi}\biggl(13\ell_{2}^{3}+2\ell_{2}^{2}\left(35q^{2}-3\left(\omega^{2}+3\right)\right)+4\ell_{2}\left(5q^{2}\left(4\omega^{2}+3\right)+30q^{4}-9\omega^{4}-2\omega^{2}+1\right)\nonumber \\
 & \quad{}+8\left(30q^{4}\omega^{2}-10q^{2}\omega^{4}-20q^{6}+\omega^{6}\right)\biggr)+\dots\\
\nonumber \\
T:\quad & \frac{\zeta\left(4,\ell'_{c}+1\right)}{16\pi}q\ell_{2}+\frac{\zeta\left(6,\ell'_{c}+1\right)}{128\pi}q\ell_{2}\left(6\ell_{2}+20q^{2}-9\omega^{2}-8\right)+\dots\\
\nonumber \\
E:\quad & \frac{\zeta\left(2,\ell'_{c}+1\right)}{32\pi}\left(\ell_{2}-4q^{2}+2\omega^{2}\right)+\frac{\zeta\left(4,\ell'_{c}+1\right)}{256\pi}\left(2\ell_{2}^{2}+\ell_{2}\left(12q^{2}+\omega^{2}-2\right)+8q^{2}\left(3\omega^{2}-4\right)-24q^{4}-2\omega^{2}\left(2\omega^{2}+3\right)\right)\nonumber \\
 & \quad{}+\frac{\zeta\left(6,\ell'_{c}+1\right)}{4096\pi}\biggl(11\ell_{2}^{3}+4\ell_{2}^{2}\left(20q^{2}-3\left(\omega^{2}+2\right)\right)+4\ell_{2}\left(2q^{2}\left(5\omega^{2}-78\right)+60q^{4}-8\omega^{4}-6\omega^{2}+3\right)\nonumber \\
 & \quad{}-16\left(q^{4}\left(80-30\omega^{2}\right)+q^{2}\left(10\omega^{4}-21\omega^{2}-88\right)+20q^{6}-\omega^{2}\left(\omega^{4}+5\omega^{2}+5\right)\right)\biggr)+\dots\\
\nonumber \\
B:\quad & \frac{\zeta\left(2,\ell'_{c}+1\right)}{32\pi}\left(\ell_{2}-4q^{2}+2\omega^{2}\right)+\frac{\zeta\left(4,\ell'_{c}+1\right)}{256\pi}\left(5\ell_{2}^{2}+\ell_{2}\left(12q^{2}+7\omega^{2}-4\right)+8q^{2}\left(3\omega^{2}-4\right)-24q^{4}-2\omega^{2}\left(2\omega^{2}+3\right)\right.\nonumber \\
 & \quad{}+\frac{\zeta\left(6,\ell'_{c}+1\right)}{4096\pi}\biggl(41\ell_{2}^{3}+4\ell_{2}^{2}\left(50q^{2}-3\left(\omega^{2}+8\right)\right)+4\ell_{2}\left(q^{2}\left(70\omega^{2}-208\right)+60q^{4}-28\omega^{4}-14\omega^{2}+17\right)\nonumber \\
 & \quad{}-16\left(q^{4}\left(80-30\omega^{2}\right)+q^{2}\left(10\omega^{4}-21\omega^{2}-88\right)+20q^{6}-\omega^{2}\left(\omega^{4}+5\omega^{2}+5\right)\right)\biggr)+\dots
\end{align}
This dependence on $\omega^{2}$and $\ell(\ell+1)$ was
also observed for global monopoles in the context of the $O(N)$ model~\citep{pufu_monopoles_2013}

\section{Only zero modes contribution in the kernels\label{sec:Only-zero-modes}}

The contribution of the zero modes in the Green's function~(\ref{eq:GF_monopole_harmonics_1})
is

\begin{align}
G_{q;0}(r,r^{\prime}) & =\frac{i}{2}\sum_{m=-q}^{q-1}\left[0\quad S_{q,q,m}^{-}\right]\left(\text{sgn}(\tau-\tau^{\prime})\begin{pmatrix}-1 & 0\\
0 & 1
\end{pmatrix}\right)\left[\begin{array}{c}
0\\
\left(S_{q,q,m}^{-}\right)^{\dagger}
\end{array}\right].
\end{align}
We focus on the contribution of this function to kernel coefficients.
For instance, for the scalar-scalar kernel coefficient~(\ref{eq:Dlo}),
we have
\begin{align}
D_{\ell}^{q}(\omega) & =\frac{4\pi}{2\ell+1}\int_{r}e^{i\omega\tau}\text{tr}\left[G_{q;0}(r,r^{\prime})G_{q;0}^{\dagger}(r,r^{\prime})\right]\sum_{m}Y_{\ell m}^{*}(\hat{n})Y_{\ell m}(\hat{n}^{\prime})+\dots
\end{align}
where the ellipses indicates terms including non-zero modes contributions
that have already been incorporated in the main text computations.
This ``zero-zero mode contribution'' has no fermion energy $\sqrt{\ell^{2}-q^{2}}\to0$
and the Green's function can be factorized as
\begin{equation}
G_{q;0}\left(r,0\right)=\text{sgn}\left(\tau\right)\tilde{G}_{q;0}\left(\hat{n},0\right).
\end{equation}
The ``zero-zero mode contribution '' to the kernel coefficient then
simplifies to 
\begin{equation}
\frac{4\pi}{2\ell+1}\times2\pi\delta(\omega)\int d\hat{n}\text{tr}\left[\tilde{G}_{q;0}\left(\hat{n},0\right)\tilde{G}_{q;0}^{\dagger}\left(\hat{n},0\right)\right]\sum_{m}Y_{\ell m}^{*}(\hat{n})Y_{\ell m}(\hat{n}^{\prime}).
\end{equation}
Let's then write kernel as 
\begin{equation}
D_{\ell}^{q}(\omega)=C_{\ell}^{q}\delta(\omega)+\text{regular terms}
\end{equation}
Hence, looking only at the scalar-scalar kernel, the contribution
around $\omega=0$ is \footnote{The contribution of the $q=0$ kernel in the denominator doesn't matter
since it can be isolated and it vanishes $\lim_{\epsilon\to0}\int_{-\epsilon}^{\epsilon}\frac{d\omega}{2\pi}\ln D_{\ell}^{0}(\omega)\to0$.}
\begin{align}
\frac{1}{2}\int_{-\epsilon}^{\epsilon}\frac{d\omega}{2\pi}\left(\sum_{\ell=0}^{\infty}(2\ell+1)\ln\left[D_{\ell}^{q}(\omega)\right]\right) & =\frac{1}{2}\int_{-\epsilon}^{\epsilon}\frac{d\omega}{2\pi}\left(\sum_{\ell=0}^{\infty}(2\ell+1)\ln\left[C_{\ell}^{q}\delta(\omega)+\text{``reg.''}\right]\right) \nonumber \\
 & =\frac{1}{2}\int_{-\epsilon}^{\epsilon}\frac{d\omega}{2\pi}\left(\sum_{\ell=0}^{\infty}(2\ell+1)\ln\left[1+\frac{C_{\ell}^{q}\delta(\omega)}{\text{``reg.''}}\right]\right) \nonumber\\
 & =\frac{1}{2}\mathcal{J}\int_{-\epsilon}^{\epsilon}\frac{d\omega}{2\pi}\left(\sum_{\ell=0}^{\infty}(2\ell+1)\ln\left[1+\delta(\omega)\right]\right)
\end{align}
where we changed variable and $\mathcal{J}$ is the resulting Jacobian.
We also used $\lim_{\epsilon\to0}\int_{-\epsilon}^{\epsilon}\frac{d\omega}{2\pi}\ln\text{``reg.''}\to0$.
It turns out that the remaining term also vanishes
\begin{equation}
\int_{-\epsilon}^{\epsilon}\frac{d\omega}{2\pi}\ln\left[1+\delta(\omega)\right]=0,\label{eq:integral_00}
\end{equation}
as we show in what follows.

The logarithm in Eq.~(\ref{eq:integral_00}) can be rewritten as
an integral 
\begin{align}
\int d\omega\ln\left[1+\delta(\omega)\right] & =\int d\omega\int_{0}^{1}dt\frac{\delta(\omega)}{1+t\delta(\omega)}.
\end{align}
We then exchange the order of integration and obtain a vanishing integral
\begin{equation}
\begin{split}\int d\omega\ln\left[1+\delta(\omega)\right]=\int_{0}^{1}dt\int d\omega\left[\delta(\omega)\times\frac{1}{1+t\delta(\omega)}\right] & =\int_{0}^{1}dt\frac{1}{1+t\delta(0)}\\
 & =\int_{0}^{1}dt\begin{cases}
1 & t=0\\
0 & t\neq0
\end{cases}=0.
\end{split}
\end{equation}

\subsection*{Generalization }

The kernel coefficients, with the ``zero-zero mode contribution'' explicitly
included, can be written as
\begin{align}
D_{\ell}^{q}(\omega) & =C_{D}\delta(\omega)+\text{reg}_{D},\\
H_{\ell}^{q,T}(\omega) & =C_{HT}\delta(\omega)+\text{reg}_{FT},\\
K_{\ell}^{q,E}(\omega) & =C_{KE}\delta(\omega)+\text{reg}_{KE},\\
K_{\ell}^{q,B}(\omega) & =0+\text{reg}_{KB}.
\end{align}
In fact, it turns out that $C_{D}=-iC_{HT}=-C_{KE}$, but this is
not necessary for the argument that follows. Again, let us consider
the calculation of the scaling dimension in $\QEDGN$ near $\omega=0$.
Once again, we can ignore the denominator:

\begin{equation}
\frac{1}{2}\int_{-\epsilon}^{\epsilon}\frac{d\omega}{2\pi}\left\{ \left[\ln D_{0}^{q}(\omega)\right]+\sum_{\ell=1}^{\infty}\left(2\ell+1\right)\ln\left[K_{\ell}^{q,B}(\omega)\left(D_{\ell}^{q}(\omega)K_{\ell}^{q,E}(\omega)+\left(1+\frac{\omega^{2}}{\ell(\ell+1)}\right)\left|F_{\ell}^{q,T}(\omega)\right|{}^{2}\right)\right]\right\} .
\end{equation}
The $K_{\ell}^{q,B}(\omega)$ is also regular and can be
removed. Also, we already showed that $\int_{-\epsilon}^{\epsilon}\frac{d\omega}{2\pi}\ln D_{\ell}^{q}(\omega)=0$,
we can use this to eliminate the $\ell=0$ contribution. We are left
with 
\begin{equation}
\frac{1}{2}\int_{-\epsilon}^{\epsilon}\frac{d\omega}{2\pi}\left(\sum_{\ell=1}^{\infty}(2\ell+1)\ln\left[D_{\ell}^{q}(\omega)K_{\ell}^{q,E}(\omega)+\left(1+\frac{\omega^{2}}{\ell(\ell+1)}\right)\left|F_{\ell}^{q,T}(\omega)\right|{}^{2}\right]\right).
\end{equation}
Let us now consider the argument of the logarithm
\begin{align}
D_{\ell}^{q}(\omega)K_{\ell}^{q,E}(\omega)+\left(1+\frac{\omega^{2}}{\ell(\ell+1)}\right)\left|F_{\ell}^{q,T}(\omega)\right|{}^{2} & =\left[C_{D}\text{reg}_{KE}+C_{D}\text{reg}_{KE}+2\Re\left(C_{FT}\text{reg}_{FT}^{*}\right)\right]\delta(\omega)\nonumber \\
 & \quad+\left[C_{D}C_{KE}+|C_{FT}|^{2}\right]\delta(\omega)^{2}+\left[\text{reg}_{D}\text{reg}_{KE}+\left(1+\frac{\omega^{2}}{\ell(\ell+1)}\right)\left|\text{reg}_{FT}\right|^{2}\right]\nonumber \\
 & \equiv a\delta(\omega)^{2}+b\delta(\omega)+c\nonumber \\
 & =a\left[\delta(\omega)-f_{1}\right]\left[\delta(\omega)-f_{2}\right].
\end{align}
Then, the scaling dimension correction near $\omega=0$ can be written
as 
\begin{equation}
\frac{1}{2}\sum_{\ell=1}^{\infty}(2\ell+1)\int_{-\epsilon}^{\epsilon}\frac{d\omega}{2\pi}\left(\ln\left[\delta(\omega)-f_{1}\right]+\ln\left[\delta(\omega)-f_{2}\right]\right)=0+0.
\end{equation}

\section{Fitting procedure for anomalous dimensions\label{sec:fitting}}

\begin{figure*}
\hfill{}\centering\subfigure[]{\includegraphics[height=5.25cm]{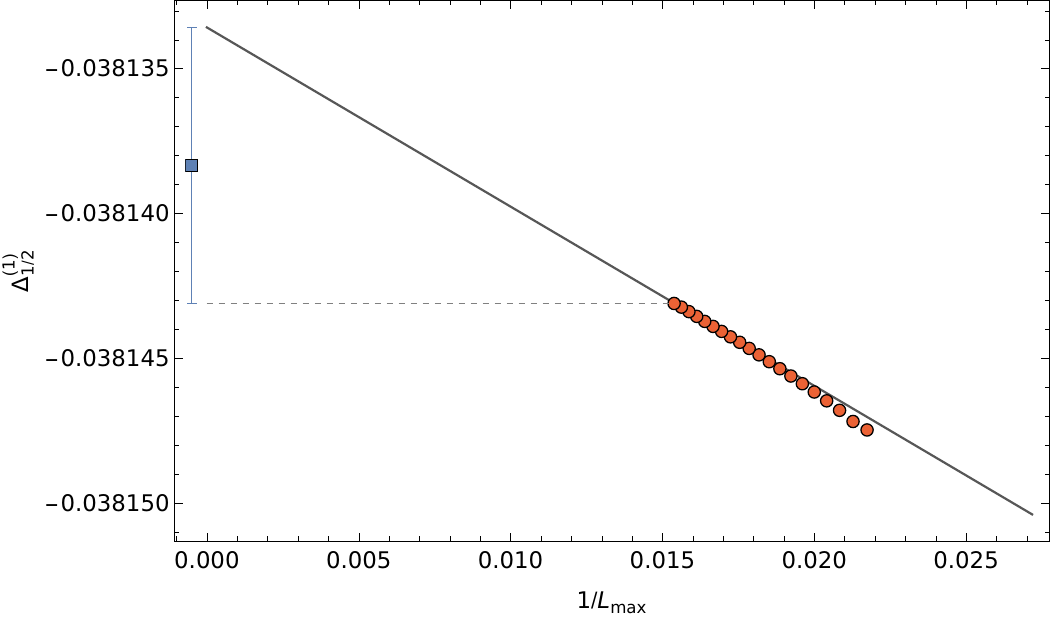}}\hfill{}\subfigure[]{\includegraphics[height=5.25cm]{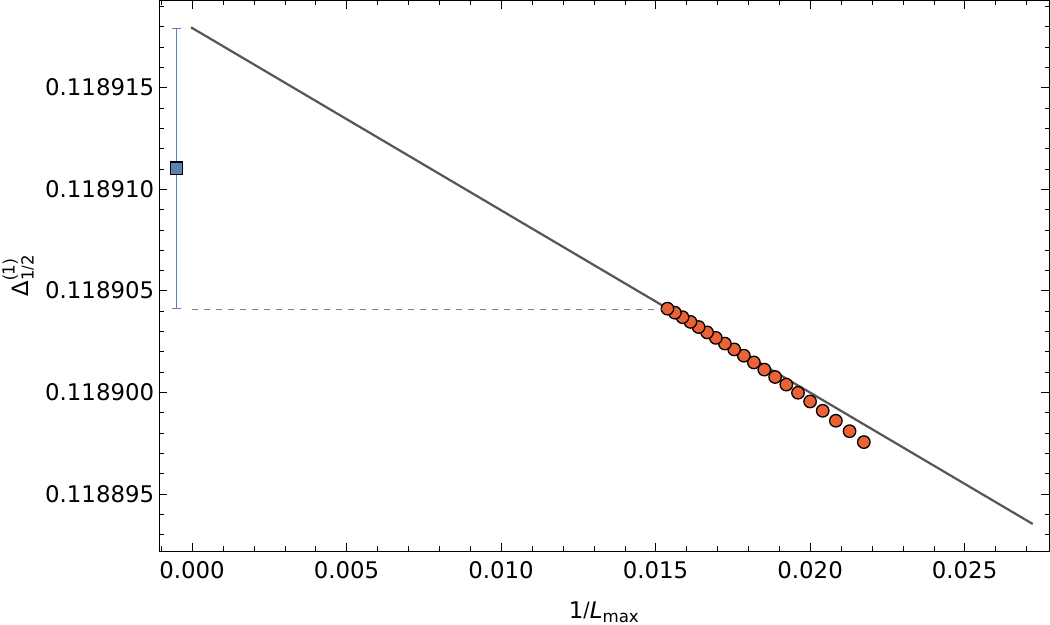}}\hfill{}

\caption{\label{fig:method_charge-1}Fitting procedure of anomalous dimensions
of the $q=1/2$ monopole $\Delta_{1/2}^{(1)}$ in (a) $\protect\QED$
; (b) $\protect\QEDGN$. The points are obtained with fits $\sum_{i=0}^{k}c_{i;k;L_{\max}}L^{-i}$
with $L\in[L_{\max}-5, L_{\max}]$. The solid line is a linear fit
in $1/L_{\max}$ with $L_{\max}\in${[}61,  65{]}. The point with the error bar corresponds to the
anomalous dimension computed with Eq.(\ref{eq:delta_mean}) and shown
in Eq.\ (\ref{eq:delta_mean_half})}
\end{figure*}

The method used to determine the monopole anomalous dimensions and
estimate the errors is described here for $\QED$ and $\QEDGN$ models. To estimate the error, we vary
the maximal cutoff $L_{\max}$~(\ref{eq:rel_cutoff}) of the dataset
used to extrapolate the anomalous dimension to $L\to\infty$.\footnote{The cutoff $\ell_{c}^{\prime}$ we use has a negligible
contribution to the uncertainty, thanks to the very precise expansion,
up to $1/\ell^{\prime18}$, of the remainder. } 
\begin{enumerate}
\item Compute the anomalous dimension up to the cutoff $L_{\max}$ (for instance, this is $L_{\max}=65$ for $q=1/2$)
\item Extrapolate the behaviour as $L\to\infty$ with a fit $\sum_{i=0}^{k}c_{i;k;L_{\max}}L^{-i}$
with polynomial order $k=4$ (for $q=1/2$, we use $L\in[L_{\max}-10, L_{\max}]$).
\item Repeat step 2 with smaller values of $L_{\max}$, computing four more
times the scaling dimension. For instance, with $q=1/2$, we repeat  with $L_{\max}\in[61,  65]$.
\item Extrapolate the behaviour as $L_{\max}\to\infty$ with a linear fit
in $1/L_{\max}$, $\tilde{c}_{0}+\tilde{c}_{1}/L_{\max}$, using the
five anomalous dimensions obtained. The fit and the anomalous dimensions are shown in Fig.~\ref{fig:method_charge-1}
for the case $q=1/2$. Additional points down to $L_{\max}=45$ are also displayed in Fig.~\ref{fig:method_charge-1} to discuss the behaviour later on.
\item Compare the anomalous dimension obtained with the maximal value of $L_{\max}$, that we note in what follows as $L_{\max}^*$ (for $q=1/2$, $L_{\max}^*=65$)  and the extrapolated value at $L_{\max}\to\infty$
and estimate the anomalous dimension as 
\begin{equation}
\Delta_{q}^{(1)}=\frac{1}{2}\left(\left.\Delta_{q}^{(1)}\right|_{L_{\max}=L_{\max}^*}+\left.\Delta_{q}^{(1)}\right|_{L_{\max}\to\infty}\right)\pm\frac{1}{2}\left(\left.\Delta_{q}^{(1)}\right|_{L_{\max}=L_{\max}^*}-\left.\Delta_{q}^{(1)}\right|_{L_{\max}\to\infty}\right).\label{eq:delta_mean}
\end{equation}
This result with
 the error bar is shown in Fig.~\ref{fig:method_charge-1}
for the case $q=1/2$ and is given by:
\begin{equation}
\Delta_{1/2,\QED}^{(1)}=-0.038138(5),\quad\Delta_{1/2,\QEDGN}^{(1)}=0.118911(7).\label{eq:delta_mean_half}
\end{equation}
\end{enumerate}
We emphasize that the data used for the extrapolation in step 4 was
itself the result of the extrapolation in step 2 (and step 3). The
extrapolated value at $L_{\max}\to\infty$ is therefore used as a guiding value. Had we taken a dataset with smaller values of $L_{\max}$, the fitted line would simply overshoot the one currently presented in Fig.~\ref{fig:method_charge-1}. The anomalous dimension in Eq.~\ref{eq:delta_mean} would have more extreme values and thus a greater error. 

\begin{figure*}
\hfill{}\centering\subfigure[]{\includegraphics[height=5.3cm]{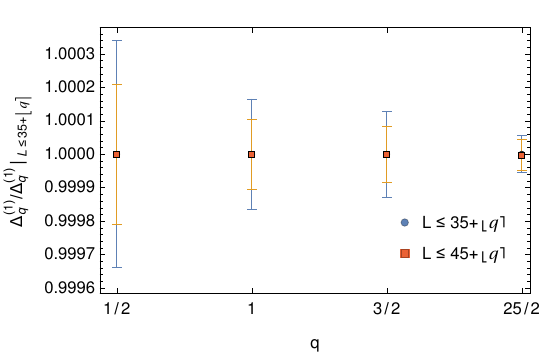}}\hfill{}\subfigure[]{\includegraphics[height=5.3cm]{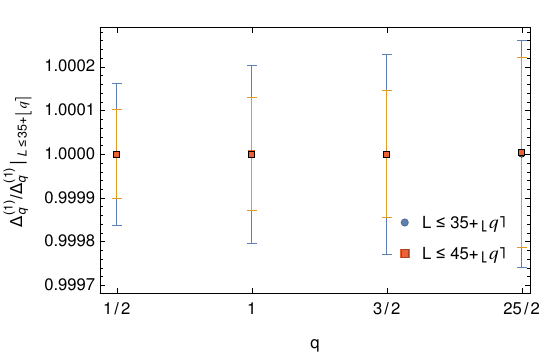}}\hfill{}

\caption{\label{fig:method_charge-1_L35_L45}Normalized anomalous dimensions
in (a) $\protect\QED$ ; (b) $\protect\QEDGN$. There are two sets
of scaling dimensions obtained for $L_{\max}\in[31+\lfloor q\rceil, 35+\lfloor q\rceil]$
($L\protect\leq35+\lfloor q\rceil$) and $L_{\max}\in[41+\lfloor q\rceil,45+\lfloor q\rceil]$
($L\protect\leq45+\lfloor q\rceil$). The anomalous dimensions are
normalized as $\Delta_{q}^{(1)}/\Delta_{q}^{(1)}|_{L\protect\leq35+\lfloor q\rceil}$.}
\end{figure*}

To further characterize the effect that the size of $L_{\max}$ has on the anomalous dimensions,  we also consider the cases $q=1, 3/2$ and $q=25/2$ where we used  a cutoff $L_{\max}=45+\lfloor q\rceil$. \footnote{Although we do obtain $q=25/2$ anomalous dimensions for the largest cutoff, the corresponding values  and errors presented in Tab.\ref{tab:scaling_all} are obtained with cutoff $L_{\max}=35+\lfloor q\rceil$} 
For those charges obtained with a larger cutoff, we can restrain our dataset and obtain the anomalous dimensions with  $L_{\max}=35+\lfloor q\rceil$. The results with $L_{\max} = 45+\lfloor q\rceil$
are slightly more precise and very similar to those with $L_{\max} = 35+\lfloor q\rceil$. As shown in Fig.~\ref{fig:method_charge-1_L35_L45}, the drift of the anomalous dimension as $L_{\max}$ is increased  is very small relatively  to the estimated errors, which indicates the stability of our method.

\begin{figure*}
\hfill{}\centering\subfigure[]{\includegraphics[height=5.25cm]{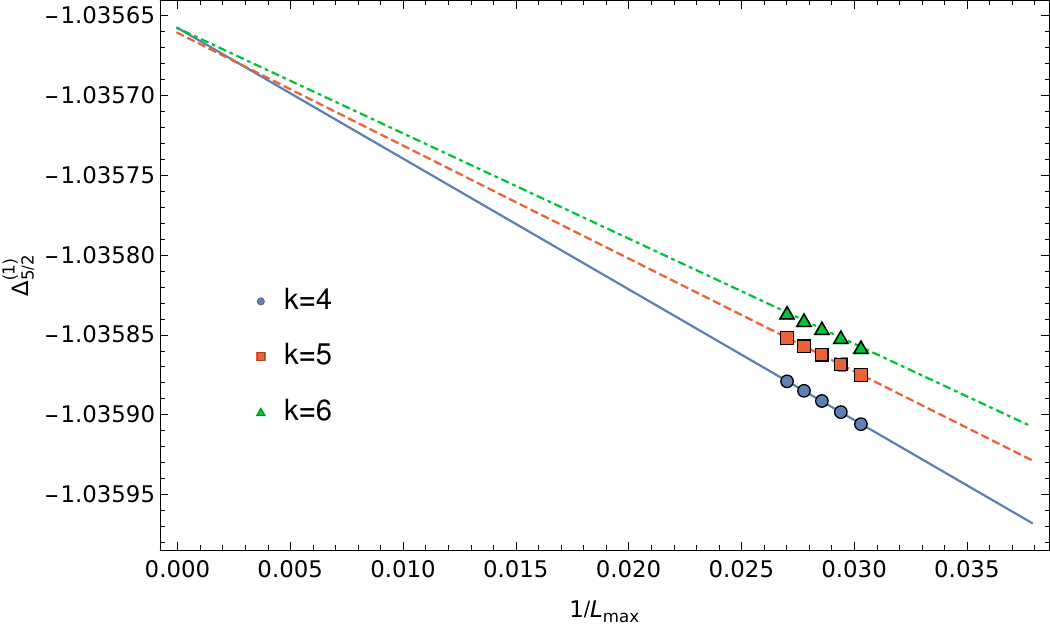}}\hfill{}\subfigure[]{\includegraphics[height=5.25cm]{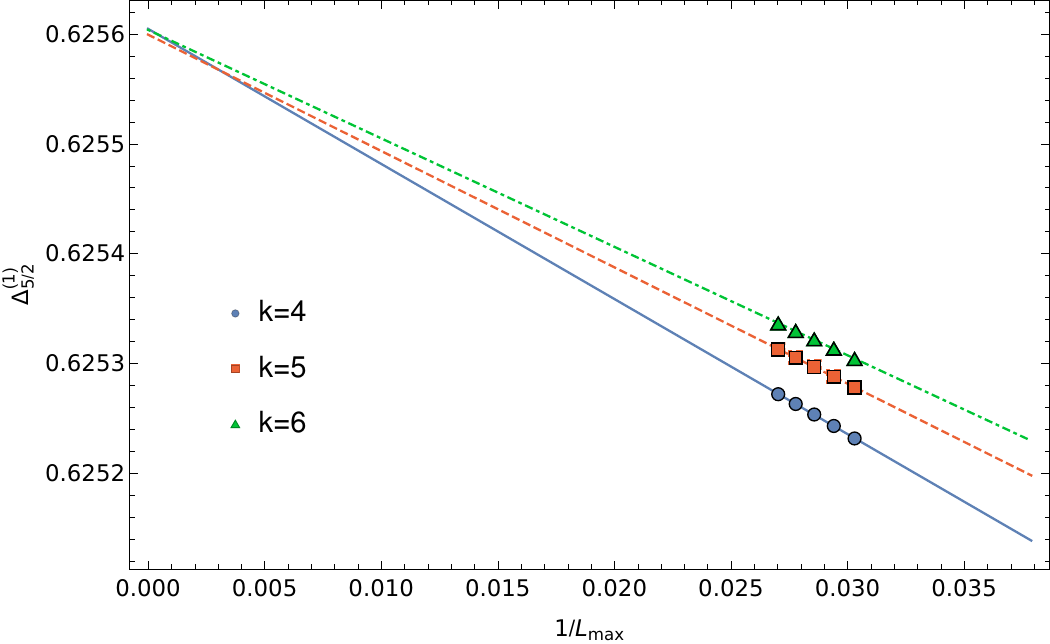}}\hfill{}

\caption{\label{fig:method_charge-1_456}Fitting procedure for anomalous dimensions
of the $q=5/2$ monopole $\Delta_{5/2}^{(1)}$ in (a) $\protect\QED$
; (b) $\protect\QEDGN$. The points are obtained with fits $\sum_{i=0}^{k}c_{i;k;L_{\max}}L^{-i}$
with $L\in[L_{\max}-\delta_{k}, L_{\max}]$ with $\delta_{k}=\{5,10,14\}$
for $k=\{4,5,6$\}. Each set of five points is obtained by varying
$L_{\max}\in${[}33, 37{]}. Solid, dashed, and dot-dashed lines are
linear fits in $1/L_{\max}$ of the $k=4,5,6$ results. }
\end{figure*}

The same procedure was also used for different fitting functions $\sum_{i=0}^{k}c_{i;k;L_{\max}}L^{-i}$
with higher polynomial order $k=5,6$, as shown in Fig.~\ref{fig:method_charge-1_456}.  We find a similar behaviour
and more precise results. However, these fits demand a larger dataset.
For larger $q$ (and therefore larger maximal cutoff since the cutoff increases with $\lceil q \rceil$), a similar behaviour remains. However, the size of datasets needs to
be increased. This was also observed for $q=1/2$ when comparing relativistic cutoffs $L_{\max}$ of different sizes. It may indicate that errors are overfitted for smaller
datasets with higher-order fits, as the effect is less important for
$k=4$. We used the quartic fit for all of the anomalous dimensions
quoted in this work.

\section{Monopole scaling dimensions for $1/2\leq q\leq 13$ \label{sec:scaling_13}}

\begin{table}[H]
\caption{\label{tab:scaling_all}Scaling dimension of monopole operators at
leading-order and next-to-leading order in $1/N$ in $\protect\QED$, 
$\protect\QEDGN$ and $\protect\QEDZ$ models. The leading-order result is the same in
all models. The scaling dimension is $2N\Delta_{q}^{(0)}+\Delta_{q}^{(1)}$.}

\begin{minipage}[t]{0.475\columnwidth}%
\begin{ruledtabular}%
\begin{tabular}{ccccc} 
 $q$ & 
$\Delta_{q}^{(0)}$ & 
$\Delta_{q,\QED}^{(1)}$ & 
$\Delta_{q,\QEDGN}^{(1)}$ \vspace{0.25em} &
$\Delta_{q,\QEDZ}^{(1)}$  \\ \hline  
$1/2$ & $0.26510$ & $-0.038138(5)$ & $0.118911(7)$ & $0.102846(9)$ \\ 
$1$ & $0.67315$ & $-0.19340(3)$ & $0.23561(4)$ & $0.18663(4)$ \\ 
$3/2$ & $1.18643$ & $-0.42109(4)$ & $0.35808(6)$ & $0.26528(7)$ \\ 
$2$ & $1.78690$ & $-0.70482(9)$ & $0.4879(2)$ & $0.3426(2)$ \\ 
$5/2$ & $2.46345$ & $-1.0358(2)$ & $0.6254(2)$ & $0.4202(3)$ \\ 
$3$ & $3.20837$ & $-1.4082(2)$ & $0.7705(3)$ & $0.4989(3)$ \\ 
$7/2$ & $4.01591$ & $-1.8181(2)$ & $0.9229(3)$ & $0.5789(4)$ \\ 
$4$ & $4.88154$ & $-2.2623(3)$ & $1.0824(4)$ & $0.6605(4)$ \\ 
$9/2$ & $5.80161$ & $-2.7384(3)$ & $1.2488(4)$ & $0.7439(5)$ \\ 
$5$ & $6.77309$ & $-3.2445(3)$ & $1.4218(5)$ & $0.8290(6)$ \\ 
$11/2$ & $7.79338$ & $-3.7788(4)$ & $1.6013(5)$ & $0.9160(6)$ \\ 
$6$ & $8.86025$ & $-4.3401(4)$ & $1.7869(6)$ & $1.0048(7)$ \\ 
$13/2$ & $9.97175$ & $-4.9269(4)$ & $1.9786(7)$ & $1.0955(8)$ \\ 
\end{tabular}
\end{ruledtabular}%
\end{minipage}$\hfill$%
\begin{minipage}[t]{0.47\columnwidth}%
\begin{ruledtabular}%
\begin{tabular}{ccccc} 
 $q$ & 
$\Delta_{q}^{(0)}$ & 
$\Delta_{q,\QED}^{(1)}$ & 
$\Delta_{q,\QEDGN}^{(1)}$ \vspace{0.25em} &
$\Delta_{q,\QEDZ}^{(1)}$  \\ \hline  
$7$ & $11.12616$ & $-5.5384(5)$ & $2.1762(7)$ & $1.1881(8)$ \\ 
$15/2$ & $12.32195$ & $-6.1735(5)$ & $2.3794(7)$ & $1.2825(9)$ \\ 
$8$ & $13.55772$ & $-6.8314(5)$ & $2.5882(8)$ & $1.3787(9)$ \\ 
$17/2$ & $14.83223$ & $-7.5113(6)$ & $2.8024(9)$ & $1.477(1)$ \\ 
$9$ & $16.14432$ & $-8.2125(6)$ & $3.0219(9)$ & $1.577(2)$ \\ 
$19/2$ & $17.49296$ & $-8.9345(6)$ & $3.2466(9)$ & $1.678(2)$ \\ 
$10$ & $18.87719$ & $-9.6766(7)$ & $3.476(1)$ & $1.781(2)$ \\ 
$21/2$ & $20.29609$ & $-10.4383(7)$ & $3.711(2)$ & $1.886(2)$ \\ 
$11$ & $21.74886$ & $-11.2191(7)$ & $3.950(2)$ & $1.993(2)$ \\ 
$23/2$ & $23.23472$ & $-12.0186(7)$ & $4.195(2)$ & $2.102(2)$ \\ 
$12$ & $24.75294$ & $-12.8363(8)$ & $4.444(2)$ & $2.212(2)$ \\ 
$25/2$ & $26.30286$ & $-13.6719(8)$ & $4.697(2)$ & $2.323(2)$ \\ 
$13$ & $27.88383$ & $-14.5249(8)$ & $4.955(2)$ & $2.437(2)$ \\ 
\end{tabular}\end{ruledtabular}%
\end{minipage}
\end{table}

\twocolumngrid

\bibliographystyle{apsrev}
\bibliography{main}

\end{document}